\documentclass[]{jfm}
\usepackage{graphicx}
\graphicspath{{figures}}
\usepackage{epstopdf,epsfig}
\usepackage{newtxtext}
\usepackage{newtxmath}
\usepackage{natbib}
\usepackage{hyperref}

\usepackage{multirow}
\usepackage{epstopdf, epsfig}
\usepackage[dvipsnames]{xcolor}

\usepackage{tikz,siunitx} 
\usepackage{pgfplots}

\usepackage{longtable}
\usepackage{mwe} 
\setcitestyle{authoryear,square,semicolon,sort}
\usepackage{amsmath, amssymb}
\usepackage{mathrsfs}
\usepackage{placeins}
\usepackage{upgreek}
\usepackage{xkeyval,xcolor}
\usepackage{natbib}
\newcommand\cites[1]{\citeauthor{#1}'s\ (\citeyear{#1})}

\usepackage{mathtools}

\usepackage{dirtytalk}

\usepackage[utf8]{inputenc} 

\DeclareUnicodeCharacter{00A0}{~}

\hypersetup{
    colorlinks = true,
    urlcolor   = blue,
    citecolor  = black,
}
\newcommand{\RomanNumeralCaps}[1]
\linenumbers

\pgfplotsset{compat=1.17} 

\def\mathclap#1{\text{\hbox to 0pt{\hss$\mathsurround=0pt#1$\hss}}}

\makeatletter
\newlength{\sfp@hseplen}\newlength{\sfp@vseplen}
\define@cmdkey{subfigpos}[sfp@]{pos}[ul]{}
\define@cmdkey{subfigpos}[sfp@]{font}[\small]{}
\define@cmdkey{subfigpos}[sfp@]{vsep}[1\baselineskip]{\setlength{\sfp@vseplen}{\sfp@vsep}}
\define@cmdkey{subfigpos}[sfp@]{hsep}[35pt]{\setlength{\sfp@hseplen}{\sfp@hsep}}
\newcommand{\subfigimg}[3][,]{%
  \setkeys{Gin,subfigpos}{pos,font,vsep,hsep,#1}
  \setbox1=\hbox{\includegraphics{#3}}
  \ifnum\pdfstrcmp{\sfp@pos}{ul}=0
    \leavevmode\rlap{\usebox1}
    \rlap{\hspace*{\sfp@hsep}\raisebox{\dimexpr\ht1-\sfp@vsep}{\sfp@font{#2}}}
    \phantom{\usebox1}
  \else\ifnum\pdfstrcmp{\sfp@pos}{ur}=0
    \leavevmode\usebox1
    \llap{\raisebox{\dimexpr\ht1-\sfp@vsep}{\sfp@font{#2}}\hspace*{\sfp@hsep}}
  \else\ifnum\pdfstrcmp{\sfp@pos}{lr}=0
    \leavevmode\usebox1
    \llap{\raisebox{\sfp@vsep}{\sfp@font{#2}}\hspace*{\sfp@hsep}}
  \else
    \leavevmode\rlap{\usebox1}
    \rlap{\hspace*{\sfp@hseplen}\raisebox{\sfp@vsep}{\sfp@font{#2}}}
    \phantom{\usebox1}
  \fi\fi\fi
}
\makeatother

\newcommand{\revPJ}[1]{\textcolor{black}{{#1}}}
\newcommand{\revJFM}[1]{\textcolor{black}{{#1}}}
\newcommand{\revthree}[1]{\textcolor{black}{{#1}}}

\title{Effects of porous substrates on the structure of turbulent boundary layers}

\author{P. Jaiswal\aff{1} \and B. Ganapathisubramani\aff{1}\corresp{\email{g.bharath@soton.ac.uk}}}

\affiliation{\aff{1} Aerodynamics \& Flight Mechanics Group, University of Southampton, Southampton, SO16 7QF, UK}

\begin{document}
\maketitle

\begin{abstract}

Three different porous substrates (with different pore sizes, $s$ and permeabilities, $K$) are used to examine their effect on the structure of boundary layer flow over them. The flow is characterised with single-point hot-wire measurements as well as planar Particle Image Velocimetry. In order to elucidate differences in shallow and deep flows past porous substrate, foams with two different thickness ($h$) are used (for all three substrates). A wide range of Friction Reynolds number ($2000<Re_\tau<13500$) and Permeability based Reynolds number ($1<Re_K< 50$) are attained. For substrates with $Re_K \sim 1$, the flow behaviour remains similar to flow over impermeable smooth walls and as such Townsend's hypothesis remains valid. \revPJ{Very-large-scale motions are observed~over permeable foams even when the $Re_K > 1$.} In contrast, a substantial reduction in velocity disturbances and associated length scales is achieved for \revPJ{permeable foams with intermediate values of pore density and relative foam thickness ($h/s$)}, which \revPJ{affects} outer-layer similarity. As \revPJ{permeability} is increased \revPJ{by increasing pore size}, the foam becomes sparse relative to viscous scales~\revPJ{at high Reynolds number}. For such foams, the flow conforms to outer-layer similarity and is more akin to flow over rough surfaces. \revPJ{Permeability attenuates the wavelengths associated with the outer-layer peak.}    





\end{abstract}

\begin{keywords}
Authors should not enter keywords on the manuscript, as these must be chosen by the author during the online submission process and will then be added during the typesetting process (see \href{https://www.cambridge.org/core/journals/journal-of-fluid-mechanics/information/list-of-keywords}{Keyword PDF} for the full list).  Other classifications will be added at the same time.
\end{keywords}

{\bf MSC Codes }  {\it(Optional)} Please enter your MSC Codes here


\section{\label{sec:level1}Introduction}

{\color{black}{Turbulent flow over and past porous surfaces is encountered in many engineering problems, ranging from flow over forest canopies \citep{Finnigan2000} to flows over and past river beds \citep{yovogan2013effect}. \revPJ{Porous surfaces have also been used for trailing-edge noise control \citep{carpio2019experimental}}. This makes the understanding of the flow behaviour over porous surfaces crucial. For a porous substrate, \cite{rosti2015direct1} showed that compared to porosity small changes in permeability can significantly alter the turbulence dynamics. The effects of wall-permeability for flows over and past porous foams was further studied in detail by \cite{Hahn2002,breugem2006influence}. \cite{breugem2006influence} suggested that an isotropic porous substrate could be fully defined by three length scales, which are the square root of material permeability $\sqrt{K}$, the substrate thickness $h$, and the characteristic size of the `roughness' elements composing the substrate $d_p$. \cite{breugem2006influence} stated that the effect of permeability on the flow is isolated if three conditions are \revJFM{met}: i) the wall thickness is larger than the flow penetration into the substrate, ii) the roughness Reynolds number $Re_d=d_p U_\tau/\nu$ is small ($Re_d<<70$, where $U_\tau$ is the skin-friction velocity and $\nu$ is the kinematic viscosity) and iii) the permeability Reynolds number $Re_K=\sqrt{K} U_\tau /\nu$ is high ($Re_K>>1$).

Studies by \cite{breugem2006influence} and \cite{manes2011turbulent} were able to meet the above mentioned criterion. Therefore, the effect of surface roughness can be neglected. It was shown that permeable wall can substantially alter eddy blocking, quasi-streamwise vortices and no slip at the wall \citep[see][for instance]{breugem2006influence}. The modification of these properties by permeable wall, which are trademarks of the turbulent boundary layer, leads to a departure from outer-layer similarity in velocity statistics. Furthermore, the impact of permeable wall can also be felt by large-scale structures, and leads to non-existence of logarithmic mean velocity law \citep{breugem2006influence}.  Similarly, the wall roughness, by itself, can also alter turbulence dynamics by destroying non-linear self-sustaining cycles of turbulence \citep{jimenez2004turbulent}. Although these mechanisms for permeable and rough surfaces are well reported in the literature, yet their relative contribution and interactions towards boundary-layer scales over a porous material, which is both rough and permeable, remains a matter for further investigation.

One such aspect is the existence of Townsend's outer-layer hypothesis \citep{townsend1980structure} for a porous wall. According to Townsend's hypothesis the outer layer flow is independent of the near-wall region; therefore, for a flow over a wall, the primary effect of the wall is impermeability and no-slip boundary conditions. To this end several studies have demonstrated its validity for smooth walls \citep{chung2014idealised}. Townsend's hypothesis has also been found to be valid for flow over rough walls, provided that the equivalent sand roughness height $k_s$ is small compared to the boundary-layer thickness \citep{jimenez2004turbulent}. Therefore, in contrast to wall-permeability, the surface roughness with a reasonable scale separation (low $k_s/\delta$) does not affect the logarithmic mean profiles and large-scale structures remain intact. In contrast for flows past porous surfaces, \citet{breugem2006influence} and \citet{suga2010effects} found that the outer-layer hypothesis holds for all but the wall-normal velocity component. \cite{breugem2006influence} ascribed the absence of self-similarity in the wall-normal velocity profiles to the weakening of \revPJ{wall blockage}. The weakening of \revPJ{wall blocking} opens a path for inner-outer boundary layer communications through enhanced ejections and sweep \citep{breugem2006influence} compared to a solid impermeable (smooth or rough) wall. \cite{breugem2006influence} argue that the enhanced ejections and sweep are sufficient to nullify Townsend's hypothesis, which requires the absence of inner layer scales \revPJ{influencing} \revPJ{outer layer}. However, they \citep{suga2010effects,breugem2006influence} were unable to decisively conclude if absence of outer-layer scaling is due to permeability or insufficient separation of scales because \cites{breugem2006influence} numerical simulations were performed \revJFM{at  low Reynolds number ($Re_{\tau}<500$)}.


To overcome the limitation of the low Reynolds number that can be achieved with DNS, \cite{manes2011turbulent} performed experimental measurements at a higher Reynolds number \revJFM{($Re_{\tau}>2000$)}. \cites{manes2011turbulent} data confirm the validity of Townsend's outer-layer similarity hypothesis for all the velocity components, for porous foams with negligible surface roughness. However, the thickness of their \citep{manes2011turbulent} porous substrate was much greater than the pore size. \revPJ{As such, the impact of substrate thickness to pore size ratio on overall flow dynamics saturates \citep{sharma2020_dense}. However, the thickness to pore ratio can be an important metric in  vegetated shear flows, as noted by \cite{efstathiou2018mean}. This ratio can also be an important metric also for trailing-edge noise research because the flow can transition from the thick foam limit $h/s>1$, at aerofoil mid-chord, to the thin foam limit $h/s\approx 1$, close to the aerofoil trailing edge. Furthermore, at finite thickness limit, roughness layer can dictate the efficacy of the wall permeability condition \citep{white2007shear}. Therefore, for such applications, further research is required to understand the impact of the $h/s$ ratio on turbulent flows over porous walls}.


%



\revPJ{\cite{efstathiou2018mean} were able to show the effect of substrate thickness on the turbulent boundary layer, and near-wall flow physics by investigating porous materials with different $h/s$ ratio.} 
\cite{efstathiou2018mean} showed that for foams with finite thickness, \revPJ{Townsend’s outer-layer similarity hypothesis remains valid}. The foams tested by \cite{efstathiou2018mean} had small values of permeability based Reynolds number ($Re_K$), especially those at the thick substrate limit. This suggests that permeability based scales were comparable to viscous scales in their study, as such it is unclear if the permeability played a role in setting wall-boundary conditions for the cases tested by \cite{efstathiou2018mean}. In addition to low values of $Re_K$, \cite{efstathiou2018mean} concluded the validity of Townsend's hypothesis solely based on \revPJ{streamwise} velocity statistics. The wall-normal statistics are especially sensitive to permeable surfaces as noted by \cite{breugem2006influence}.~\revPJ{Furthermore, \cite{efstathiou2018mean} have reported only the impact of frontal solidity on velocity statistics, yet as shown by \cite{placidi2018turbulent} at moderate shelter solidity the velocity fluctuations also depends upon the details of the local morphology. As such,~for flows over porous foams with moderate shelter solidity, outer-layer similarity may not hold.~Therefore the validity of Townsend's outer-layer hypothesis for porous (rough and permeable) foams is an open question.}
Is the flow over such porous surfaces analogous to flows over rough surfaces away from the wall? If so, does the outer-layer similarity in velocity statistics holds for such porous foams? Thus the primary objective of the current paper is to test Townsend's outer-layer hypothesis at a high Reynolds number for turbulent flows past porous foam with varying thicknesses, permeability and roughness.


%

A potential similarity between flows past canopies \citep{sharma2020_dense} and foam \citep{efstathiou2018mean} is that as the pore size is increased, a thin substrate limit is achieved where the velocity profile becomes fuller, ultimately resulting in loss of the inflection point. This ensures absence of any Kelvin-Helmholtz instability, which results in reduction in \revPJ{streamwise} velocity spectra compared to cases where \revPJ{the Kelvin-Helmholtz (KH) type flow instability is present.} \revPJ{\cite{kuwata2017direct} also report the presence of KH-type flow instability, which led to pressure fluctuations being correlated in the spanwise direction.} On the one hand, numerical simulations \citep{motlagh2016pod,kuwata2017direct} have been performed at much lower Reynolds numbers compared to experimental studies, on the other hand, experimental studies \citep{efstathiou2018mean} have reported this based only single-point \revPJ{velocity statistics}. Therefore, in the current study, PIV measurements were carried out for each wall topology. PIV inherently shows the flow structures, and one does not have to rely on the assumption of frozen turbulence to recover spatial information from temporal single-point velocity measurements. \revPJ{Therefore, the second objective of this manuscript is to unravel flow structures present in flows over porous foam with varying thicknesses, which can provide direct experimental evidence on the existence or non-existence of KH type instability, and impact of porous substrates on the structure of turbulent boundary layer. To our knowledge, this is the first study of the spatial structure of turbulence over porous foams at high Reynolds number ($Re_{\tau}~>{2000}$).}

As argued by \cite{finnigan2009turbulence} and \cite{manes2011turbulent} the imprint of Kelvin-\revJFM{Helmholtz} instability is best visible in streamwise velocity spectra. While \cite{manes2011turbulent} performed measurements only at the thick foam limits, \cite{efstathiou2018mean} were unable to report near-wall streamwise velocity spectra data due to noise. Additionally, both these measurements were performed at the dense foam limit.~\revPJ{Finally to the best of our knowledge, for flows over porous foams the presence of Very Large Scale Motions (VLSMs) \citep{kim1999very} have never been reported in the literature.~\cite{manes2011turbulent}, argued the scale separation required to detect VLSMs~\citep{hutchins2007large} was not achieved in their experiments. In contrast, \cite{efstathiou2018mean} argue that the VLSMs, like ones found over flows past rough or smooth walls, are absent for flows over porous surfaces.}~Therefore, this study fills the scientific gap by reporting streamwise energy spectra for flows over and past porous foams with varying thicknesses and pore densities at high Reynolds numbers. Thus, third and final objective of this study is to report \revPJ{streamwise} turbulent kinetic energy spectra over a wide range of foam thickness and density, and delineate their impact on the associated time scales and turbulent energy.

The paper is structured as follows: Details on porous materials, the test setup, experimental methods, and the associated measurement uncertainty can be found in section \ref{sec:level2}. \revPJ{Section \ref{sec:level3} describes the methodology while section} \ref{sec:level4} reports the experimental findings in such a way that each of its three subsections aims to investigate the three objectives of the current manuscript. In order to evaluate as to whether the established scaling and similarity laws for (impermeable) rough wall flows can also be applied for permeable wall flow, the velocity statistics scaled with outer-layer variable are shown in section \ref{sec:level4a}. Section~\ref{sec:level4b} seeks to investigate the structure of turbulent flows over and past a porous wall, which is the second objective of this paper. Section~\ref{sec:level4c} quantifies the \revPJ{streamwise} turbulent kinetic energy spectra to elucidate the difference between shallow and deep flows past a porous foams. Section \ref{sec:level5} provides an extended discussion on the findings. Finally, conclusions and perspectives are drawn in section \ref{sec:level6}. 

}}

\section{\label{sec:level2}{Experimental {Set-up and Instrumentation}}:}

\subsection{Porous substrate}

Three foams with varying number of cells per inch, each having two thicknesses, were used in the present study. \revPJ{High porosity open-cell reticulated polyurethane foam with almost constant porosity (empty volume over total volume) $\epsilon \approx 0.97 \pm 0.01$ were used. The porosity values for all three substrates have been summarised in table \ref{tab:Boundary_hw}.} The pore size of the substrates were also obtained, and these are $s \approx 3.84,\,0.89,\,0.25\,$mm ordered from the most porous to the least porous substrate. To measure these material properties all substrates were scanned (CT-scan) with a voxel resolution of $0.056\,$mm in all three dimensions. The data was later imported into the open source FijiJ software and the commercial Avizo software to estimate total porosity and pore size respectively. Total porosity was obtained by applying a \revPJ{Otsu’s \citep{otsu1979threshold} method for thresholding to}~the 3D stack of reconstructed images and pore sizes were obtained by applying iterative threshold and image segmentation. To put the total porosity values measured into context, spherical glass beads in a `uniformly random' form have a porosity ranging from $\epsilon \approx 0.64 $ to $\epsilon \approx 0.36$, cylindrical packings have a porosity range $\epsilon \approx 0.59$ to $\epsilon \approx 0.32 $ and cylindrical fibres $\epsilon \approx 0.919$ to $\epsilon \approx 0.682$ as reviewed in \cite{Macdonald1979}. 

\subsection{Experimental test section}

\begin{figure}
\centering
\includegraphics[scale=0.25]{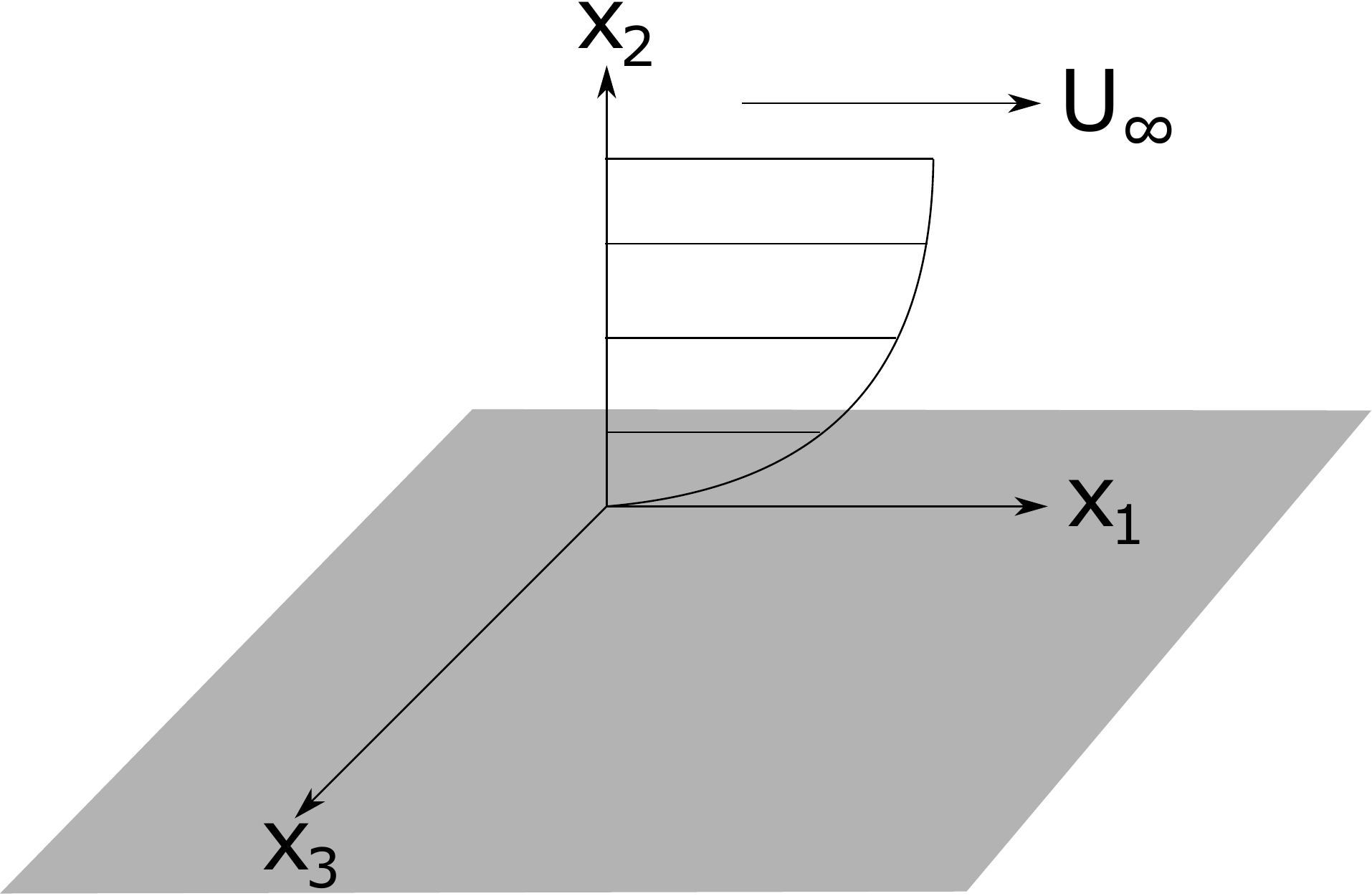}
\caption{Coordinate-system}
\label{fig:cordinates-piv}
\end{figure}

All experimental investigations were conducted at the University of Southampton, in an open-circuit suction type wind tunnel. The wind tunnel has a working section of $4.5$ \revPJ{m} in length, with a $0.9$ \revPJ{m} height, and a $0.6$ \revPJ{m} cross-plane length. Over the bottom wall of the wind tunnel, the turbulent boundary-layer has a zero-pressure-gradient. 
The bottom wall of the test section is covered with the porous substrate. In the present paper, the coordinate system is defined such that the subscripts 1, 2 and 3 are used to define entities in \revPJ{streamwise}, wall-normal and spanwise directions respectively (see figure \ref{fig:cordinates-piv}). Furthermore, upper case letters are used to denote statistical mean, while lower case letters are used to denote the standard deviation. For instance, $U_1$ and $u_2$ denote the mean \revPJ{streamwise} velocity and the standard deviation of wall-normal velocity respectively.

\subsection{Hot-wire measurements}
Hot Wire Anemometry (HWA) measurements were performed with a single wire boundary-layer probe. This single wire, made with tungsten, has a $5~\revPJ{\mu\rm{m}}$ diameter and a sensing length of 1 \revPJ{mm}. The hot wire probe was connected to a
DANTEC Streamline Pro anemometer, which was operating in Constant Temperature Anemometry (CTA) mode, at a fixed overheat ratio of $0.8$. The signals were sampled at a rate of 20 kHz for a duration of at-least 3 minutes, which is equivalent to $\sim 20000$ boundary-layer turnover time. The hot wire measurements allow a direct comparison of the velocity profiles measured using PIV. In the present manuscript single wire measurements were used to quantify temporal scales, and single-point velocity statistics.

\subsection{Particle Image Velocimetry} \label{PIV-setup}

\begin{figure}
\centering
\includegraphics[scale=0.5]{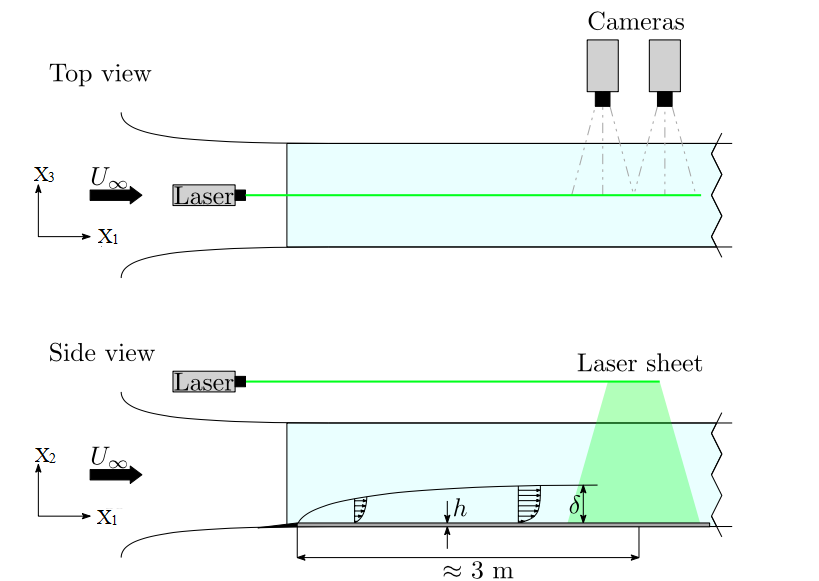}
\caption{A schematic representation Planar PIV setup.}
\label{fig:planar-piv}
\end{figure}

In order to investigate flow structures in the mean flow direction, planar particle image
velocimetry (PIV) measurements were performed. Images for the PIV measurements were taken with Lavision's $16$ Mpix CCD camera. The images were recorded in dual frame mode. For illumination, a dual pulse ND:YAG laser from Litron was used. A Magnum $1200$ fog machine, equipped with a Glycerol-water based solution, was used to generate tracer particles for PIV measurements. The average size of the resulting tracer particles was approximately $1~ \revPJ{\mu m}$. A flat laser sheet of about $1$ \revPJ{mm} was generated by placing a cylindrical lens with a negative focal length after a set of spherical doublets. All the images were processed using Lavision's commercial software Davis 8.2. In total about 2500 \revJFM{image pairs} were acquired at a sampling rate of $0.5$ Hz, which ensures that the individual velocity field is statistically independent. The maximum free stream displacement was around 7 pixels, which implies a random error of $1.5\%$ in PIV measurements. The velocity vector field were computed with a multi-grid cross-correlation scheme, which has a final window size of $24 \times 24$ \revPJ{pixels$^2$}, and an overlap of $50\%$ between the windows. Finally, PIV measurements have been performed at \revJFM{a} distance of approximately $3.0$ \revPJ{m} from the inlet, as shown in figure  \ref{fig:planar-piv}. Given the fact that the PIV measurement domain extends to almost twice the boundary-layer thickness, the streamwise averaged boundary-layer thickness is used to normalise flow quantities throughout the rest of the manuscript. 


\subsection{Skin-friction measurements}
In the present work, a floating element balance presented by \citet{ferreira2018alternative} is used to quantify skin-friction coefficient (\revPJ{C$_f$} ). On the bottom wall at approximately $3.3$ \revPJ{m} from the inlet a floating element balance \citep{ferreira2018alternative} is flush mounted onto the wind tunnel floor. \revPJ{The gap surrounding the balance is taped over to prevent leaks. The porous foams are cut with $0.1$ \revPJ{mm} precision to accommodate onto the surface of balance. Note that this precision is within the size of a pore for all surfaces and therefore its effect on the flow should be negligible. The measurement uncertainty in \revPJ{C$_f$}, for all the cases reported in this study, using the floating element is about $5\%$ \citep[see Appendix of][for instance]{gul2021revisiting}. More information on the drag balance measurements can be found in \cite{Esteban}.}

\subsection{Measurement uncertainty}

The statistical quantities, such as mean and standard deviation, are estimated based on number of independent samples. The number of independent samples are 2500 (number of images) and 20000 (boundary-layer turnover time) for PIV and HWA respectively. Based on the number of independent samples, uncertainty in the averaging of statistical quantities can be estimated. Averaging uncertainty, for all the statistical quantities reported, are expressed with $95\%$ ($20:1$ odds) confidence \citep[see][for instance]{Glegg_book}. 

\begin{table}
\caption{Uncertainty quantification for various measured quantities.}
\label{errortable}       
\begin{tabular}{p{7cm}p{6cm}}
\hline\noalign{\smallskip}
Quantity Measured  & Uncertainty ($95$ $\%$ confidence) \\
\hline
Tunnel Inlet velocity &  $1$ $ \% $ $U_\infty$ \\
Random error mean velocity Planar PIV &  $ \sim 0.1$ $ \% $ $U_\infty$  \\ 
Averaging uncertainty $R_{ij} = 0.05$ PIV & $4.0 \%$  \\ 
\revPJ{Averaging uncertainty} \revPJ{$u_i^2$}  & \revPJ{$5.6 \%$ $u_i^2$}  \\
Averaging uncertainty 
on autospectra &  $1.96\%$  \\
\revPJ{C$_f$} floating element  &  $5 \%$ \\
\hline
\noalign{\smallskip}
\end{tabular}
\end{table}

Finally, the statistical uncertainty in estimating two-point zero time delay correlation $R_{ij}$ \citep[see][for instance]{benedict1996towards}, $\epsilon_{R_{ij}}$, can be estimated with $95\%$ confidence by: 

\begin{equation} \label{eq:un_lscale}
     \epsilon_{R_{ij}}   =  \frac{2}{ \sqrt{N}}  \times (1-R_{ij}^2)
\end{equation}
where, $N$ is the number of independent samples. 
Finally, the measurement uncertainty for all the flow quantities have been summarised in table \ref{errortable} \revPJ{while the table \ref{tab:Boundary_hw} provides a
summary of experimental conditions}.

\begin{table}
\begin{center}
\resizebox{\columnwidth}{!}{%
\begin{tabular}{c c c c c c c c c c c c c c c} 
 $s$ & \revPJ{$\epsilon$} & $h$ & $K$& $U_\infty$ & $\delta_{99}$ & $U_\tau$ & $Re_\tau$ & $Re_K$  &$s^+$ & $\delta_{99}/|y_d|$ & $k_{s}^+$ & \revPJ{$k_{s}/\delta_{99}$} & $h/s$\\
 (mm) &  & (mm) & ($10^{-9}\,$m$^2$) &(m$/$s) & (cm) & (m$/$s) & & & & & & & \\ \hline
 \multirow{1}{*}{0.245 (90 PPI)}
 & \multirow{1}{*}{0.96} 
 & \multirow{1}{*}{3} & \multirow{1}{*} {2.63} & 10.4 & 7.41 & 0.494 & 2349& 1.63 &7.76 & 2470 & 50.73 & 0.021 & \multirow{1}{*}{12.2}\\
  \hline

 \multirow{2}{*}{0.245 (90 PPI)}
 & \multirow{2}{*}{0.96} 
 & \multirow{2}{*}{15} & \multirow{2}{*}{2.63} & 10.3 & 9.09 &0.479 & 2831 & 1.60 &7.63 & 313.45 & 79.7& 0.028 & \multirow{2}{*}{61.2}\\

 &  & &  & 17.8 &  10.67 &0.926 & 6756&3.25& 15.51 & 1333 & 161.9 & 0.024 &\\ \hline

 \multirow{2}{*}{0.89 (45 PPI)}
 & \multirow{2}{*}{0.97} 
 & \multirow{2}{*}{3} & \multirow{2}{*}{36.2} & 10.3 & 8.44 &0.528 & 2871 & 6.47& 30.27 & 401.9 & 118.3& 0.041 & \multirow{2}{*}{3.3}\\
 &  & & & 26.5 &  9.45 &1.407 & 8545 & 17.20& 80.47 & 525 & 314.4& 0.036 &\\ \hline

 \multirow{2}{*}{0.89 (45 PPI)}
 & \multirow{2}{*}{0.97} 
 & \multirow{2}{*}{15} &  \multirow{2}{*}{36.2} & 9.8 & 9.66 &0.572 & 3716 & 7.32 & 34.23 & 112.3&301.6& 0.081 &\multirow{2}{*}{16.8}\\

 &  & &  & 17.8 &  10.44 &1.060 & 7436 & 13.55&63.39 & 99.4&557.9& 0.075 &\\ \hline

 \multirow{1}{*}{3.84 (10 PPI)}
 & \multirow{1}{*}{0.98} 
 & \multirow{1}{*}{3} &  \multirow{1}{*}{245} &9.8 & 9.21 &0.591 & 3644 & 19.58 &151.93 & 118&367.7 & 0.1 &\multirow{1}{*}{0.78}\\ \hline

 \multirow{2}{*}{3.84 (10 PPI)}
 & \multirow{2}{*}{0.98} 
 & \multirow{2}{*}{15} & \multirow{2}{*}{245} &10.3 & 14.74 &0.781 & 7417 & 24.90 & 193.22 & 30.8&2100& 0.283 & \multirow{2}{*}{3.9}\\

 & & &  &18.4 &  14.71 &1.410 & 13367 & 44.98 & 378.94 & 29.5&3791& 0.283 &\\ \hline

\end{tabular}
}
\end{center}
\caption{Details of the porous-wall experimental data. The friction velocity ($U_\tau$) is obtained from the direct measure of skin friction from the floating element drag balance.} 
\label{tab:Boundary_hw}
\end{table}

\revPJ{\section{Methodology\label{sec:level3}}}

\revPJ{In the present paper, both transitionally and fully rough flows above porous foams will be investigated at high Reynolds number. As the increase in permeability is achieved by increasing the pore size of the foams, the thickness and pore-size ratio range from  $h/s=0.7$ to $h/s=60$. This allows investigation of differences between shallow and deep flows as the thickness is varied. At the same time, increasing or decreasing pores per inch should also permit one to cover the dense and sparse foam limits. If the nominal pore size (s) is taken as the characteristic length scale to compute the roughness Reynolds number $s^+=s U_\tau/\nu$ \citep[see][for instance]{efstathiou2018mean}, then $s^+$ for all but one case appears to be way beyond the condition (Reynolds number based on roughness$<<70$) to decouple permeability from roughness \citep{breugem2006influence}. Therefore, with the possible exception of thicker foam with highest number of cells at lowest free-stream velocity ($U_{\infty}$) tested, all the test cases should experience both permeable and roughness effects. The effects of porous wall and its overarching influence on the structure of turbulent boundary layer will be quantified using single-point and two-point statistics, the spatio-temporal scales and energy spectra. The friction-based Reynolds number ($Re_\tau = U_\tau \delta_{99}/\nu$ where $U_\tau$ is the skin friction velocity and $\delta_{99}$ is the boundary layer thickness) for the present study, was in the range $Re_\tau \approx 2000 - 13500$. The permeability Reynolds number ($Re_K = U_\tau \sqrt{K}/\nu$) was in the range $Re_K \approx 1 - 50$, see the table \ref{tab:Boundary_hw} for details.} 

\revPJ{The equivalent sandgrain roughness ($k_s$) was calculated following the procedure outlined by \cite{Esteban}. Briefly, the equivalent sand grain roughness in wall units ($k_s^+$) using:}

{\begin{equation} 
\revJFM{\Delta U^+ = \frac{1}{\kappa}ln(k_s^+) + {B - B^{\prime}_{FR}}}.
  \label{equivalent_ks}     
\end{equation}}

\revPJ{Here, $\Delta{U}^+$ is the roughness function, which is the downward shift in the log region compared to smooth wall. $B - B^{\prime}_{FR}$ was taken to be \revthree{$-3.5$} following \cite{jimenez2004turbulent}. \cite{Esteban} had calculated the values of equivalent sandgrain roughness assuming a “universal” value of the von Karman constant ($\kappa = 0.39$) in order to decouple the effects of permeability from roughness. This procedure of calculating $ k_s^+$; therefore, amounts to shifting the $\Delta{U}^+$-$k_s^+$ plot until it coincides with the fully rough asymptote, which was obtained for rough pipe flows by \cite{Nikuradse1933}. Following this procedure \citep{Esteban} report that at lowest flow speeds the data does not coincide with the aforementioned fully rough asymptote \citep[see figure 7 (\textit{b})][]{Esteban} for $90$ and $45$ PPI foams despite attaining high values of $k_s^+$, i.e. $k_s^+>70$. This is because the $k_s^+$ defined using the above methodology has contributions from both surface roughness and permeability, as already shown by \cite{Esteban} and \cite{wangsawijaya2023towards_JFM}. Therefore, some of the data reported here correspond to a transitionally rough cases. Finally, the equivalent sand grain roughness normalised by the inner ($k_s^+$), and outer ($k_s/\delta_{99}$) wall units are reported in table \ref{tab:Boundary_hw}.}

\revPJ{As the overall goal is to quantify the effect of increasing wall-permeability  and relative foam thickness on the turbulent boundary-layer. Therefore, wall-permeability  (for a given substrate thickness) was systematically increased at fixed inlet velocity. Although this ensures that the Reynolds number based on fetch ($Re_{x_1}$) is the same for all the cases, the Reynolds number based on inner scales or the K\'arm\'an number is different. This is because for the cases tested, the permeability and roughness based Reynolds number increase simultaneously, and the inner velocity scales with the latter. Nevertheless, HWA measurements were performed at several flow speeds, which permits a broad coverage of parameter space and some iso-K\'arm\'an number data is also available. As evidenced from table \ref{tab:Boundary_hw}, cases over a wide range of Reynolds number have been investigated. \revPJ{Furthermore, some additional HWA and drag-balance measurements were performed at higher free-stream velocities in order to match $Re_{\tau}$ and asses the impact of $Re_{k}$ and $s^{+}$ on velocity statistics}. Large values of $\delta_{99}/|y_d|$ for the three porous cases reported confirms a large separation between inner and outer scales for permeable walls \citep{clifton2008shear}. For references, $|y_d|$ corresponds to the absolute value of zero plane position \citep[see][]{Esteban}. The lowest Reynolds number reported, in the present paper, is higher than most of the previous investigations \citep[compared to][for instance]{efstathiou2018mean}, which permits a clear separation of scales and extending the study to both the transitionally and fully rough regimes.  }


\section{\label{sec:level4}Results}

\begin{figure*}
  \centering
  \begin{tabular}{@{}p{0.5\linewidth}@{\quad}p{0.5\linewidth}@{}}
  \hspace{3.5 cm}
  \subfigimg[width=65 mm,pos=ul,vsep=20pt,hsep=32pt]{}{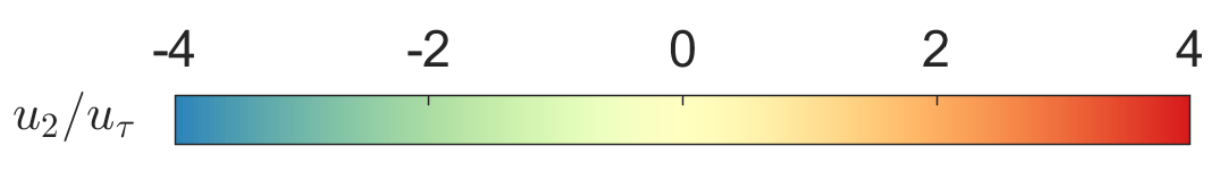} 
    \\
    \subfigimg[width=70 mm,pos=ur,vsep=2pt,hsep=38pt]{(a)}{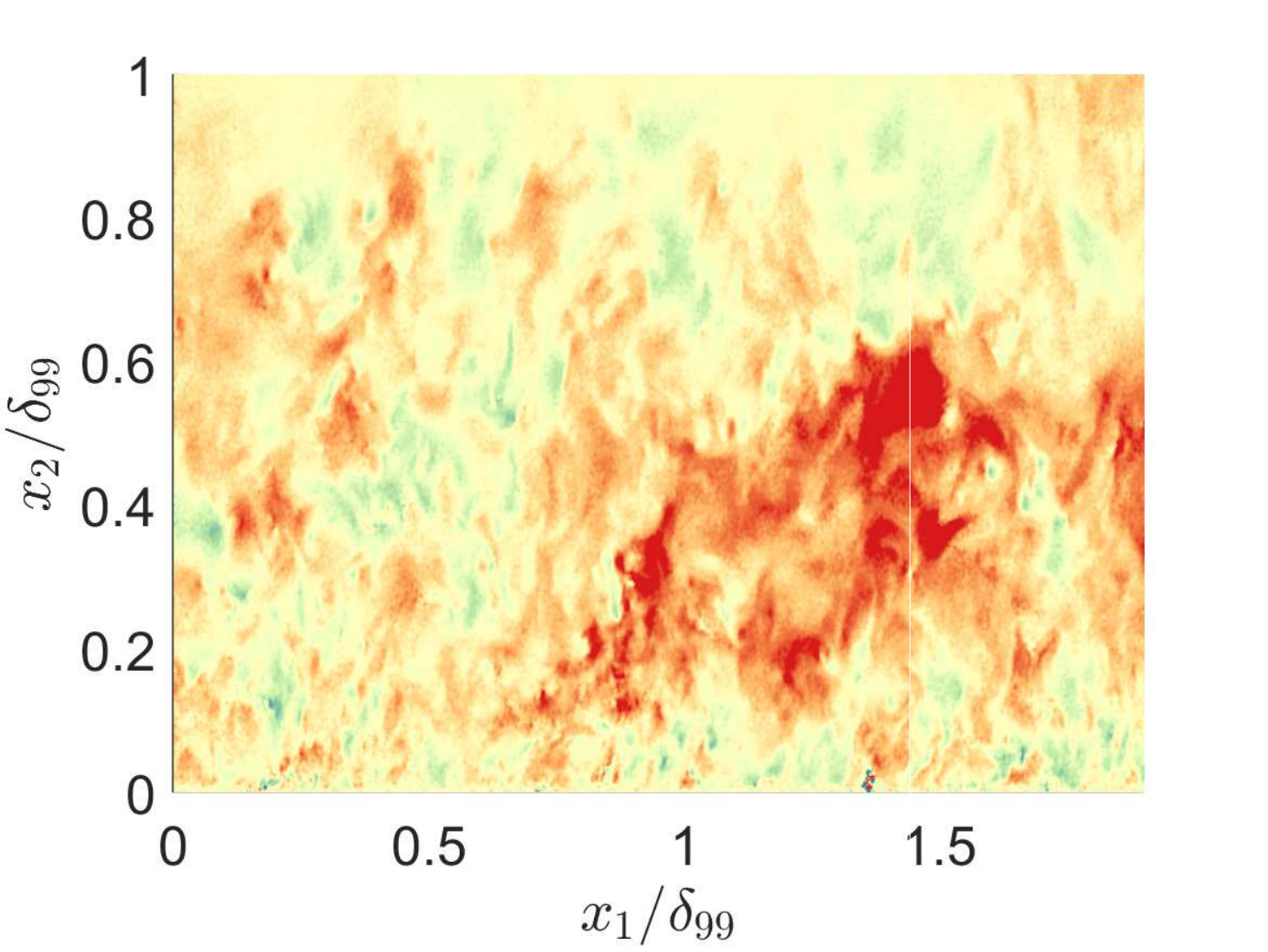} &
    \subfigimg[width=70 mm,pos=ur,vsep=2pt,hsep=38pt]{(b)}{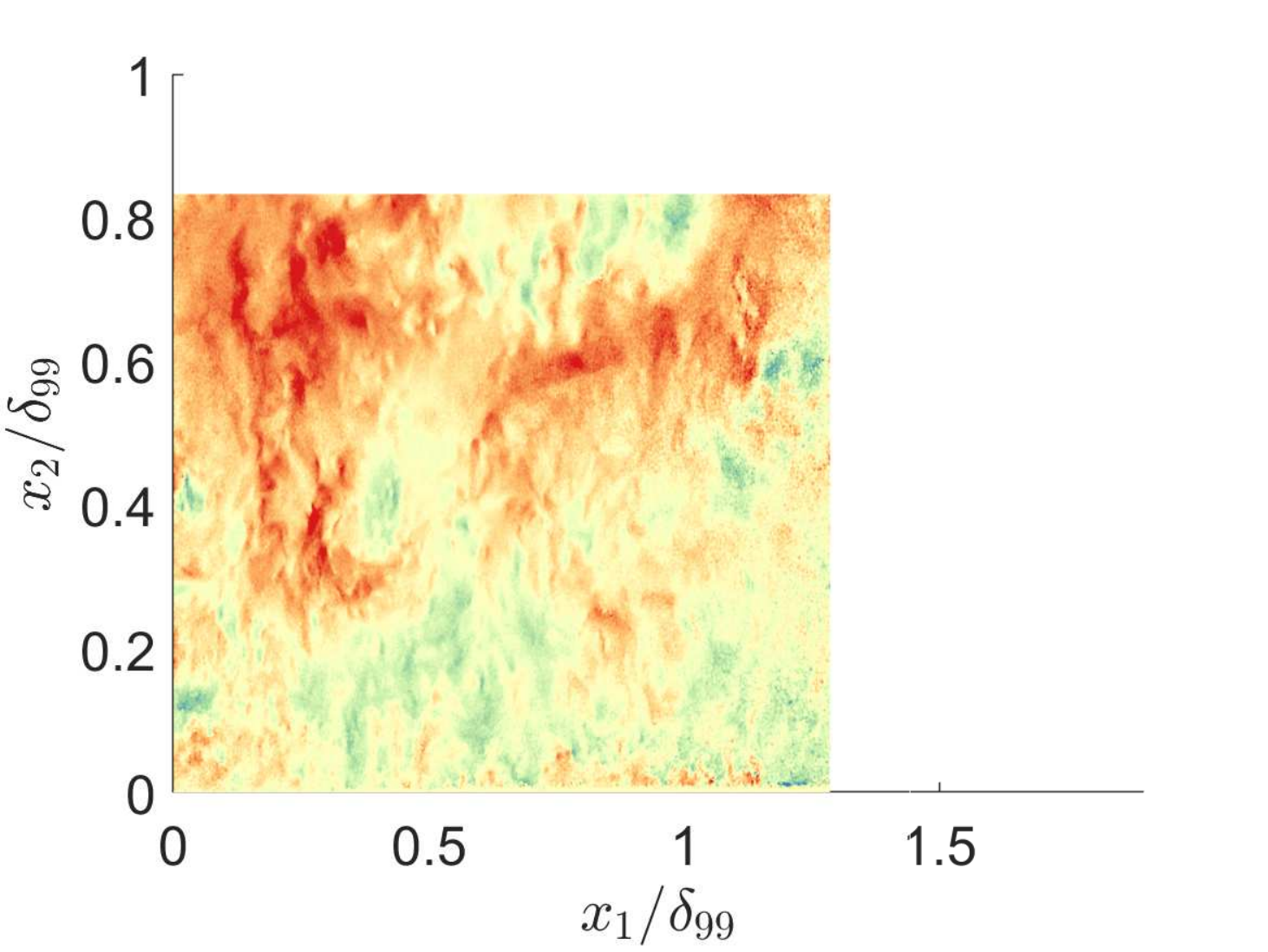} \\
     \subfigimg[width=70 mm,pos=ur,vsep=2pt,hsep=38pt]{(c)}{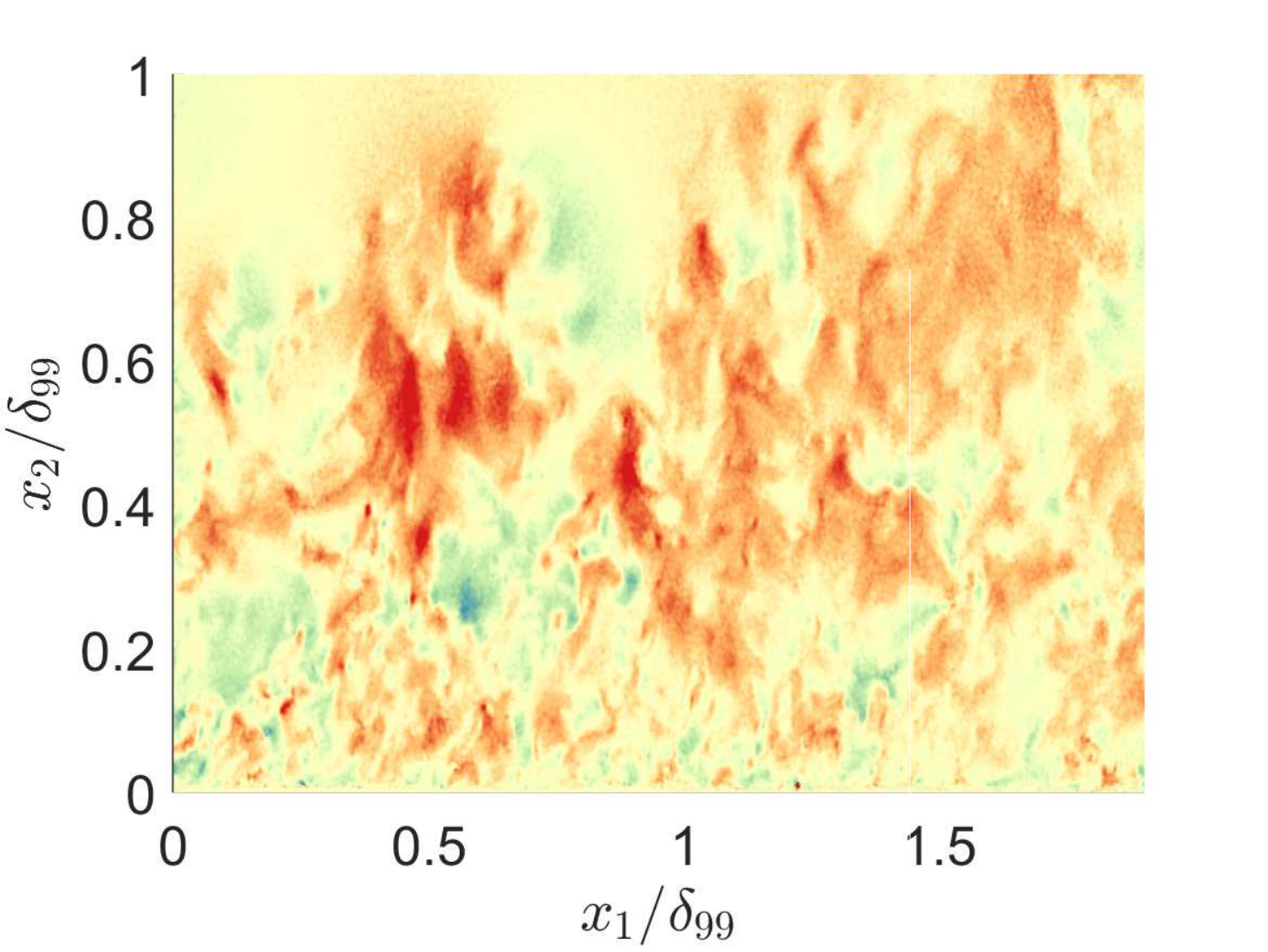} &
    \subfigimg[width=70 mm,pos=ur,vsep=2pt,hsep=38pt]{(d)}{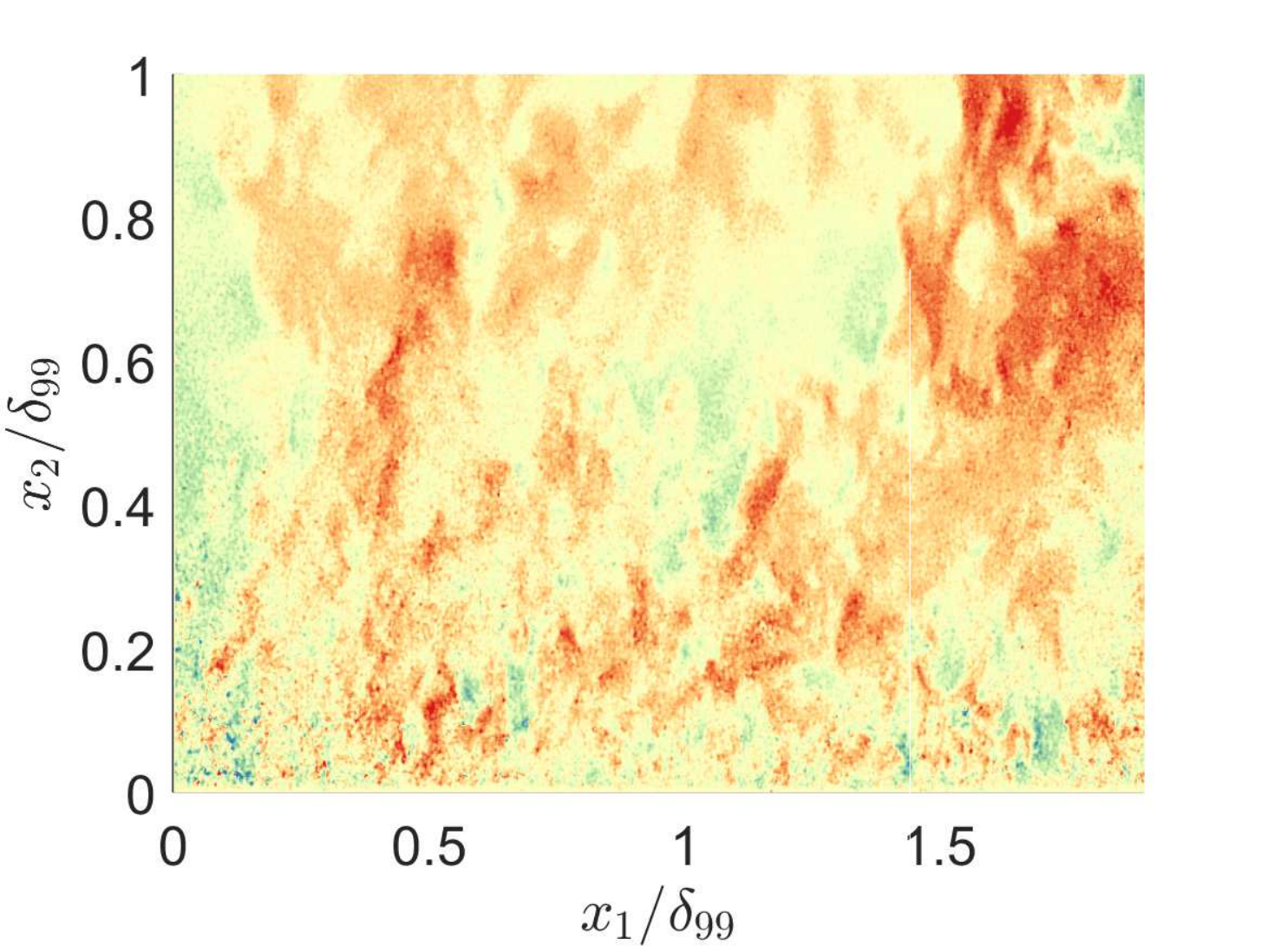} \\
     \subfigimg[width=70 mm,pos=ur,vsep=2pt,hsep=38pt]{(e)}{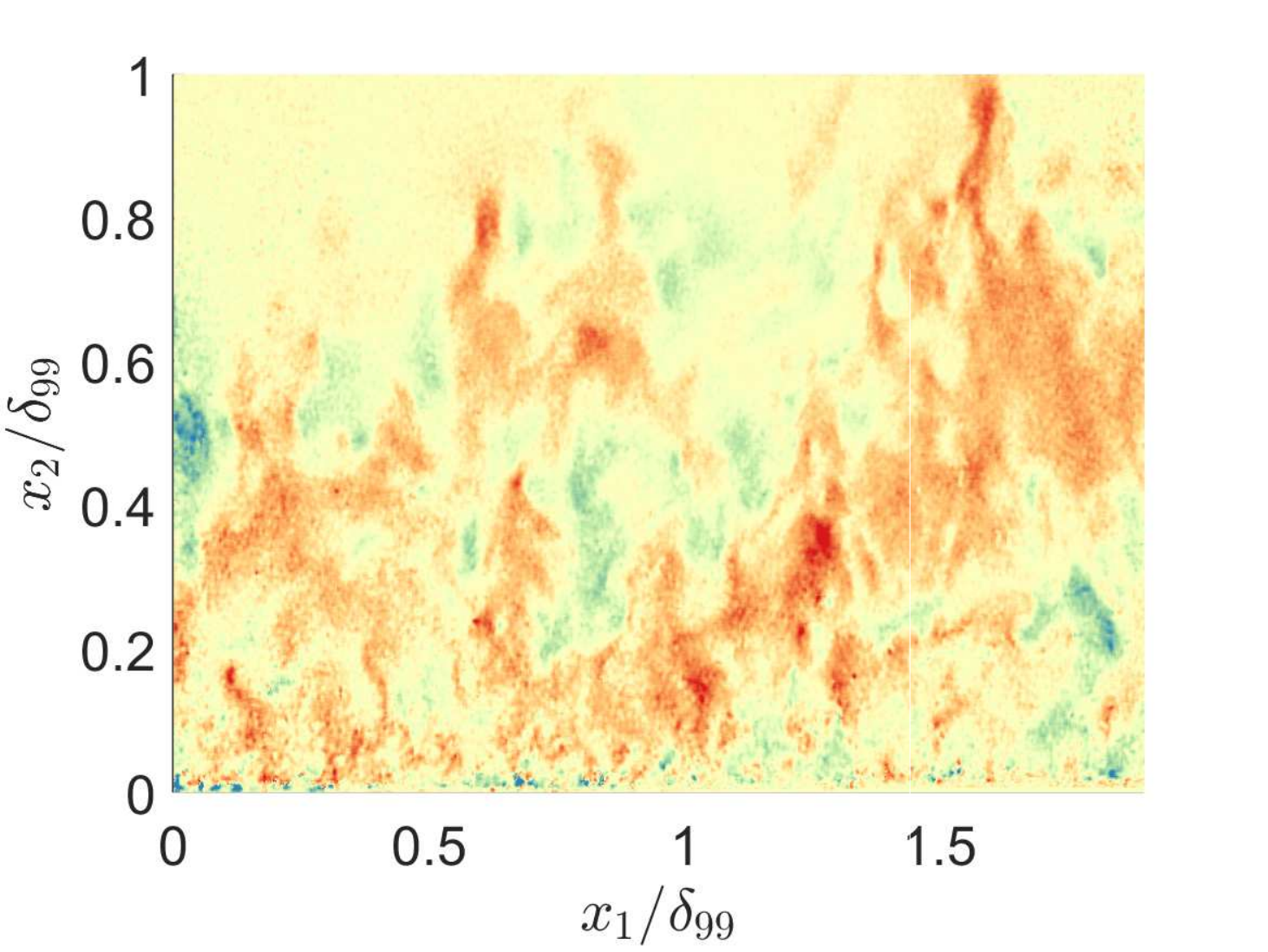} &
    \subfigimg[width=70 mm,pos=ur,vsep=2pt,hsep=38pt]{(f)}{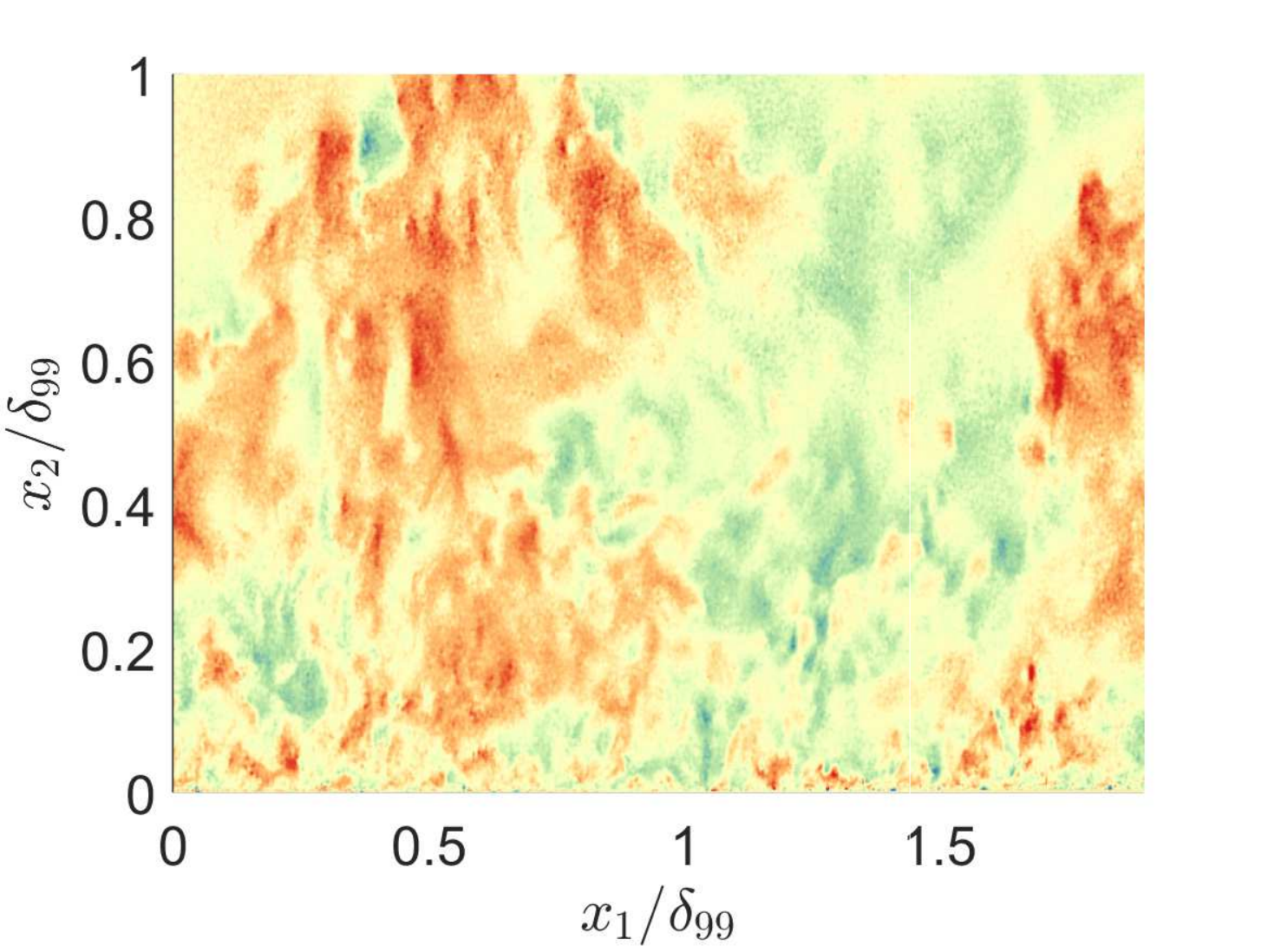} \\
  \end{tabular}
  \caption{Snapshot of wall-normal velocity fluctuations field normalised by $U_{\tau}$. \revPJ{Data in the figure correspond to measurements performed at U$_{\infty} \approx 10$ m/s.} (a) 3 \revPJ{mm} thick $10$ PPI foam, (b) 15 \revPJ{mm} thick $10$ PPI foam, (c) 3 \revPJ{mm} thick $45$ PPI foam, (d) 15 \revPJ{mm} thick $45$ PPI foam, (e) 3 \revPJ{mm} thick $90$ PPI foam, (f) 15 \revPJ{mm} thick $90$ PPI foam. }
  \label{fig:inst_velo}
\end{figure*}

\revPJ{The figure \ref{fig:inst_velo}, shows the fluctuating component of the wall-normal velocity normalised by \revJFM{$U_{\tau}$}. As the measurements were performed only over the porous substrate, $x_2=0$ corresponds to the surface of the foam. All the plots show similar contour levels indicating that a reasonable ``collapse'' of the distribution. This will be further explored in this section through more detailed statistical analysis.}

 \revPJ{In order to} cross validate PIV and HWA measurements, wall-normal profiles of \revPJ{streamwise} mean velocity and its root mean squared values were compared. The mean velocity, obtained from PIV and HWA measurements, show a good agreement; therefore, to keep the manuscript succinct only a comparison of the variance of velocity fluctuations will be shown.


\subsection{Outer-layer scaling\label{sec:level4a}}
\begin{figure}
\centering
\includegraphics[scale=0.5]{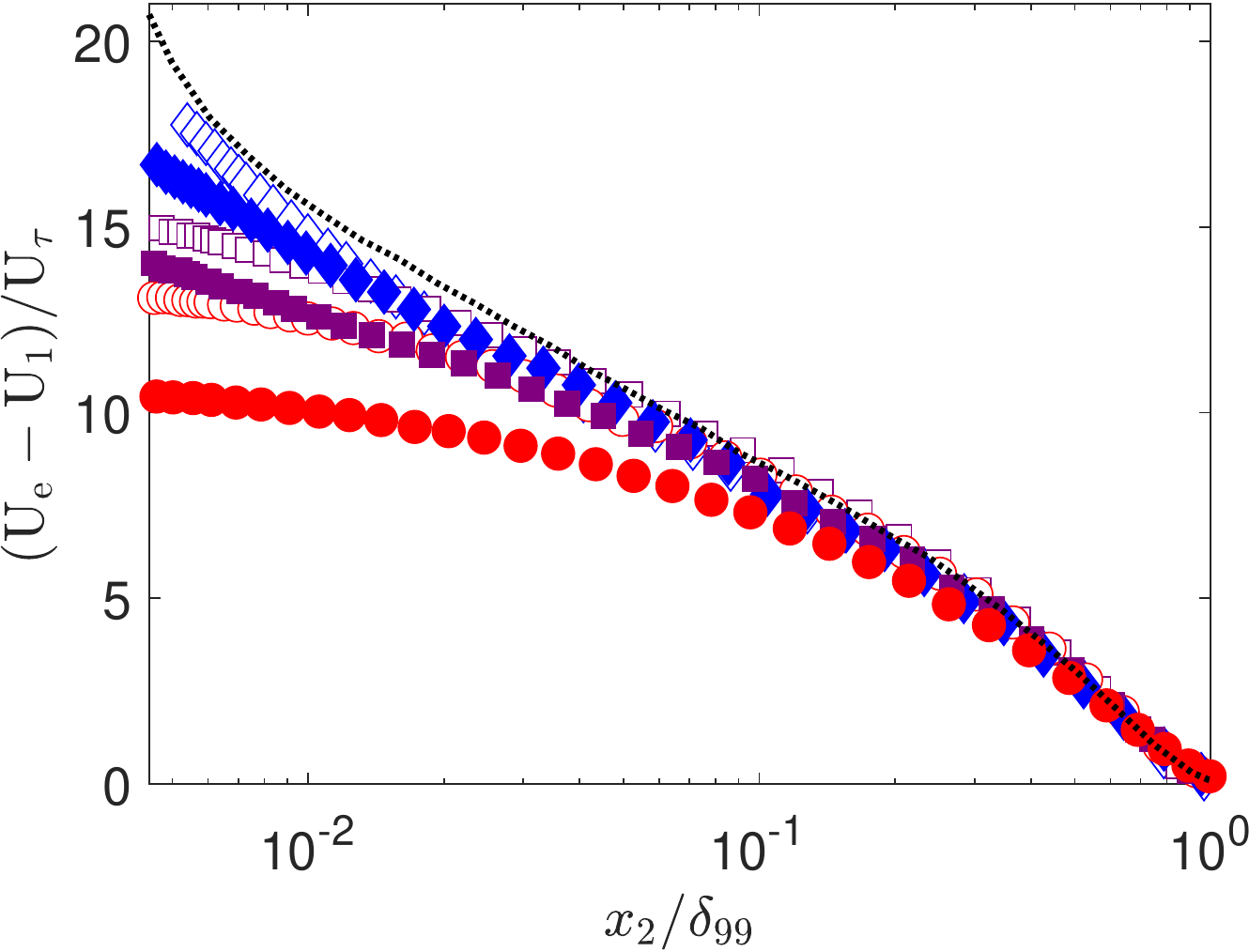}
\caption{Mean \revPJ{streamwise} velocity deficit normalised by inner velocity. \revPJ{Data in the figure correspond to measurements performed at U$_{\infty} \approx 10$ m/s.} Legends: \revJFM{PIV data;~Red circles 10 PPI, purple squares 45 PPI, blue diamonds 90 PPI foam substrate.} Filled symbols correspond to the 15 \revPJ{mm} thick substrate while hollow correspond to the 3 \revPJ{mm} thick substrate. \revPJ{Black dotted line corresponds to smooth wall data at $Re_{\tau} \approx7000$.}}
\label{fig:velocity_Deficit}
\end{figure}

In order to validate \cites{townsend1980structure} outer-layer hypothesis, first and second order velocity statistics are plotted in outer-layer scaling, e.g. $\delta_{99}$. The mean \revPJ{streamwise} velocity in the defect form is shown in figure \ref{fig:velocity_Deficit}. The boundary-layer thickness $\delta_{99}$ is used to scale the wall-normal distance while the inner-velocity $U_\tau$ is used to scale the \revPJ{streamwise} velocity $U_1$. The figure \ref{fig:velocity_Deficit} shows good collapse beyond $x_2/\delta_{99}=0.3$, as has been reported by earlier studies \citep{breugem2006influence,manes2011turbulent,efstathiou2018mean}. In the present form (figure \ref{fig:velocity_Deficit}), the velocity deficit increases with increasing cells per inch in a porous substrate and the thickness of the substrate, and is consistent with the observations of \cite{breugem2006influence}. However, \cite{efstathiou2018mean} had reported a slightly non-monotonic behaviour in velocity deficit. \cite{efstathiou2018mean} had attributed this non-monotonic behaviour to the transition from deep to shallow flow over porous substrate. Furthermore, they had reported similar non-monotonic trend in higher velocity statistics.



\begin{figure*}
  \centering
  \begin{tabular}{@{}p{0.5\linewidth}@{\quad}p{0.5\linewidth}@{}}
 
    \subfigimg[width=72 mm,pos=ul,vsep=2pt,hsep=38pt]{}{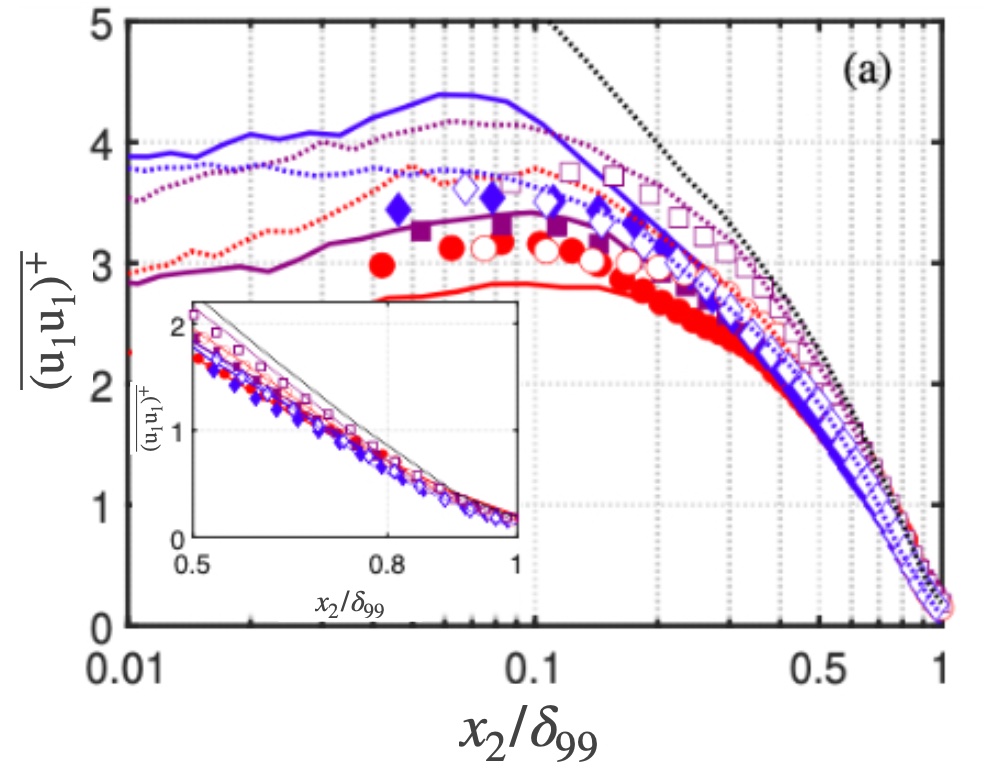} &
    \subfigimg[width=72 mm,pos=ul,vsep=2pt,hsep=38pt]{}{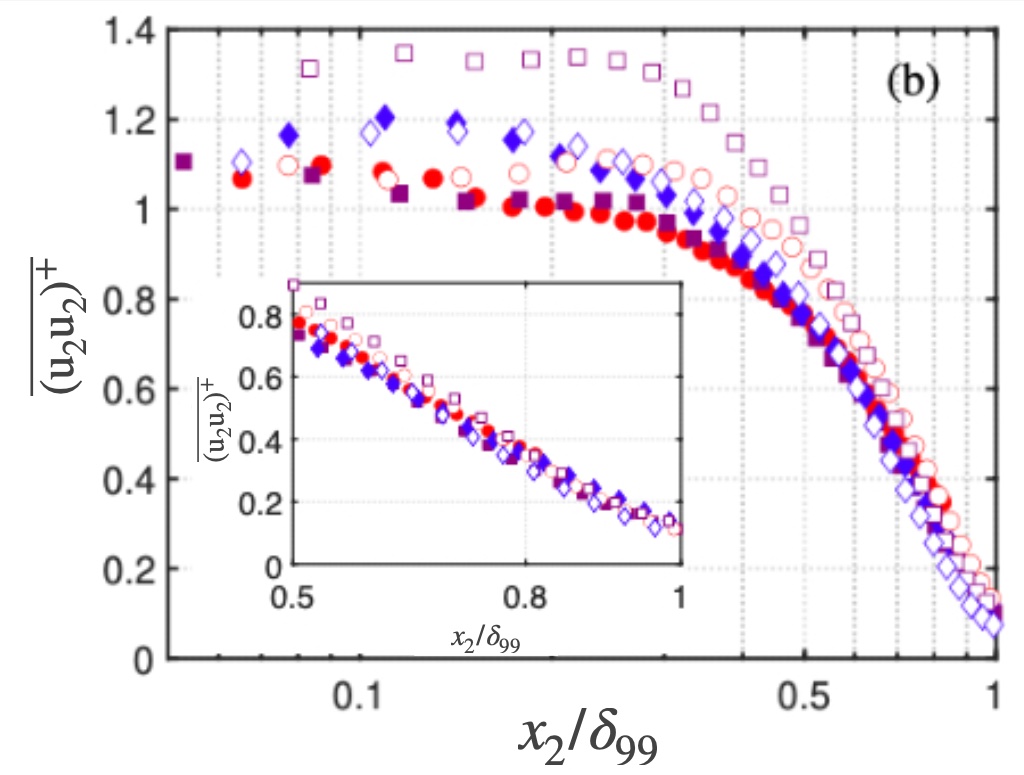} \\
    
    \hspace{3.6 cm}
    \subfigimg[width=72 mm,pos=ul,vsep=2pt,hsep=38pt]{}{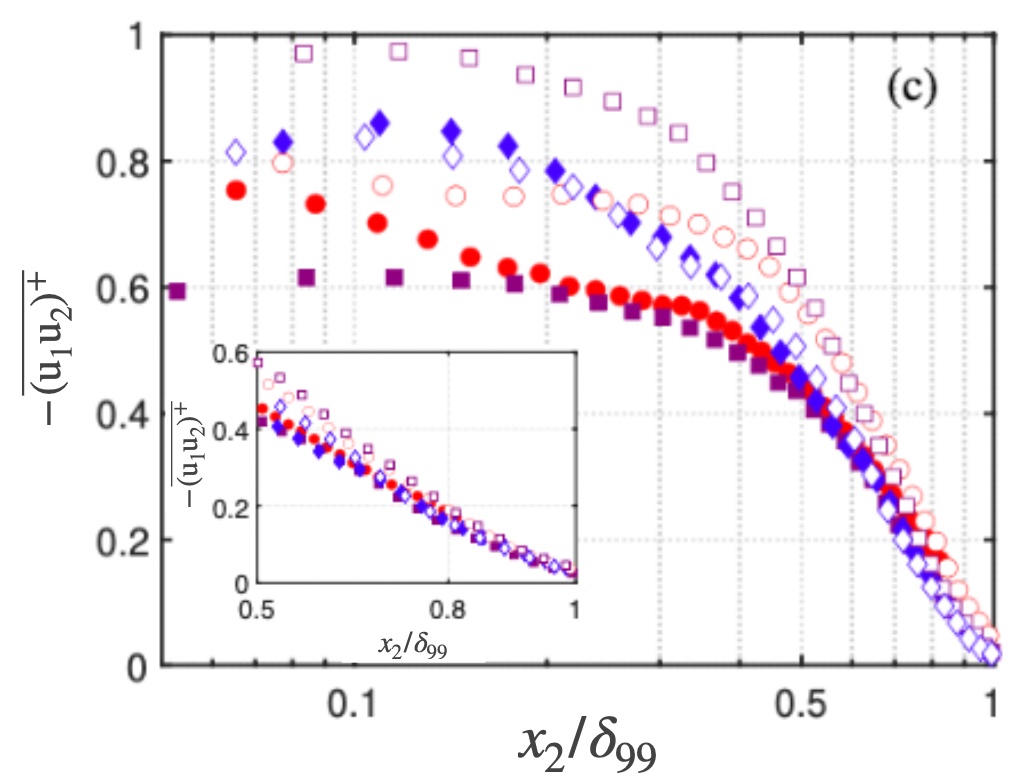}
    \\
  \end{tabular}
  \caption{Outer-layer scaling of \revPJ{the \revJFM{normalised} Reynolds stress tensor components \revJFM{$\overline{u_1 u_1}^+$, $\overline{u_2 u_2}^+$, and $\overline{-u_1 u_2}^+$} in the $x_1-x_2$ plane.} \revPJ{Data in the figure correspond to measurements performed at U$_{\infty} \approx 10$ m/s.} Velocity fluctuations are normalised by the inner velocity (${u_{\tau}}^2$). $x_2/\delta_{99}=0$ corresponds to the flow-substrate interface. Legends: PIV data; \revPJ{10 PPI: red circles, 45 PPI: purple squares, 90 PPI: blue diamonds.} Open symbols are for $3$ \revPJ{mm} thick substrate, while filled symbols correspond to $15$ \revPJ{mm} substrate. HWA data; 10 PPI:red lines, 45 PPI:purple lines, 90 PPI:blue lines.  Solid lines correspond to 15 \revPJ{mm} thick substrate, while the dotted lines correspond to 3 \revPJ{mm} thick substrate. \revPJ{Black dotted line corresponds to smooth wall data at $Re_{\tau} \approx7000$. Figures in inserts show the plot in linear axis.} 
 } 
  \label{fig:wall_parallel}
\end{figure*}

The variance of \revPJ{streamwise} velocity disturbance (${u_1}^{2}$ \revPJ{$\rm{m}^2/\rm{s}^2$}) normalized by friction velocity ($U_{\tau}^2$), $u_1^+$, is shown in figure \ref{fig:wall_parallel}. A good collapse between PIV and HWA measurements are obtained except in the near-wall region. Near wall PIV measurements are compromised by modulation error \citep{spencer2005correcting} because the window of interrogation is larger than the near wall structures. The near wall data is also compromised due to laser light reflections, therefore the near-wall PIV data ($\sim{2}$ \revPJ{mm}) were omitted. 
For the $15$ \revPJ{mm} thick substrate, reduction in \revPJ{streamwise} turbulence intensity scales with an increase in wall-permeability. This trend is consistent with the observations made by \cite{manes2011turbulent}. Furthermore, the peak in ${u_1}^{+}$ for thin foam is considerably closer to the wall than the thick foam, which highlights the importance of permeability based Reynolds number $Re_{K}$ and roughness based Reynolds number $s^+$. For all the substrates tested over a broad range of Reynolds numbers ($Re_{K}$ and $s^+$), a good collapse of \revPJ{streamwise} velocity fluctuations is obtained in the outer-layer region when plotted against outer-layer variable ($\delta_{99}$). 

The turbulent fluctuations for \revPJ{Reynolds shear stress} and wall-normal velocity component are shown in figure \ref{fig:wall_parallel} (\textit{b-c}). The wall-normal velocity is especially susceptible to permeability \citep[see][for instance]{breugem2006influence}. Slightly away from the wall $x_2/\delta_{90}\sim 0.1$, the $45$ PPI foams shows largest differences in the wall-normal velocity disturbances possibly signaling increased permeability effects. The $10$ PPI foam shows classic flat near wall-velocity fluctuations, as has been reported for fully-rough flows. The wall-normal component is associated with active motions, i.e. turbulent motion that contribute to Reynolds shear stress. The \revPJ{Reynolds shear stress}, which is comprised of both active and inactive motions, also shows a significant spread in the outer layer for the 45 PPI foam. Therefore, impact of relative foam thickness ($h/s$) on velocity statistics is quiet substantial for this case. It is important to note that the spread in $u_1u_2^{+}$ profiles in the present study is similar to spread in wall-normal velocity variance reported by \cite{manes2011turbulent}. Therefore, the existence of outer-layer similarity for wall-normal velocity profiles is questionable even though the present study has been performed at very high Reynolds number. The $90$ PPI foam has a very low permeability based Reynolds number $Re_{K} \sim 1$ and a large separation between zero plane position $y_d$ and boundary-layer thickness.


 

\begin{figure*}
  \centering
  \begin{tabular}{@{}p{0.5\linewidth}@{\quad}p{0.5\linewidth}@{}}
  
    \subfigimg[width=65 mm,pos=ul,vsep=15pt,hsep=32pt]{(a)}{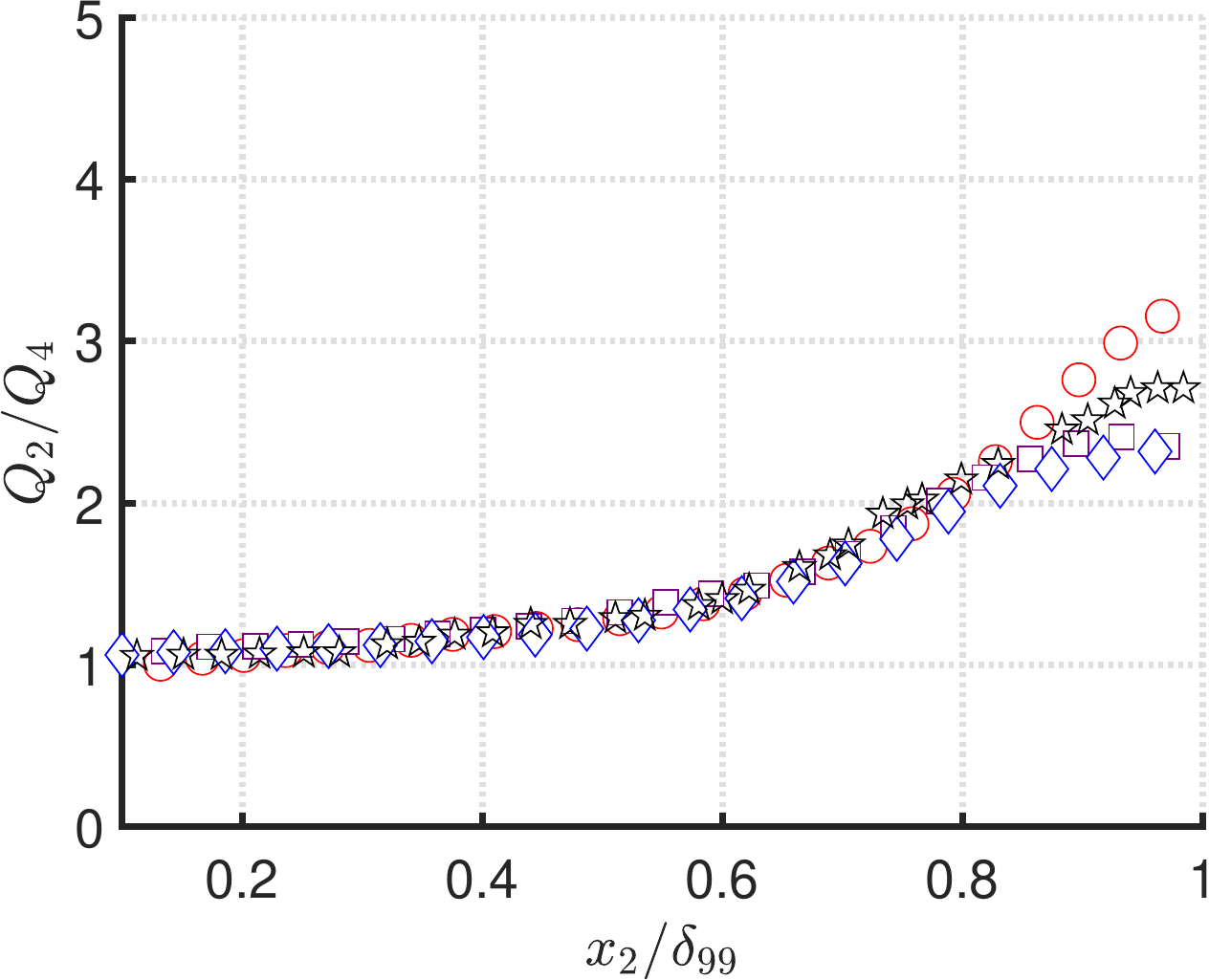} &
    \subfigimg[width=65 mm,pos=ul,vsep=15pt,hsep=32pt]{(b)}{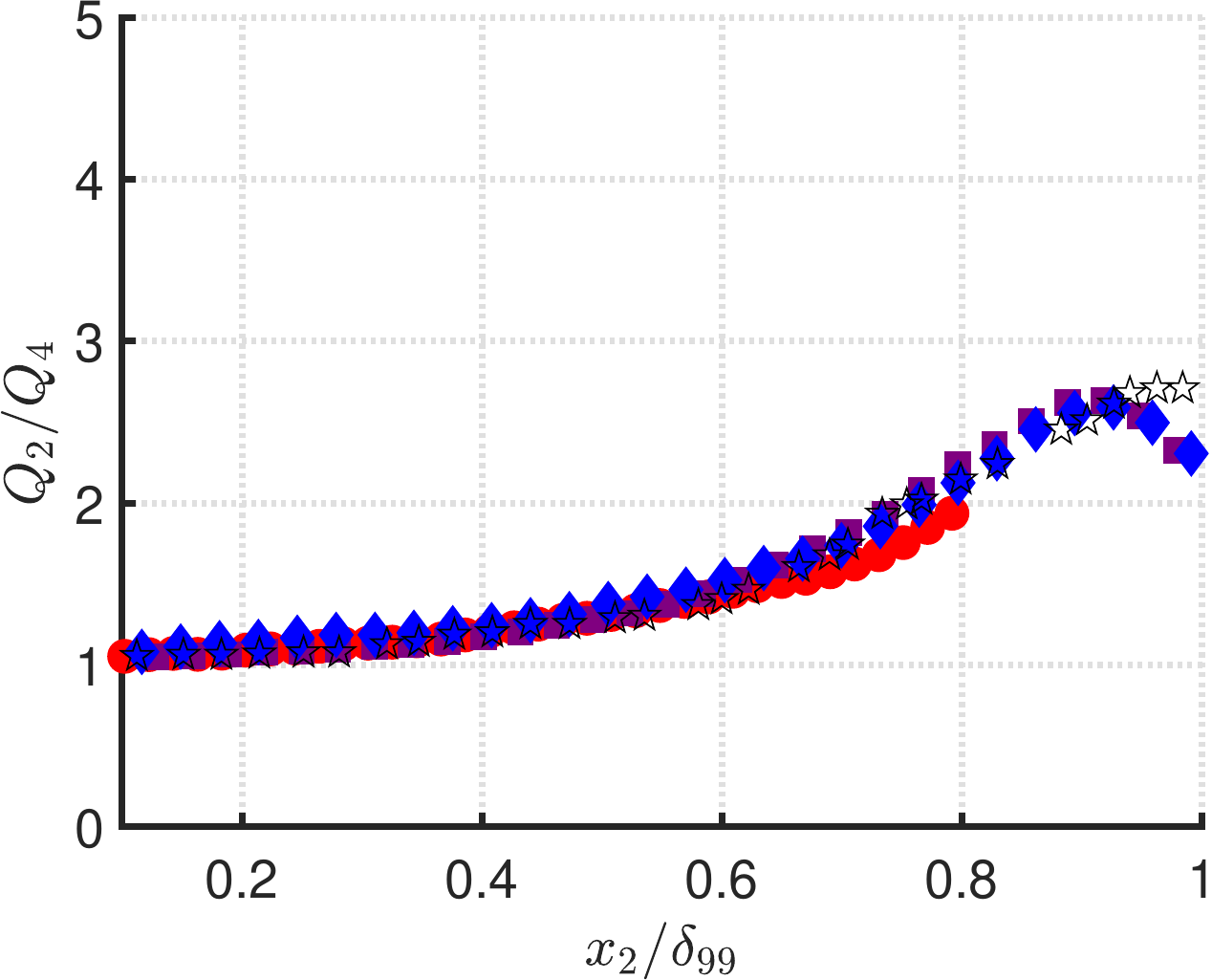}

    \end{tabular}
  \caption{Ratio of Reynolds-shear-stress contributions from Q2 and Q4 events, estimated using PIV measurements performed at U$_{\infty}=10$ m/s, as a function of wall-normal distance. (a) 3 mm thick substrate (b)~15 mm thick substrate. Legends: Red circles are 10 PPI, purple squares 45 PPI, blue diamonds 90 PPI foam substrate. The black pentagrams are \cites{wu2007outer} smooth-wall measurements at $Re_{\tau}=3470$. } 
  \label{fig:quadrant_ratio}
\end{figure*}

\begin{figure*}
  \centering
  \begin{tabular}{@{}p{0.5\linewidth}@{\quad}p{0.5\linewidth}@{}}
  
    \subfigimg[width=65 mm,pos=ul,vsep=15pt,hsep=32pt]{(a)}{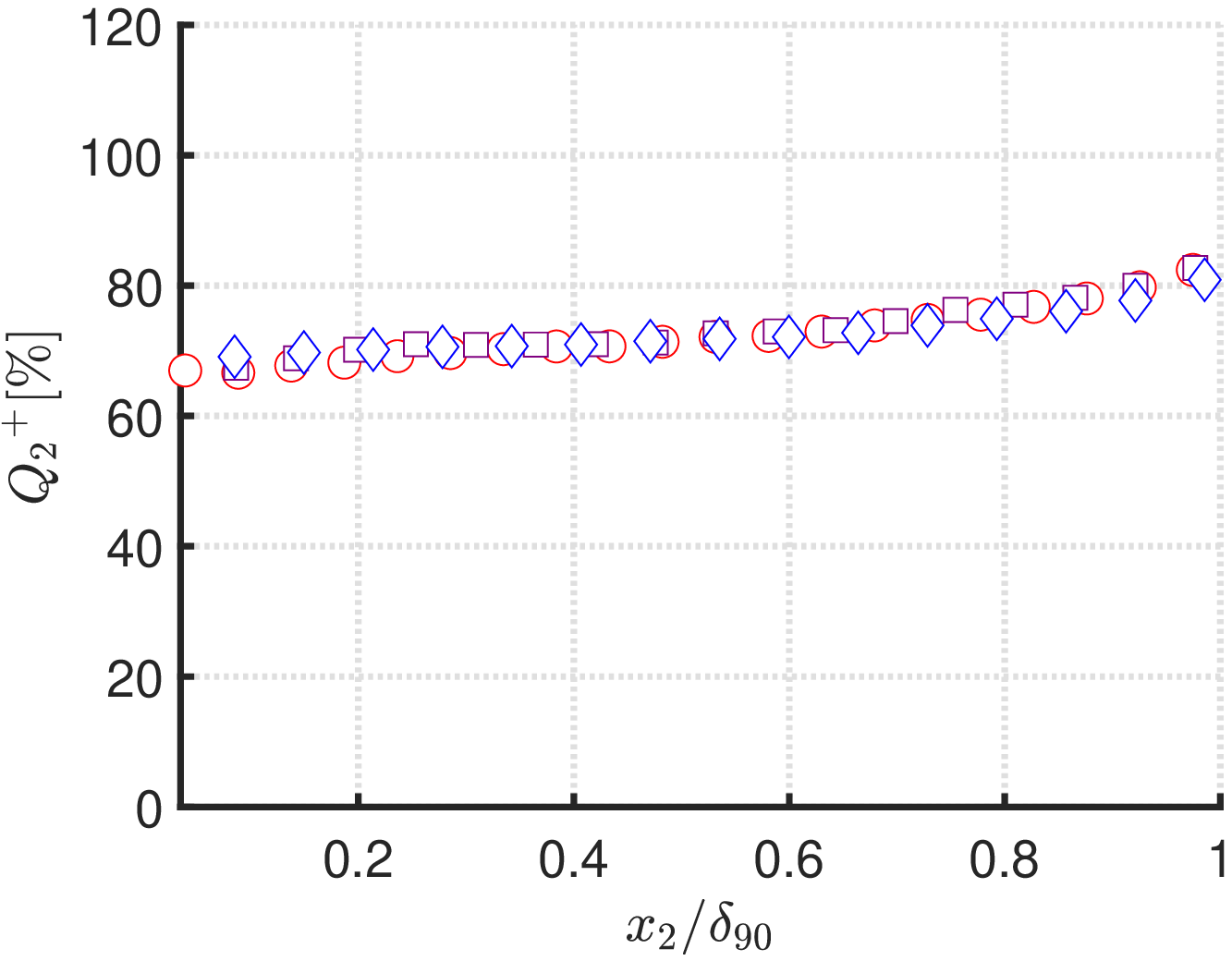} &
    \subfigimg[width=65 mm,pos=ul,vsep=15pt,hsep=32pt]{(b)}{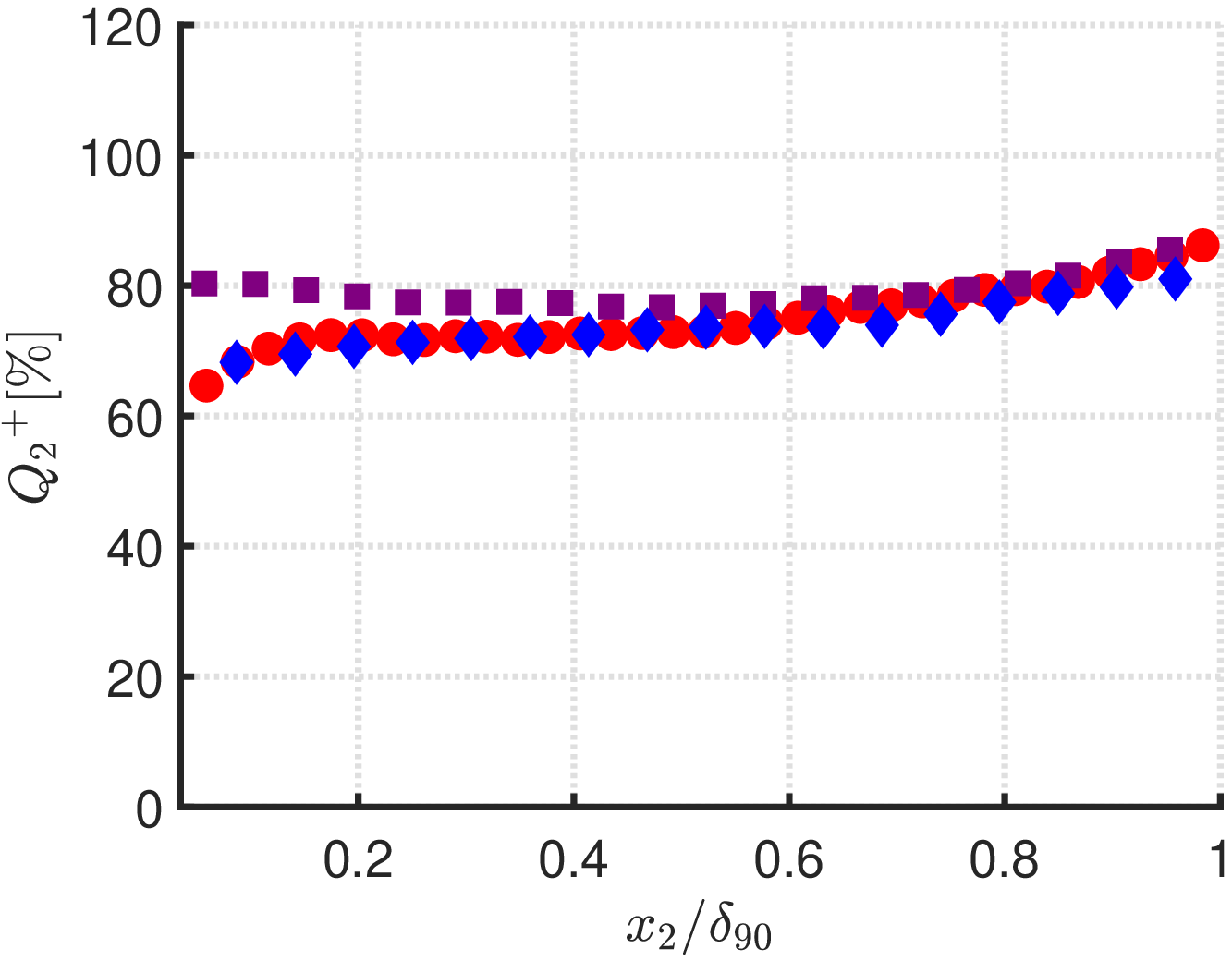} \\
    
     \subfigimg[width=65 mm,pos=ul,vsep=15pt,hsep=32pt]{(c)}{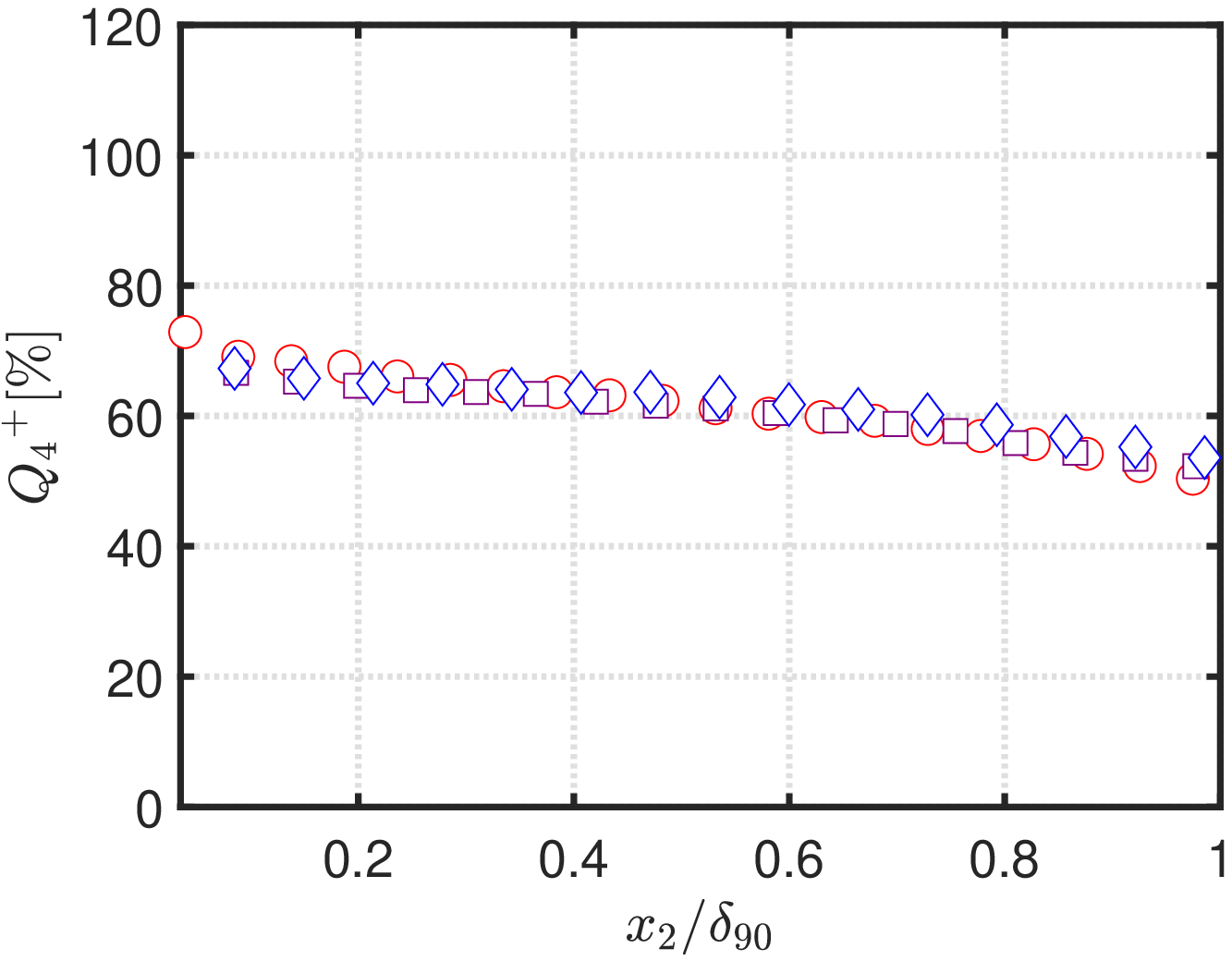} &
    \subfigimg[width=65 mm,pos=ul,vsep=15pt,hsep=32pt]{(d)}{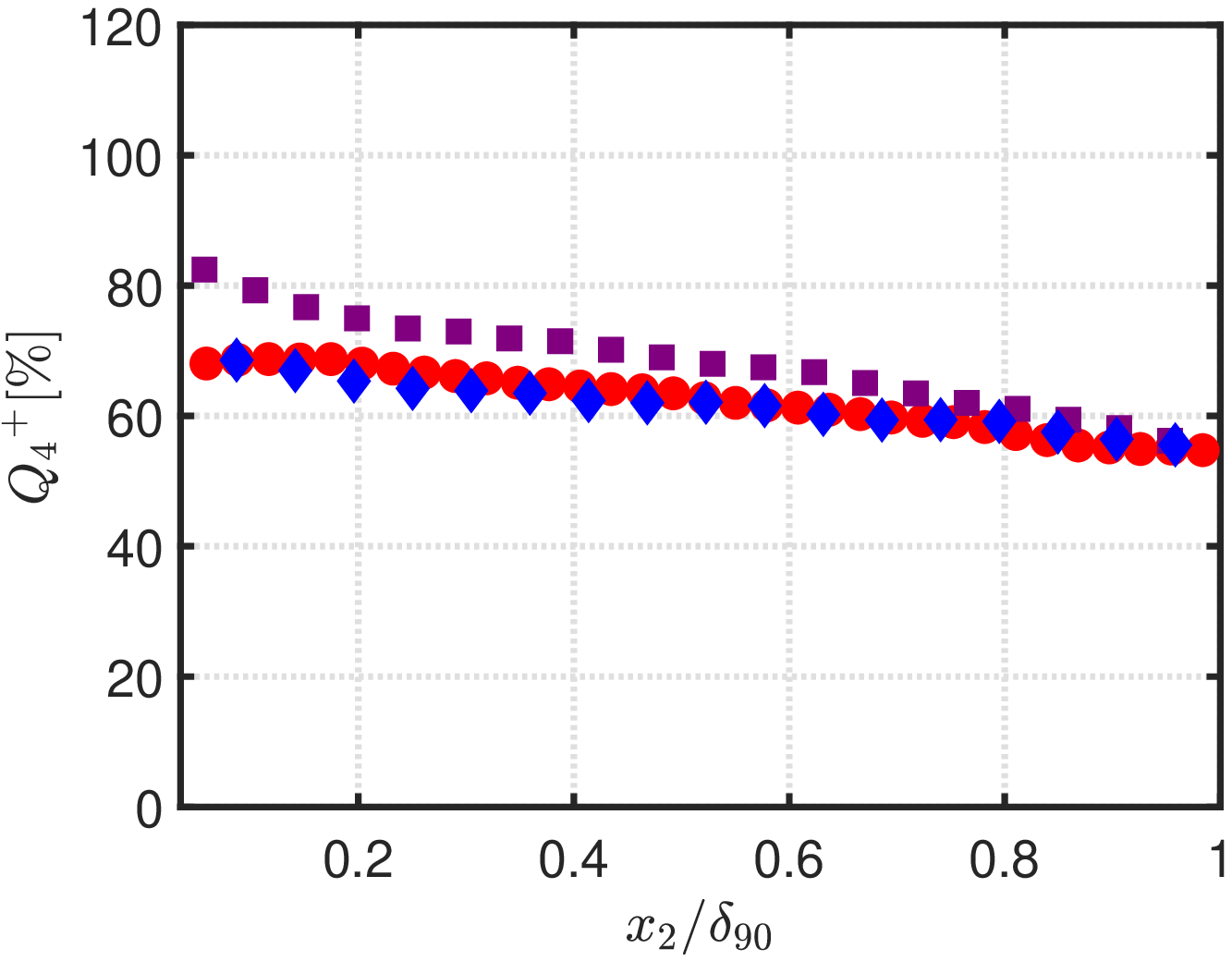} \\
    
    \end{tabular}
  \caption{ \revPJ{Quadrant analysis from PIV measurements performed at U$_{\infty} \approx 10$ m/s.}\revJFM{~(a) and (b)~Relative contribution from \revJFM{${Q_2}^+/-\overline{u_1 u_2}^+$} events for ~3 \revPJ{mm} and 15 \revPJ{mm} thick substrate respectively. (c) and (d)~Relative contribution from \revJFM{${Q_4}^+/-\overline{u_1 u_2}^+$} events for ~3 \revPJ{mm} and 15 \revPJ{mm} thick substrate respectively.} Legends: \revJFM{Red circles are 10 PPI, purple squares 45 PPI, blue diamonds 90 PPI foam substrate. Open symbols are for $3$ \revPJ{mm} thick substrate, while filled symbols correspond to $15$ \revPJ{mm} substrate}.} 
  \label{fig:quadrant}
\end{figure*}

Although \revPJ{Reynolds shear stress} tensor component ($-u_i u_j$) are statistical indicator of momentum transfer in the form of Reynolds shear stress, a more efficient dichotomy of outward-inward transport of momentum by turbulence can be obtained \revJFM{by} quadrant analysis \citep{wallace2016quadrant}. \revJFM{Figure \ref{fig:quadrant_ratio} shows the ratio of the contributions from the $Q_2$ events to the
contributions from the $Q_4$ events. $Q_2$, referred as ejections, marks the instances when a low speed fluid parcel is transported away from the wall. In contrast $Q_4$, referred as sweep, is transport of high speed fluid parcel towards the wall. As such, the ratio $Q_2/Q_4$ quantifies the relative importance of these events at a given wall-normal location. A good collapse between the different cases and smooth walls is obtained except at the edge of boundary layer, where extremely small magnitudes of Reynolds shear stress are expected, as already argued by \cite{wu2007outer}.}

Since the~\revPJ{Reynolds shear stress} tensor's component $-\overline{u_1 u_2}$ is less than zero for well developed turbulent \revJFM{boundary layer} past a wall, only \revJFM{the relative contributions from negative quadrants} Q2 and Q4 \revJFM{are} shown in figure \ref{fig:quadrant}. \revthree{As a note of caution to the reader, higher values of $Q_2^+$ or $Q_4^+$ does not imply higher overall levels of \revPJ{Reynolds shear stress} $-\overline{u_1 u_2}$ as it is normalised by the later. As such, the percentage, plotted in figure \ref{fig:quadrant}, is the contribution of these events to Reynolds shear stress, rather than the number of Q2 and Q4 events among all events.} Notice that the outer-layer variables are now non-dimensionalised with $\delta_{90}$ \revPJ{(based on $90\%$ of the free-stream velocity)} because PIV measurements for $10$ PPI $15$ \revPJ{mm} substrate is unable to fully capture the boundary layer thickness $\delta_{99}$. \revPJ{As such, boundary-layer thickness based on $90\%$ of the free-stream velocity ($\delta_{90}$) was used instead of the frequently defined value based on $99\%$ of the free-stream velocity.} Nevertheless, it was verified that normalizing the plots with $\delta_{90}$ or $\delta_{99}$ had no impact on the outer-layer scaling. Therefore, subsequent plots will be normalised by $\delta_{90}$. Figure \ref{fig:quadrant} shows the relative contributions of $Q_2^+$ and $Q_4^+$ quadrants as function of wall-normal distance. 

While for the 3 \revPJ{mm} substrates (\ref{fig:quadrant} (\textit{a} and \revPJ{\textit{c}})) a good collapse is achieved irrespective of foam permeability, for the 15 \revPJ{mm} thick porous substrate weaker collapse among the various foams can be seen. These differences exist well into the outer layer for the case of $45$ PPI foam, which shows an increased $Q_2^+$ and $Q_4^+$ events. It is known that wall-permeability in absence of surface roughness, opens the path between near and outer wall regions \citep{breugem2006influence} and can invalidate the Townsend's hypothesis. Similarly, \cite{carpio2019experimental} have reported an increase in Q2 and Q4 events with an increase in permeability. However, both these studies were performed at a low to moderate Reynolds numbers. \revPJ{While in our case, where both surface roughness and permeability are present, we see that increase in $Q_2^+$ and $Q_4^+$ events are only seen by foam with intermediate permeability.} \revPJ{This can be due to local foam morphology. In particular, with an increase in pore size, the shelter solidity \citep{placidi2018turbulent}, $\lambda_{s}$, decreases and vice-versa. For foams with intermediate pore size, such as the 45 PPI foam, the $\lambda_{s}$ remains at intermediate range, compared to 10 and 90 PPI foams, for which local morphology can influence turbulence statistics, as shown by \cite{placidi2018turbulent}.} \revPJ{Alternatively, the increase in $Q_4^+$ events for the 45 PPI and 15 \revPJ{mm} thick substrate, can perhaps be explained by relaxation in the wall-blocking condition. Finally}, in order to fulfil the continuity condition, the $Q_2^+$ events need to rise accordingly \citep{krogstad1992comparison}. More importantly, it appears that $Q_2^+$ and $Q_4^+$ events do scale with wall permeability, for $s^+ \sim 30$, and that permeability opens a path of increased sweep events close to the wall. Therefore, the effects of porous walls extent to the outer-layer regions, this is sufficient to invalidate the Townsend's outer-layer hypothesis. At shallow and deep substrate limits the permeability and pore size based Reynolds number are similar, the only noticeable difference are in the values of $k_s^+$ and $y_d$.

To conclude, figures \ref{fig:velocity_Deficit} and \ref{fig:wall_parallel} show \revPJ{streamwise} mean and variance collapse in the outer-layer when the velocity scales are normalised by $U_{\tau}$ and wall-normal distance by $\delta_{99}$. However, as the substrate thickness and permeability  is increased, the collapse for wall-normal component in the outer-layer becomes less evident. Furthermore, a good collapse in quadrants $Q_2^+$ and $Q_4^+$ is observed in figure \ref{fig:quadrant} for the thinner foam substrate. For thick substrates, collapse is achieved either when the substrate has permeability based Reynolds number comparable to viscous scales ($90$ PPI foam) or when the substrate is sparse and \revPJ{$Re_{\tau} \ge 7000$}, i.e $10$ PPI foam. These results cast doubts on the validity of Townsend's outer-layer hypothesis for turbulent flow past porous wall with varying thicknesses. Therefore, a detailed investigation on flow-structures will be performed in the following section. 

\subsection{\label{sec:level4b}The structure of turbulent boundary-layer over porous walls}
As mentioned in the introduction, for flow over and past porous foams no previous study at high Reynolds ($Re_{\tau} \sim 2000$) have reported multi-point correlation analysis, instead only single point statistics have been reported \citep{manes2011turbulent,efstathiou2018mean}. Therefore, in the current manuscript, two-point velocity correlation will be used to study the spatial structure of turbulence convecting over porous foams. 

In the present work, {two-point correlation} is denoted by{:}

{\begin{equation} 
\revPJ{R_{ij}({x_1},{x_1}^{\prime},x_2,{x_2}^{\prime},x_3,{x_3}^{\prime}) =   \frac{\overline{{u_{i}}(x_1,x_2,x_3) {u_{j}}({x_1}^{\prime},{x_2}^{\prime},{x_3}^{\prime}) }}{\sqrt{\overline{{u_i}^2(x_1,x_2,x_3}}) \times \sqrt{\overline{{u_j}^2({x_1}^{\prime},{x_2}^{\prime},{x_3}^{\prime})}}}  }
  \label{two_point}         
\end{equation}}

where $ {u_i}(x_1,x_2,x_3) $ is the {$i$-th} component of the velocity fluctuation at the fixed or reference location while ${u_j}({x_1}^{\prime},{x_2}^{\prime},{x_3}^{\prime})$ denotes the {$j$-th} component of the velocity fluctuations at the moving point. The terms \revPJ{$\sqrt{\overline{{u_i}^2(x_1,x_2,x_3}}$  and  $\sqrt{\overline{{u_j}^2({x_1}^{\prime},{x_2}^{\prime},{x_3}^{\prime})}}$ are standard deviation of turbulent fluctuations}, at the fixed and moving point respectively.
Equation~(\ref{two_point}), 
assumes that the flow is inhomogeneous in all three spatial directions. In the current study, we only treat the wall-normal location as the inhomogeneous direction.



\begin{figure*}
  \centering
  \begin{tabular}{@{}p{0.5\linewidth}@{\quad}p{0.5\linewidth}@{}}
    \subfigimg[width=65 mm,pos=ul,vsep=15pt,hsep=32pt]{(a)}{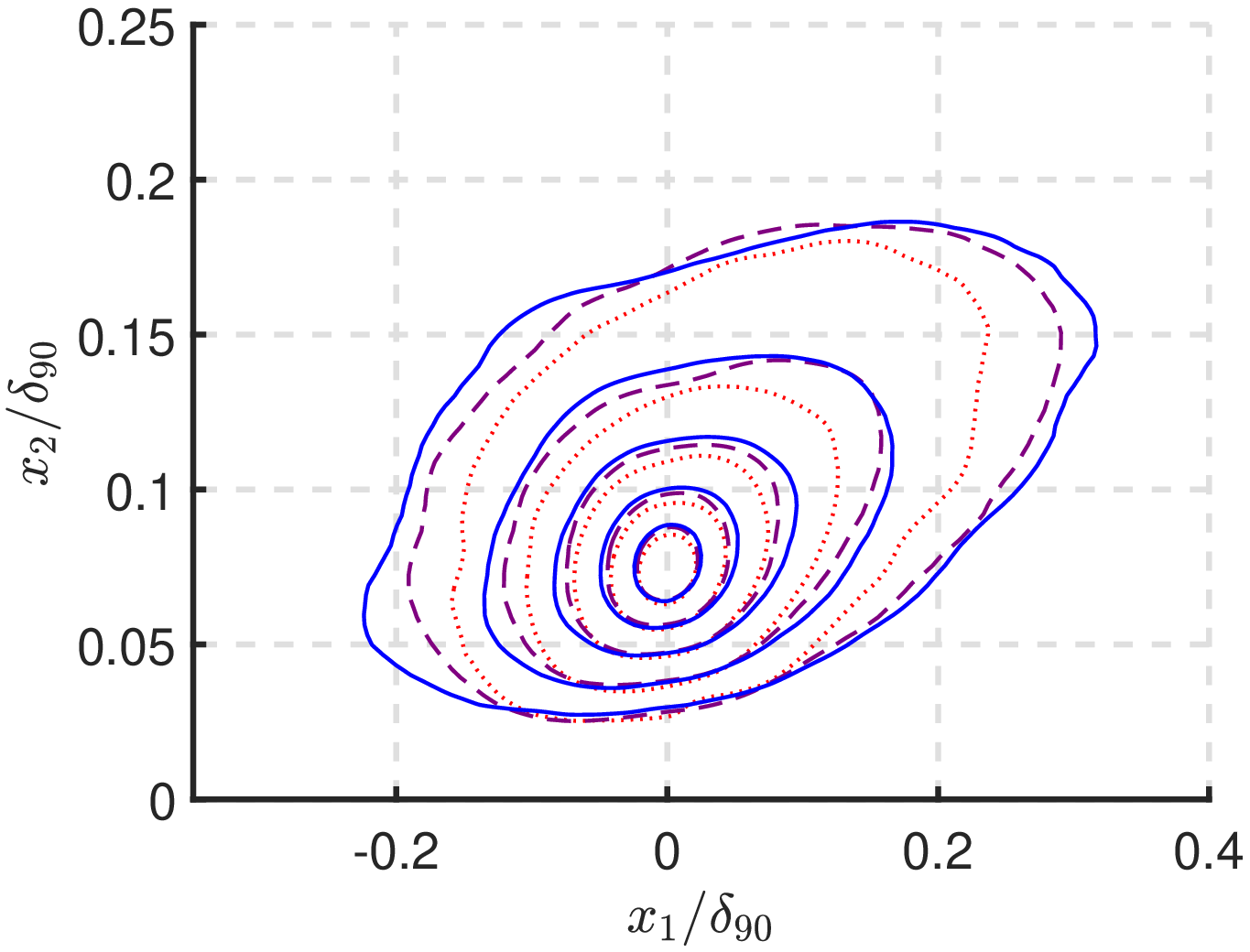} &
    \subfigimg[width=65 mm,pos=ul,vsep=15pt,hsep=32pt]{(b)}{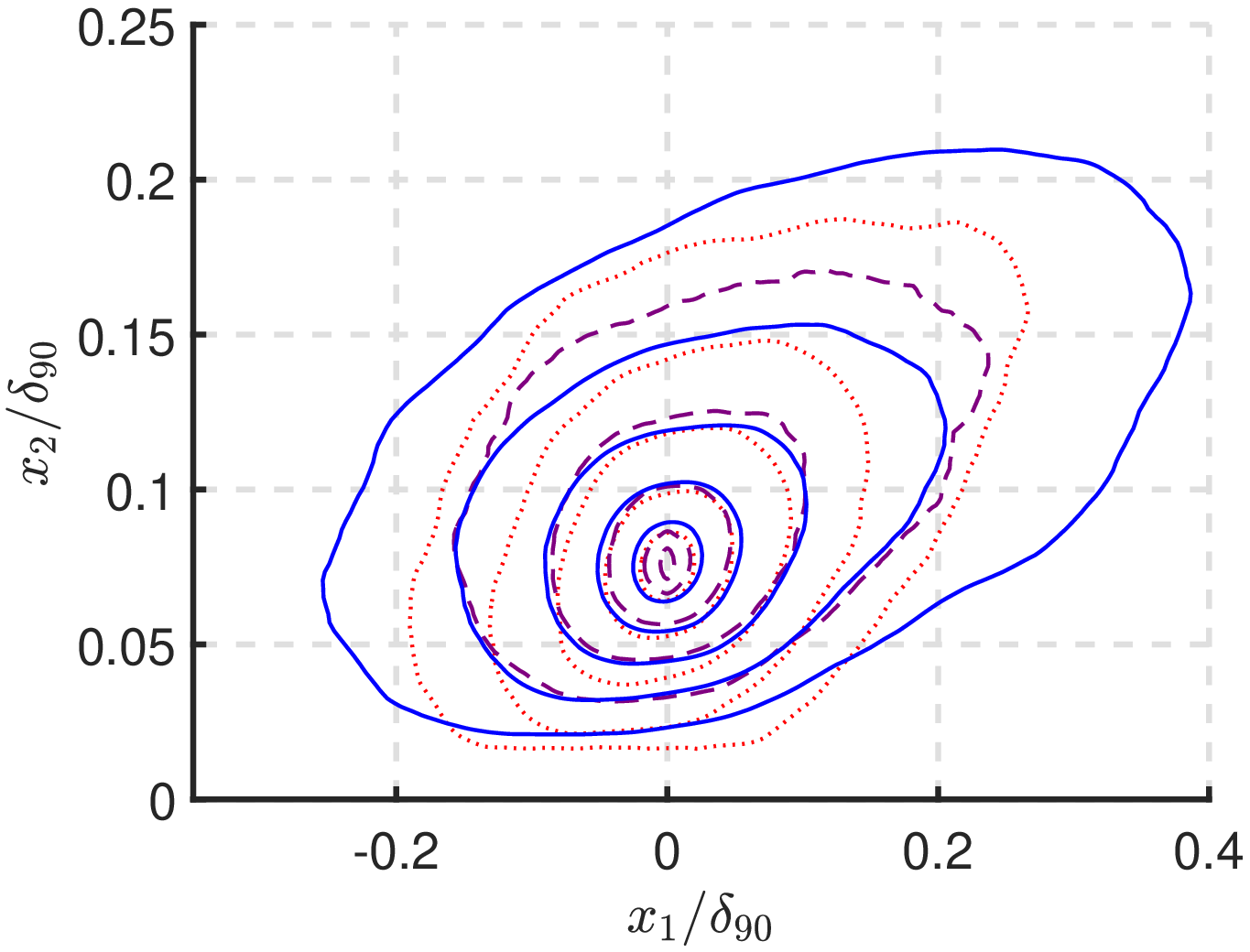} \\
    \subfigimg[width=65 mm,pos=ul,vsep=15pt,hsep=32pt]{(c)}{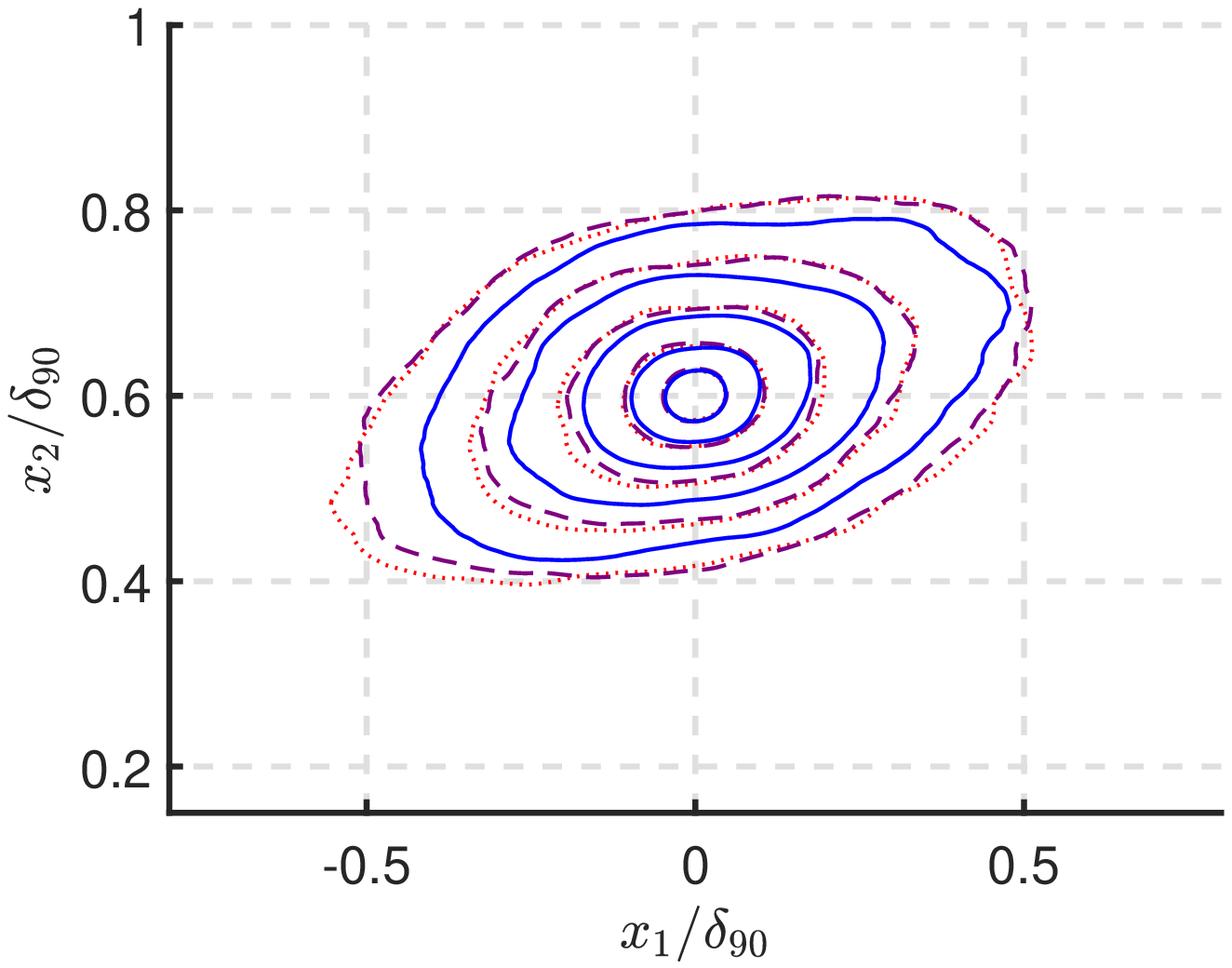} &
    \subfigimg[width=65 mm,pos=ul,vsep=15pt,hsep=32pt]{(d)}{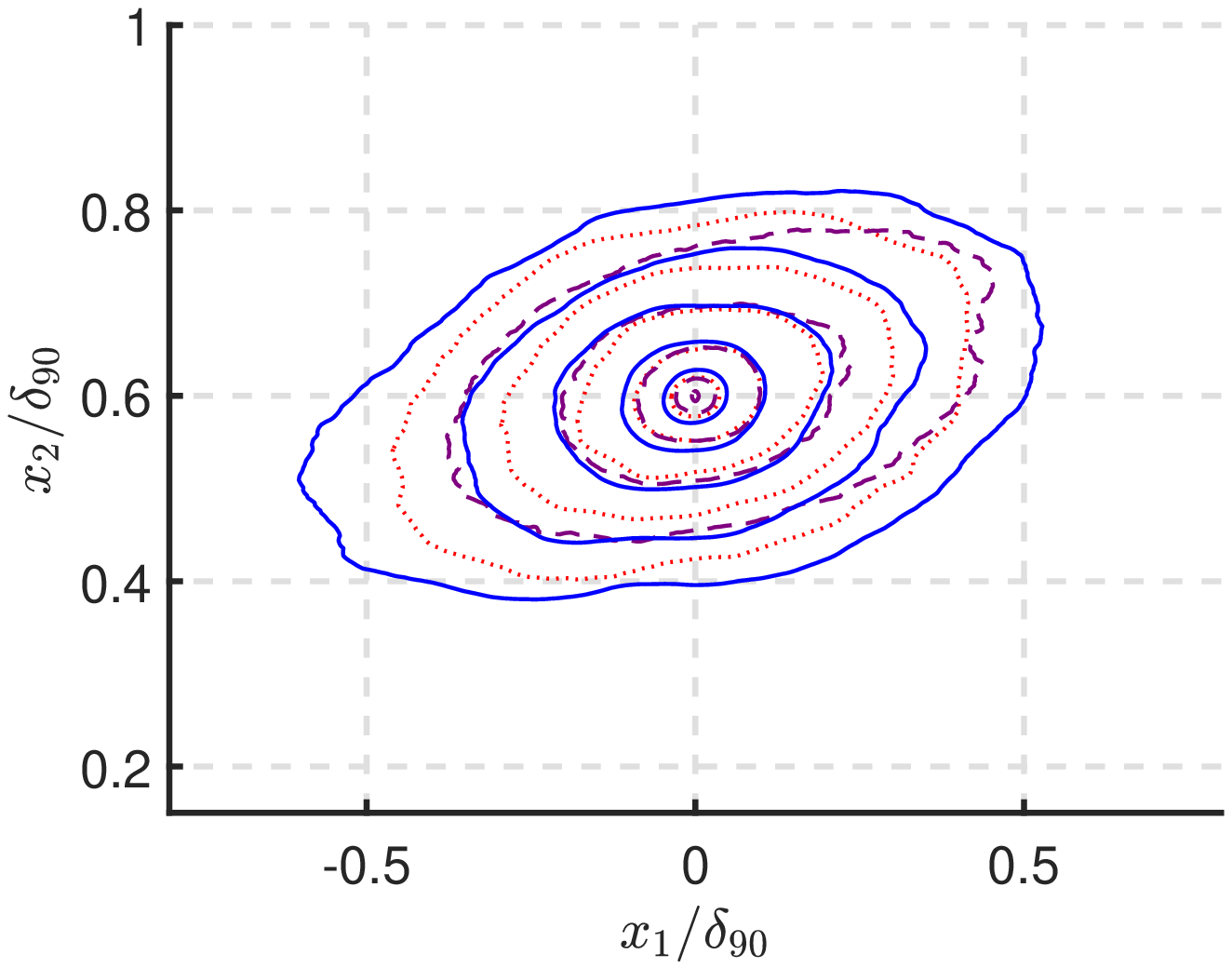} \\
    
  \end{tabular}
  \caption{\revPJ{streamwise} velocity two-point zero time delay correlation, R$_{11}(x_1,x^{\prime}_1,x_2,x^{\prime}_2,x_3,x_3)$. \revPJ{Data in the figure correspond to measurements performed at U$_{\infty} \approx 10$ m/s.} Plots on the left are for 3 \revPJ{mm} thick substrate while plots on the right are for 15 \revPJ{mm} thick substrate. (a) Fixed point at ($0.07 \times  \delta_{90}$); (b)Fixed point at ($0.07 \times  \delta_{90}$); (c) Fixed point at ($0.6 \times \delta_{90}$); (d)~Fixed point at ($0.6 \times  \delta_{90}$). Legends: Red dotted lines for 10 PPI foam substrate, purple dashed lines 45 PPI foam substrate, and blue solid lines 90 PPI foam substrate. The iso-contour lines are from $0.5$ to $0.9$ with an increment of $0.1$.}
  \label{fig:correlation_R11}
\end{figure*}
\begin{figure*}
  \centering
  \begin{tabular}{@{}p{0.5\linewidth}@{\quad}p{0.5\linewidth}@{}}
    \subfigimg[width=65 mm,pos=ul,vsep=15pt,hsep=32pt]{(a)}{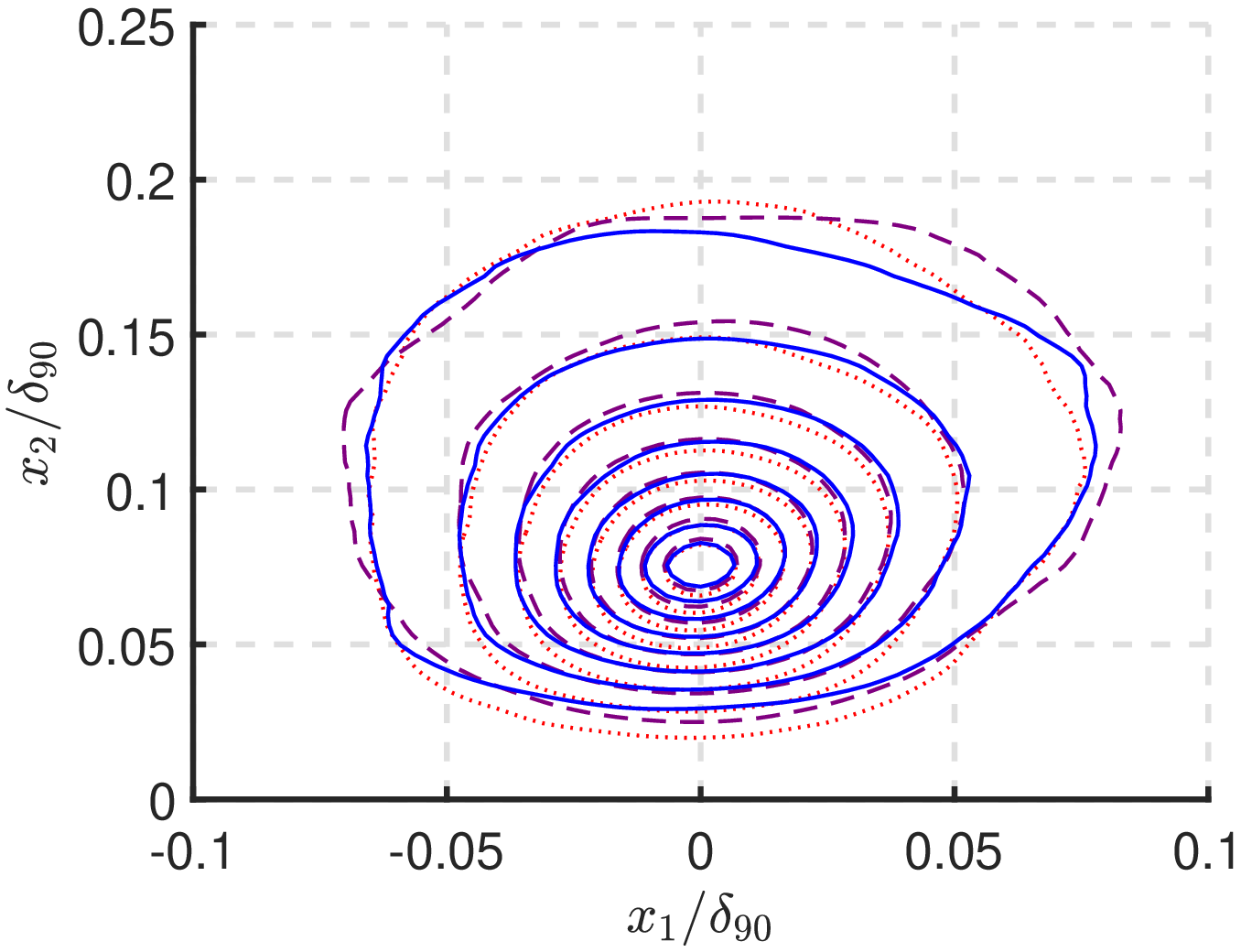} &
    \subfigimg[width=65 mm,pos=ul,vsep=15pt,hsep=32pt]{(b)}{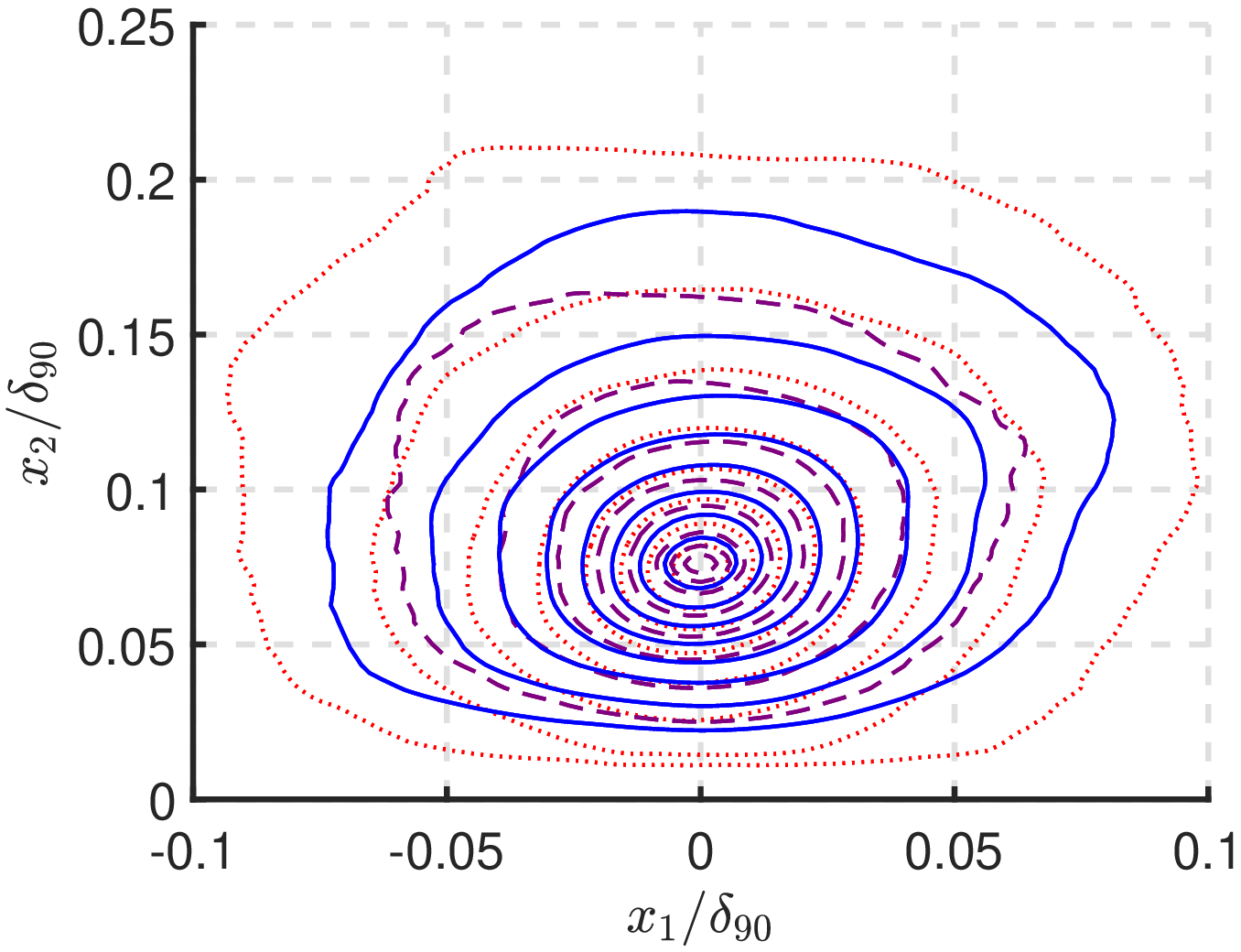} \\
    \subfigimg[width=65 mm,pos=ul,vsep=15pt,hsep=32pt]{(c)}{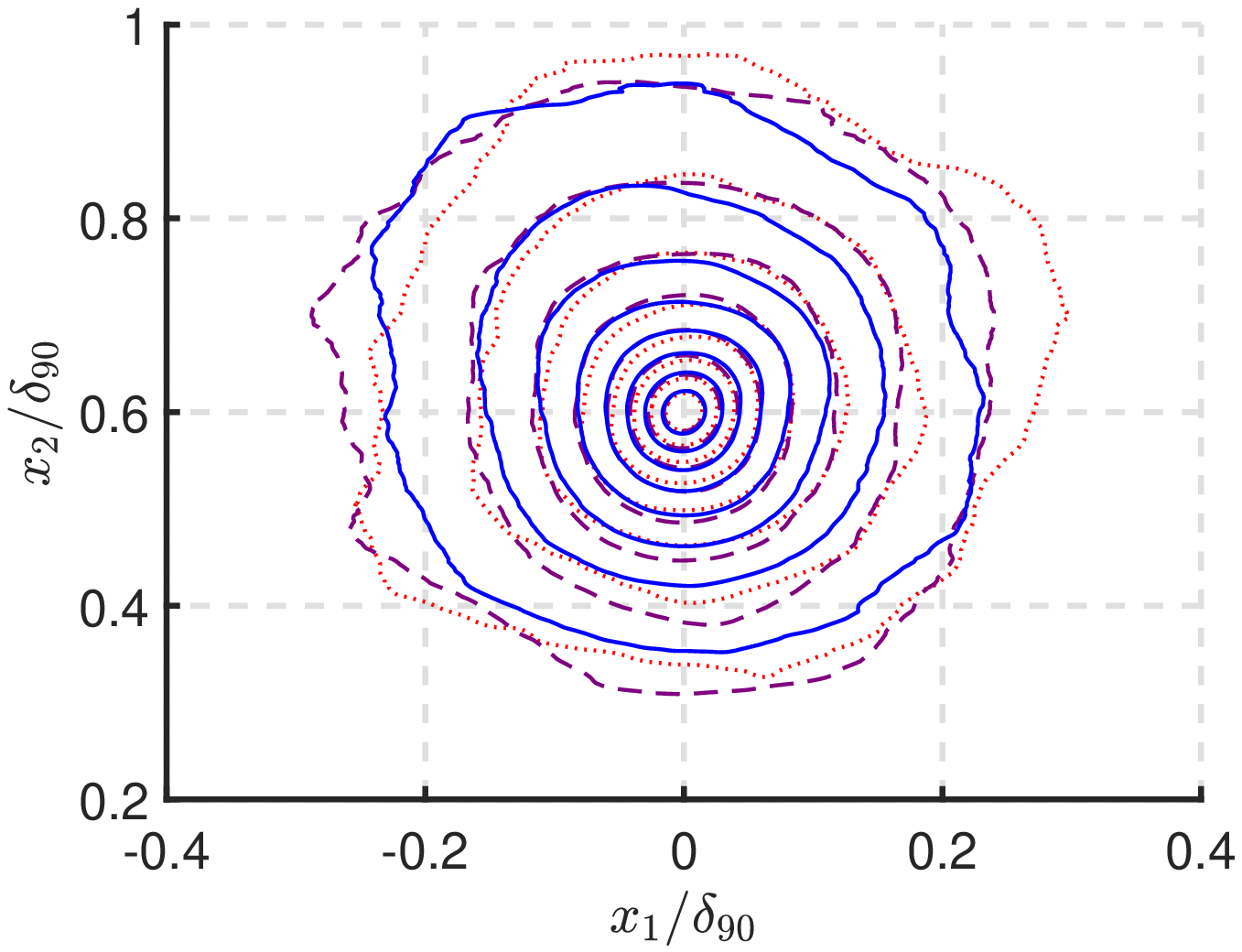} &
    \subfigimg[width=65 mm,pos=ul,vsep=15pt,hsep=32pt]{(d)}{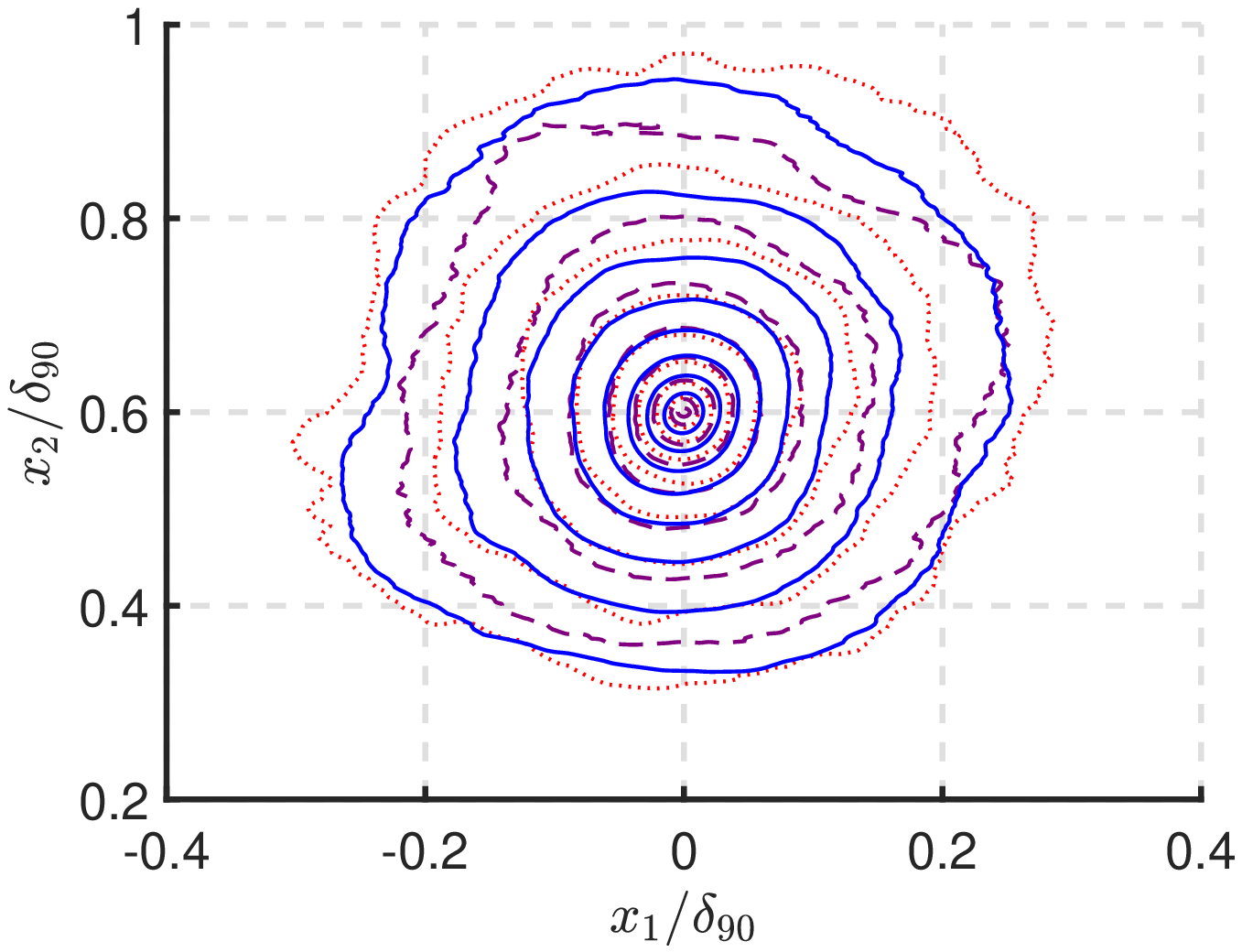} \\
    
  \end{tabular}
  \caption{Wall-normal velocity two-point zero time delay correlation, R$_{22}(x_1,x^{\prime}_1,x_2,x^{\prime}_2,x_3,x_3)$. \revPJ{Data in the figure correspond to measurements performed at U$_{\infty} \approx 10$ m/s.} Plots on the left are for ~3 \revPJ{mm} thick substrate while plots on the right are for ~15 \revPJ{mm} thick substrate. (a) Fixed point at ($0.07 ~\times ~ \delta_{90}$); (b)~Fixed point at ($0.07 ~\times ~ \delta_{90}$); (c) Fixed point at ($0.6 ~\times ~ \delta_{90}$); (d)~Fixed point at ($0.6 ~\times ~ \delta_{90}$). Legends: 10 PPI:red dotted line, 45 PPI:purple dashed, 90 PPI:blue solid. The iso-contour lines are from $0.2$ to $0.9$ with an increment of $0.1$.} 
  \label{fig:correlation_R22}
\end{figure*}
\begin{figure*}
  \centering
  \begin{tabular}{@{}p{0.5\linewidth}@{\quad}p{0.5\linewidth}@{}}
    \subfigimg[width=65 mm,pos=ul,vsep=15pt,hsep=32pt]{(a)}{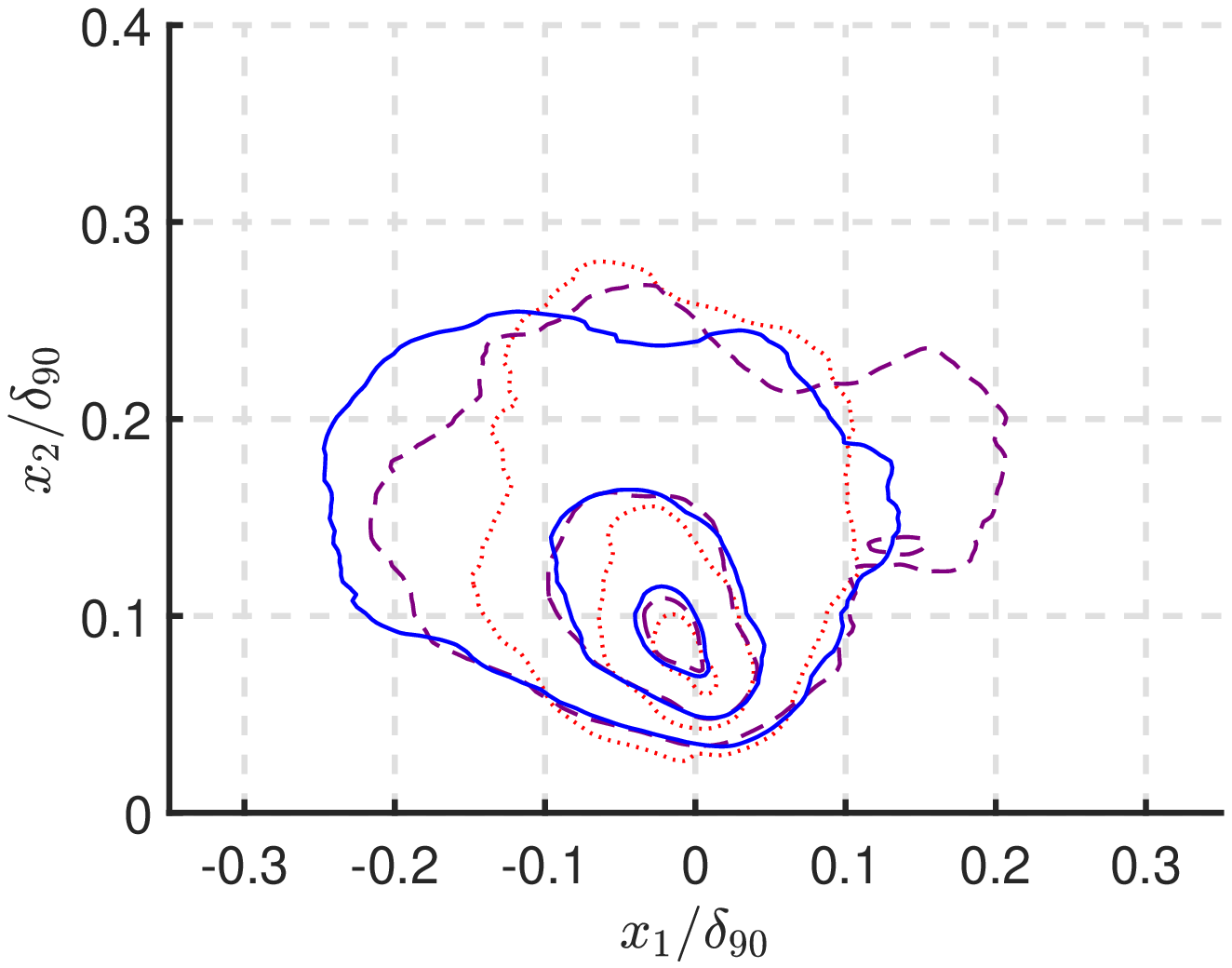} &
    \subfigimg[width=65 mm,pos=ul,vsep=15pt,hsep=32pt]{(b)}{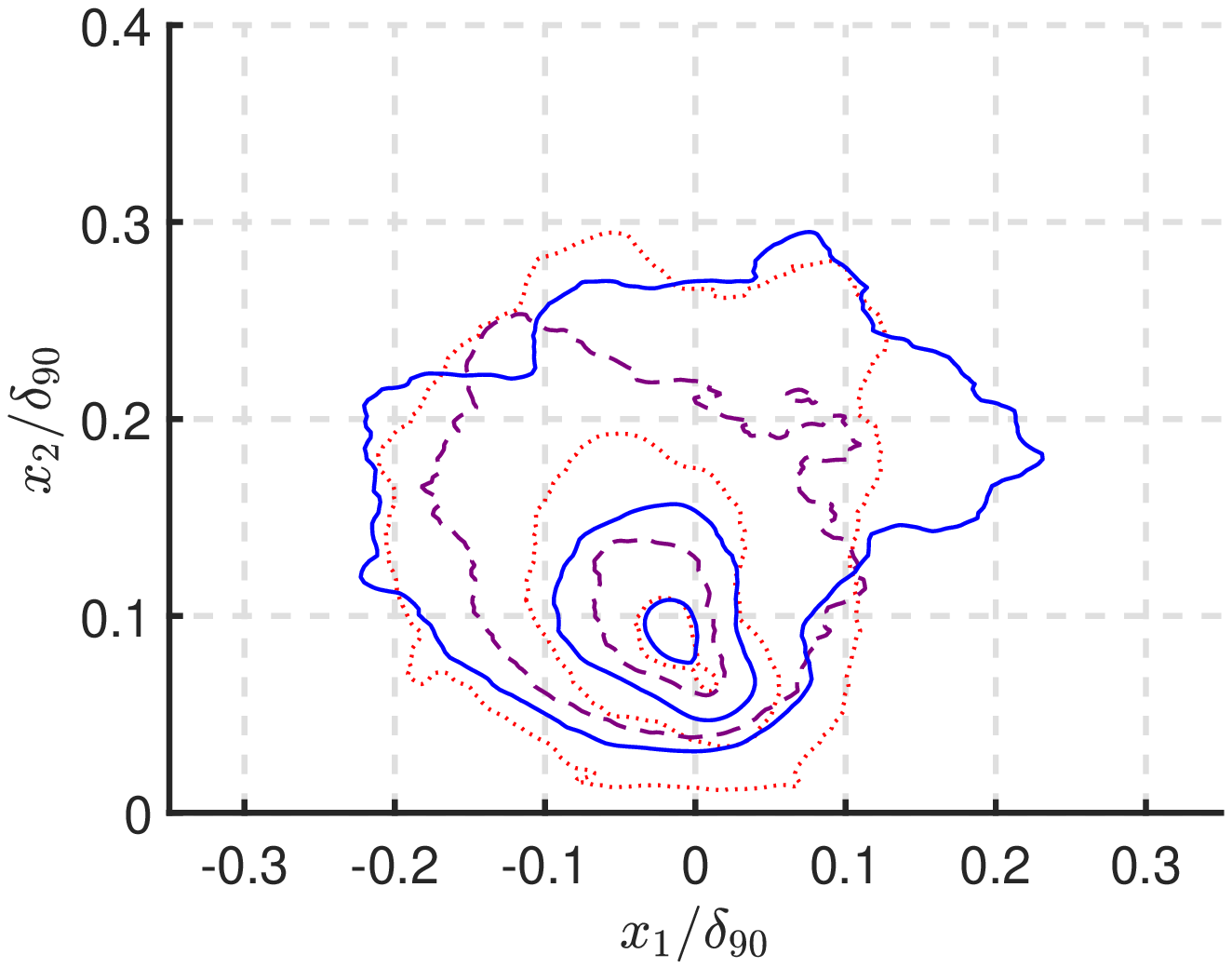} \\
   \subfigimg[width=65 mm,pos=ul,vsep=15pt,hsep=32pt]{(c)}{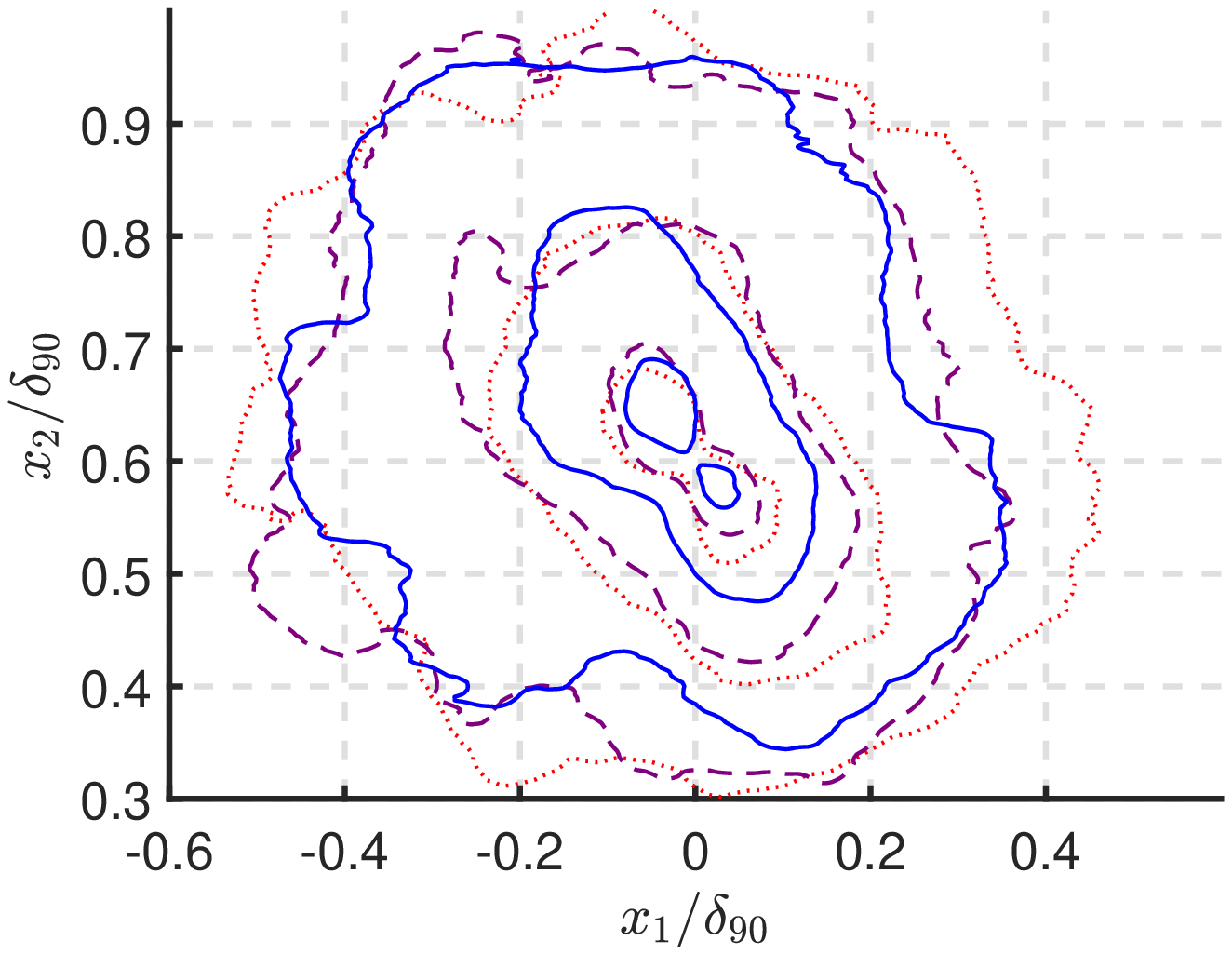} &
   \subfigimg[width=65 mm,pos=ul,vsep=15pt,hsep=32pt]{(d)}{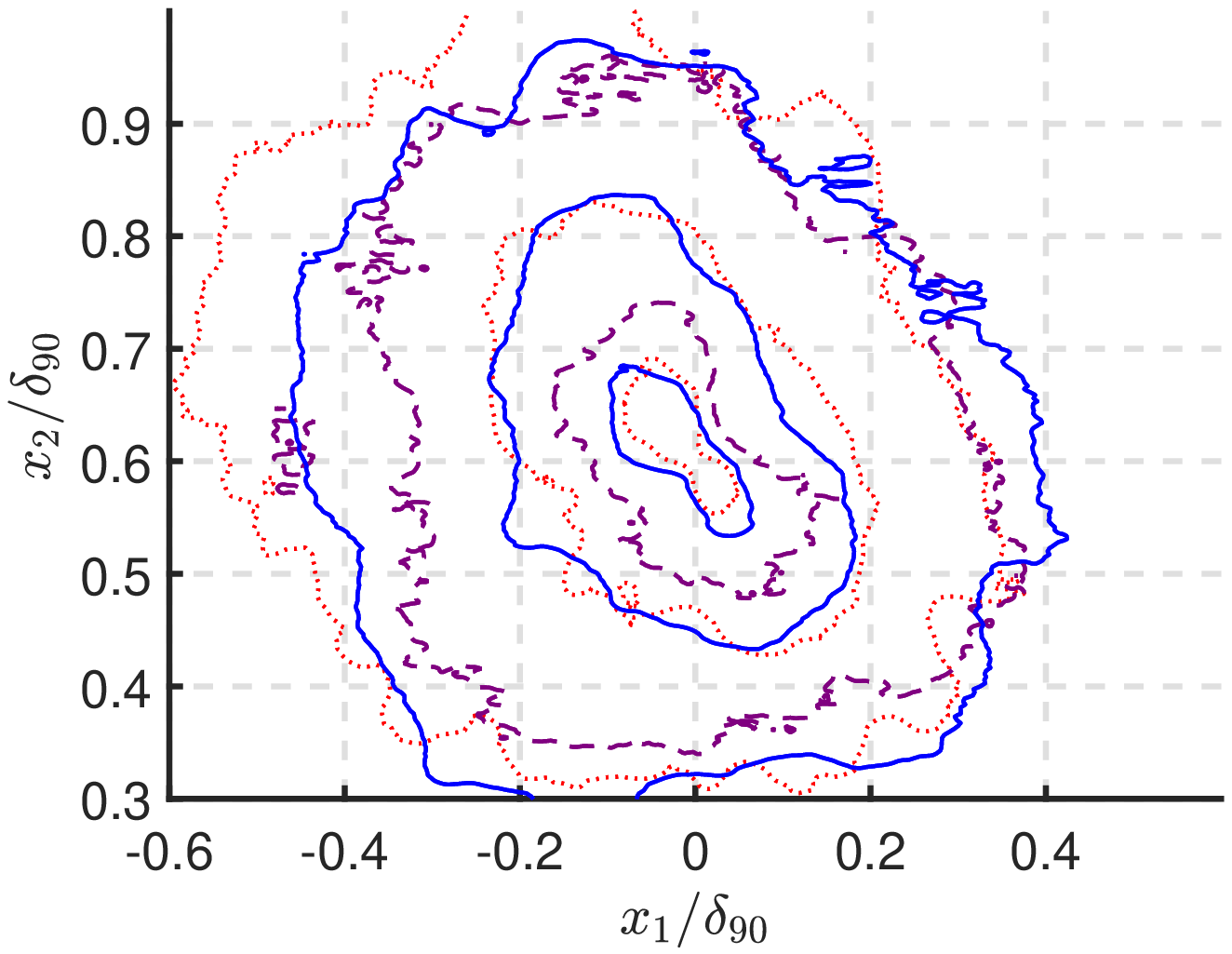} \\
    
  \end{tabular}
  \caption{~The two-point zero time delay correlation of \revPJ{the Reynolds shear stress} component R$_{12}(x_1,x^{\prime}_1,x_2,x^{\prime}_2,x_3,x_3)$. \revPJ{Data in the figure correspond to measurements performed at U$_{\infty} \approx 10$ m/s.} Plots on the left are for ~3 \revPJ{mm} thick substrate while plots on the right are for ~15 \revPJ{mm} thick substrate. (a) Fixed point at ($0.07 ~\times ~ \delta_{90}$); (b)~Fixed point at ($0.07 ~\times ~ \delta_{90}$); (c) Fixed point at ($0.6 ~\times ~ \delta_{90}$); (d)~Fixed point at ($0.6 ~\times ~ \delta_{90}$). Legends: 10 PPI:red dotted line, 45 PPI:purple dashed, 90 PPI:blue solid. The three iso-contours levels correspond to the values of $-0.4$, $-0.35$  $-0.30$. } 
  \label{fig:correlation_R12}
\end{figure*}

As explained in the previous sections, near wall PIV data ($\sim 2$ \revPJ{mm}) could not be used due to modulation error and reflections close to the wall. Furthermore, it must be remembered that PIV measurements truncate both large and small scales. On the one hand, the size of the camera sensor sets the upper limit on the largest scale that can be imaged. While on the other hand, the smallest scale that can be captured is directly proportional to the final window size \citep{foucaut2004piv}. Nevertheless, PIV measurements inherently show the spatial structure of turbulence without invoking Taylor's hypothesis. The two-point correlation maps, obtained using PIV measurements, are plotted in figures \ref{fig:correlation_R11}, \ref{fig:correlation_R22}, and \ref{fig:correlation_R12}. 

Figure \ref{fig:correlation_R11} shows the two-point zero time delay correlation for the \revPJ{streamwise} velocity correlation in the wall-normal plane ($x_1-x_2$). Plots on left correspond to the 3 \revPJ{mm} thick porous substrate and on the right correspond to 15 \revPJ{mm} thick substrate. The correlation maps for near-wall fixed points (\ref{fig:correlation_R11} (\textit{b}-\textit{d})) shows a poor collapse \revPJ{in the outer layer}. \revPJ{Note that we are using $\delta_{90}$ as the scaling variable instead of $\delta_{99}$ because the full-extent of the boundary layer is not captured for the 10 PPI and 15 mm thick substrate case, as mentioned earlier}. It is important to note that the entire extent of the \revPJ{streamwise} velocity correlation could not be captured; therefore, only values of correlation above $0.5$ are shown. As can be seen from figure \ref{fig:correlation_R11}, for any given point downstream of the fixed point, $R_{11}$ appears to be inclined away from the wall. The characteristic inclination of $R_{11}$ is linked to the statistical mean inclination of the hairpin structures with respect to the wall \citep{ganapathisubramani2005investigation}. In particular, a slight increase in angle could result in better access to higher momentum for hairpin structures. \cite{sillero2014two} reports the characteristic inclination of these hairpin structures are $\sim 10^{\circ}$. Surface roughness is known to increase the inclination angle of $R_{11}$. While \cite{volino2007turbulence,wu2010spatial} have reported a slight increase ($\sim{15^{\circ}}$) in the inclination angle of $R_{11}$ compared to smooth walls, \cite{krogstad1994structure} report almost a four-fold increase. In the present manuscript, average inclination were calculated following \cites{volino2007turbulence} procedure of fitting a line in a least square sense that passes through the iso-contours of $R_{11}$. The resulting angle close to the wall were found to be a function of the pore size (see table \ref{tab:angle}). In the present case, where both $k_s^{+}$ and $Re_{K}$ increase simultaneously, the $10$ PPI has the highest inclination ($\sim 20^{\circ}$) compared to other surfaces. It is important to mention that other studies  (\citealt{volino2007turbulence,wu2010spatial}) had tested surfaces with varying roughness but the inclination was found to be independent of $k_s^{+}$. The inclination appears to be independent of the thickness of foam, but scales with pore density. Therefore, the increased inclination could due to deeper penetration (filling up) of flow past porous foams compared to skimming off for flow past foams with low permeability  (e.g. $90$ PPI). {This suggests that with decreasing pore density, a transition to sparse canopy like behavior is obtained.} 


In the case of $R_{22}$, one can quantify overall correlation length as the field-of-view is large enough compared to overall extent of $R_{22}$. The vertical velocity correlations, shown in figure \ref{fig:correlation_R22}, appears to be more sensitive to wall-permeability and thickness. Firstly, $R_{22}$ appears to be symmetric in the \revPJ{streamwise} direction; however, a compression is observed in the wall-normal direction. Therefore, while permeable boundary condition with finite permeability does relax the wall blocking, it is not do enough to achieve symmetry in the wall-normal planes. The vertical velocity correlations for the $90$ PPI foam remains invariant as the thickness of the substrate is increased. This clearly shows when $Re_K \sim 1$, then foams behave like a smooth wall, and permeable effects are negligible. 
Interestingly, for the thicker foam substrate, as the \revPJ{permeability}  is increased, the overall extent of $R_{22}$ first decreases ($45$ PPI) and then increases ($10$ PPI), as evidenced from figure \ref{fig:correlation_R22} (\textit{b}). In contrast, \cite{carpio2019experimental} had reported decrease in correlation in the extent of $R_{22}$ with increasing permeability. It is noteworthy that while the flow over $45$ PPI case 15 \revPJ{mm} case is \revPJ{in the dense canopy regime}, the flow over $10$ PPI case 15 \revPJ{mm} case is in \revPJ{the sparse canopy regime \citep{sharma2020_sparse}}. Therefore, it appears that the permeable effects, which lead to the reduction in the extent of $R_{22}$ is no longer dominant in sparse foams, where the \revPJ{flow transitions to a sparse canopy regime}. Furthermore, when the overall extent of $R_{22}$ is normalized by the boundary-layer thickness, a good collapse is obtained (\ref{fig:correlation_R22} (\textit{a-c})) for the thinner substrate. This indicates that reduction in the extent of wall-normal velocity correlation ($R_{22}$) with permeability is ineffective for the $3$ \revPJ{mm} thick substrate, yielding a better collapse for all the cases well into the outer layer. In contrast, for the thicker foam, the influence of permeability is present well into the outer-layer (\ref{fig:correlation_R22} (\textit{d})), provided permeability is greater than the viscous scales ($Re_K>1$) and the foam operates at a dense ($s^+<100$) and deep ($h/s>10$) limits. At sparse foam limit deeper flow penetration can be seen from figure (\ref{fig:correlation_R22} (\textit{b})), this is inline with the \say{filling up effect} remark made earlier in conjunction with mean inclination of the hairpin structures with respect to the wall.


\begin{table}
\begin{center}
\begin{tabular}{c c c } 
Foam & $h$ & Angle \\
(PPI)& (mm)&  (Degree)  \\ \hline
\multirow{2}{*}{90}   & 3   &  {13}  \\
   & 15  &{13.4}   \\
\hline
\multirow{2}{*}{45}   & 3   &  {15.8}  \\
   & 15  &{16.2}   \\
\hline
\multirow{2}{*}{10}   & 3   &  {20.6}  \\
   & 15  &{19.4}   \\
\hline

\end{tabular}
\end{center}
\caption{Angle of $R_{11}$ at $0.07 \times \delta_{90}$.}
\label{tab:angle}
\end{table}

The two-point correlation of \revPJ{Reynolds shear stress} component (the \revPJ{streamwise} and wall-normal velocity fluctuations), $R_{12}$, in $x_1-x_2$ plane, is plotted in figure \ref{fig:correlation_R12}. The correlation $R_{12}$ is representative of the extent to which the \revPJ{streamwise} velocity are associated with a single ejection or sweep event, induced by the wall-normal velocity fluctuations. Recently, \cite{gul2021revisiting} showed for rough walls, $R_{12}$ scales with the boundary-layer thickness and is independent of the $k_s^+$ or the K\'arm\'an number. For the 15 \revPJ{mm} substrate, as the \revPJ{permeability}  is increased, the $R_{12}$ first decreases ($45$ PPI) and then increases ($10$ PPI). This behaviour is similar to $R_{22}$ correlation map, as shown earlier. In particular, the extent of the correlation map $R_{12}$ seems to be shortest for $45$ PPI 15 \revPJ{mm} thick substrate. In fact, the maximum iso-contour levels plotted, e.g. $-0.4$, is visibly absent for the $45$ PPI $15$ \revPJ{mm} thick substrate case. {Therefore, in response to the first objective of the present paper, the inability of boundary-layer thickness to collapse the overall extent of the $R_{22}$ and $R_{12}$ (well into the outer-layer) suggests that Townsend's outer-layer similarity for these higher order quantities may not be valid for $15$ \revPJ{mm} thick porous substrates.}



Finally, \cite{manes2011turbulent,efstathiou2018mean} have linked improved mixing for the thickest and most permeable foams to the presence of KH instability. Furthermore, \cite{manes2011turbulent} and \cite{sharma2020_dense} state such KH type instability occur at the interface of porous substrate, and at a distance $x_2/\delta_{99}<0.1$. Indeed, KH instability are known to induce large quasi two-dimensional rollers, which leads to periodic organisation of wall-normal flow disturbances \revJFM{\citep[see figure 5.23 of][for instance]{jaiswal2020etude}}. {Although the figures \ref{fig:correlation_R22} show limited streamwise extent, no periodic structures or modes associated with KH instability were observed for $R_{22}$ or $R_{12}$ (figures \ref{fig:correlation_R22} (\textit{d})) within the measurement domain. This suggests that no KH type flow instability may be present in the cases that were investigated.}



As mentioned earlier, only the length scales associated with wall-normal can be quantified due to limited bandwidth (FOV) of our PIV measurements. In the present work, the turbulence correlation length is defined as:

{\begin{equation} 
\Lambda^{\vert k \pm}_{ij}(x_i)  =    \int_{-\infty}^{\infty} R_{ij}(x_i,\textbf{x}_{k\pm}) \, {\rm d}x_{k\pm}
\label{eq:lscale}
\end{equation}}

Here, $\textbf{x}_{k\pm}$ is the separation vector in the direction $k$. The subscript $\pm$ denotes moving point direction. For instance, if the fixed point is located above the moving points for the wall-normal velocity calculations, then the integration of \eqref{eq:lscale} will yield $\Lambda^{\vert 2-}_{22}(x_2)$. This way of defining length scale is particularly appropriate for \revJFM{inhomogeneous} turbulence \citep[see][for instance]{jaiswal2020use}. However, as noted by \cites{sillero2014two}, a clear physical interpretation of the length scale is difficult. The present study seeks to compare the effects of porous surfaces on the large scale turbulence structures, therefore a contextual interpretation of the length-scale can be used where it is a metric to quantify overall spatial correlation of velocity disturbances. $R_{ij}$ being a statistical quantity, equation \eqref{eq:lscale} cannot be used directly without accumulating averaging $\epsilon_{R_{ij}}$ errors (see equation \eqref{eq:un_lscale}). In order to avoid the accumulation of errors, length scales were estimated by fitting an exponential decay function \citep[see][for implementation details]{jaiswal2020use}.  

\begin{figure*}
  \centering
  \begin{tabular}{@{}p{0.5\linewidth}@{\quad}p{0.5\linewidth}@{}}
    \subfigimg[width=65 mm,pos=ul,vsep=15pt,hsep=32pt]{(a)}{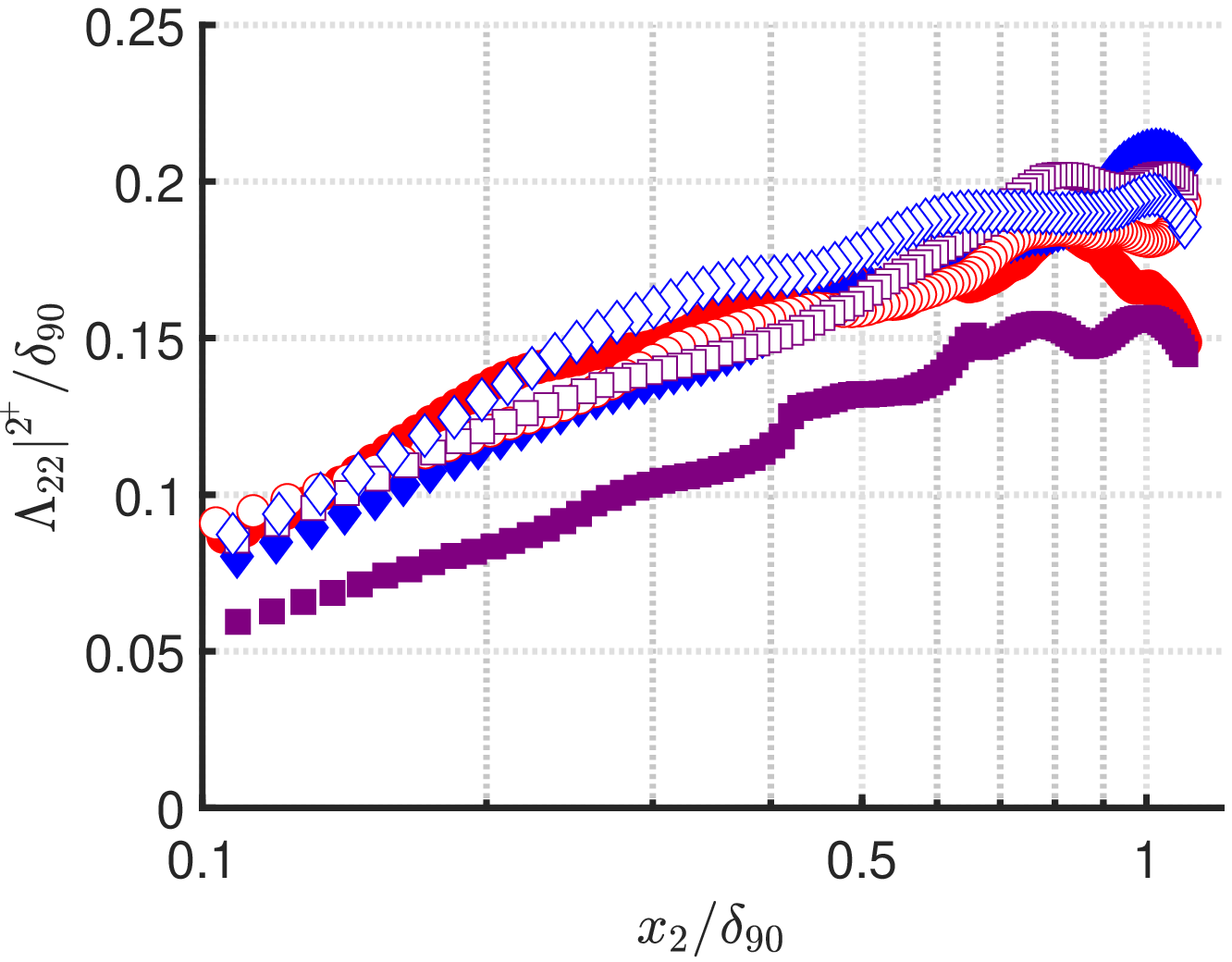} &
    \subfigimg[width=65 mm,pos=ul,vsep=15pt,hsep=32pt]{(b)}{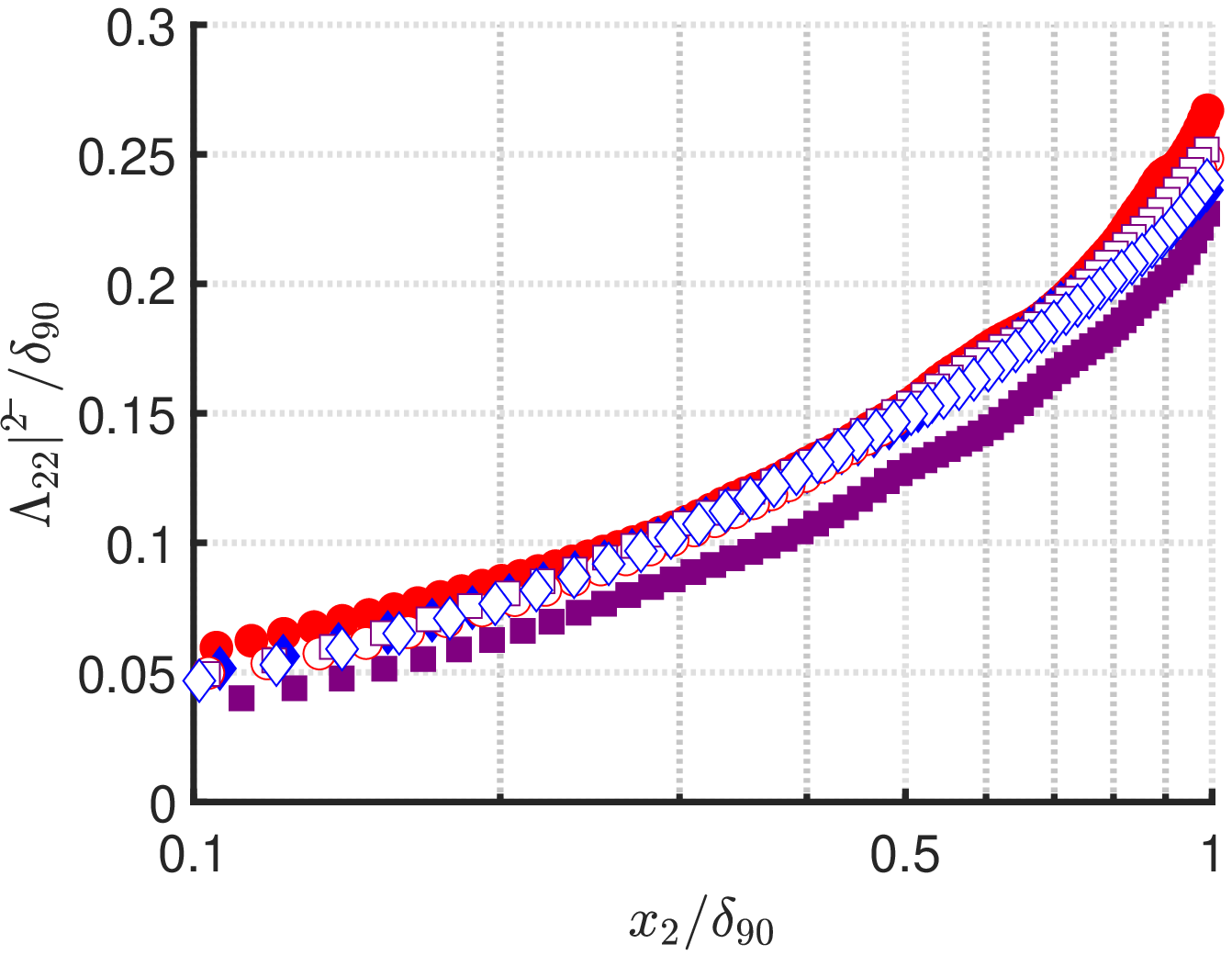} \\
    
  \end{tabular}
  \caption{Longitudinal length scales of wall-normal velocity component, \revPJ{estimated from PIV measurements performed at U$_{\infty} \approx 10$ m/s.} (a)~Integral length scale $\Lambda^{\vert 2+}_{22}(x_2)$ (b)~Integral length scale $\Lambda^{\vert 2-}_{22}(x_2)$
  Legends: Red circles 10 PPI, purple squares 45 PPI, and blue diamonds 90 PPI. Open symbols are for $3$ \revPJ{mm} thick substrate, while filled symbols correspond to $15$ \revPJ{mm} substrate.} 
  \label{fig:spatialscales}
\end{figure*}

Figure \ref{fig:spatialscales} shows the longitudinal correlation length scales for wall-normal velocity component. Figure \ref{fig:spatialscales} (\textit{a}-\textit{b}), shows the wall-normal correlation lengths, $\Lambda^{\vert 2-}_{22}(x_2)$ and $\Lambda^{\vert 2+}_{22}(x_2)$, in the $x_1$-$x_2$ plane. The length scale $\Lambda^{\vert 2-}_{22}(x_2)$ should be particularly sensitive to the blocking effects induced by the wall. 
This is not surprising because the blocking effects are pre-dominant when approaching the wall \citep[see][for instance]{jaiswal2020use}. Nevertheless, difference in length scale for thicker $15$ \revPJ{mm} porous substrate is more visible. For instance, correlation length scales ($\Lambda^{\vert 2-}_{22}(x_2)$) for the $45$ PPI substrate is smaller throughout the boundary layer. Surprisingly, the length scale $\Lambda^{\vert 2+}_{22}(x_2)$ shows even more substantial reduction for the $45$ PPI and 15 \revPJ{mm} thick foam. {Therefore, in response to the second objective of the present paper, no KH type flow instabilities were observed while substantial differences in length scales ($\Lambda^{\vert 2+}_{22}(x_2)$) are observed in the boundary-layer above the porous foam at a thick substrate limit ($h/s>10$).}

As mentioned previously, the length scale associated with \revPJ{streamwise} velocity disturbance could not be quantified due to limited field of view.
However, thanks to single-wire measurements, a high-fidelity estimation of time-scales associated with \revPJ{streamwise} velocity fluctuations is possible. Similar to length scale, the time scale can be defined as:

\begin{equation} \label{eq:tscale}
T({\textbf{x}_{\textbf{i}}})= \int_{-\infty}^{\infty} R_{11} \big( {\textbf{x}_{\textbf{i}}}(t),{\textbf{x}_{\textbf{i}}}(t+{\rm{d}}t) \big)  {\rm{d}}t
\end{equation}

In order to reduce error in the estimation of $T({\textbf{x}_{\textbf{i}}})$, the temporal correlations were fitted with an exponential decay function $\exp(-a{x}_k)$ to reduce the accumulation of errors in estimating $T({\textbf{x}_{\textbf{i}}})$.

\begin{figure*}
  \centering
  \begin{tabular}{@{}p{0.5\linewidth}@{\quad}p{0.5\linewidth}@{}}
    
    \subfigimg[width=65 mm,pos=ur,vsep=15pt,hsep=12pt]{(a)}{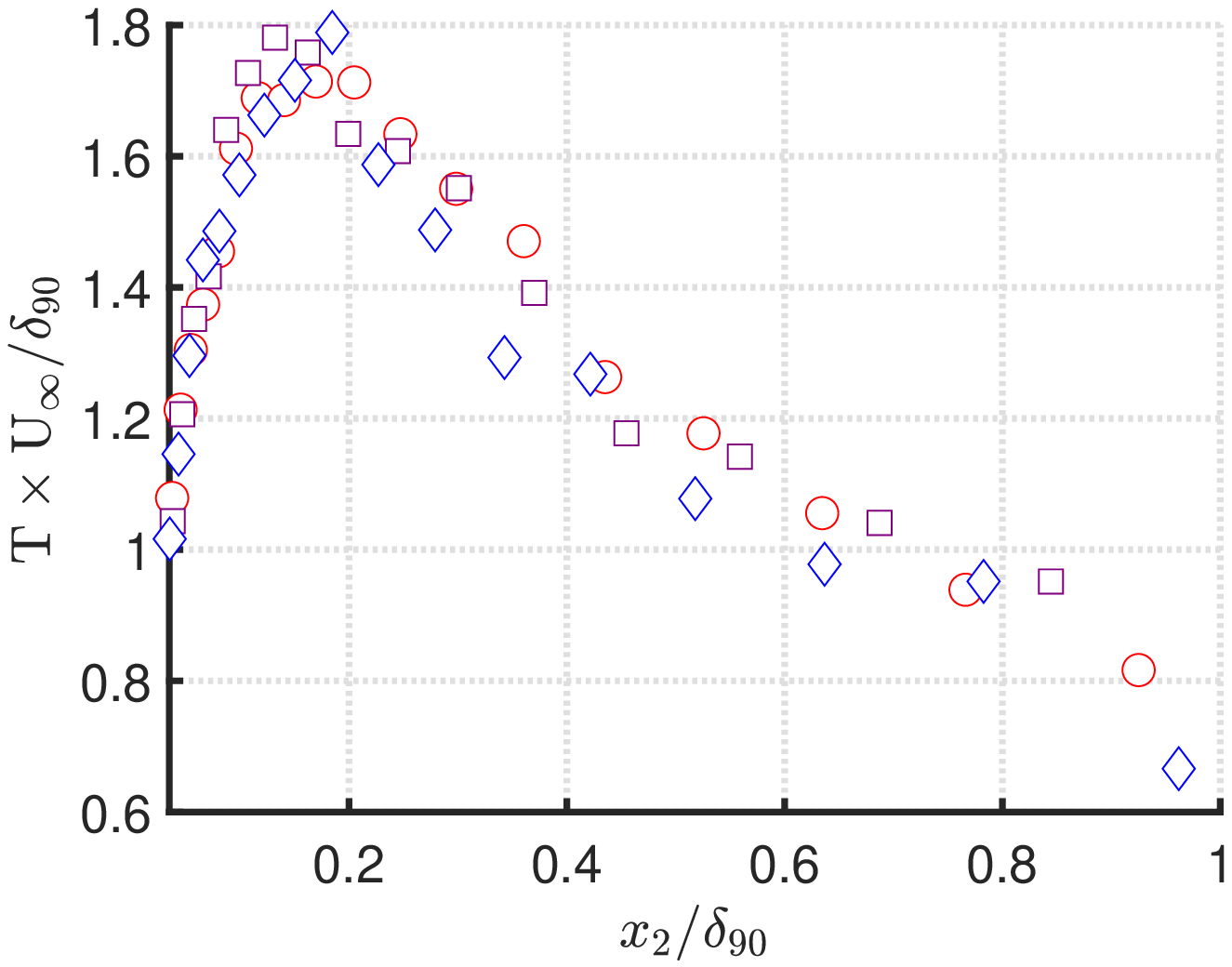} &
    \subfigimg[width=65 mm,pos=ur,vsep=15pt,hsep=12pt]{(b)}{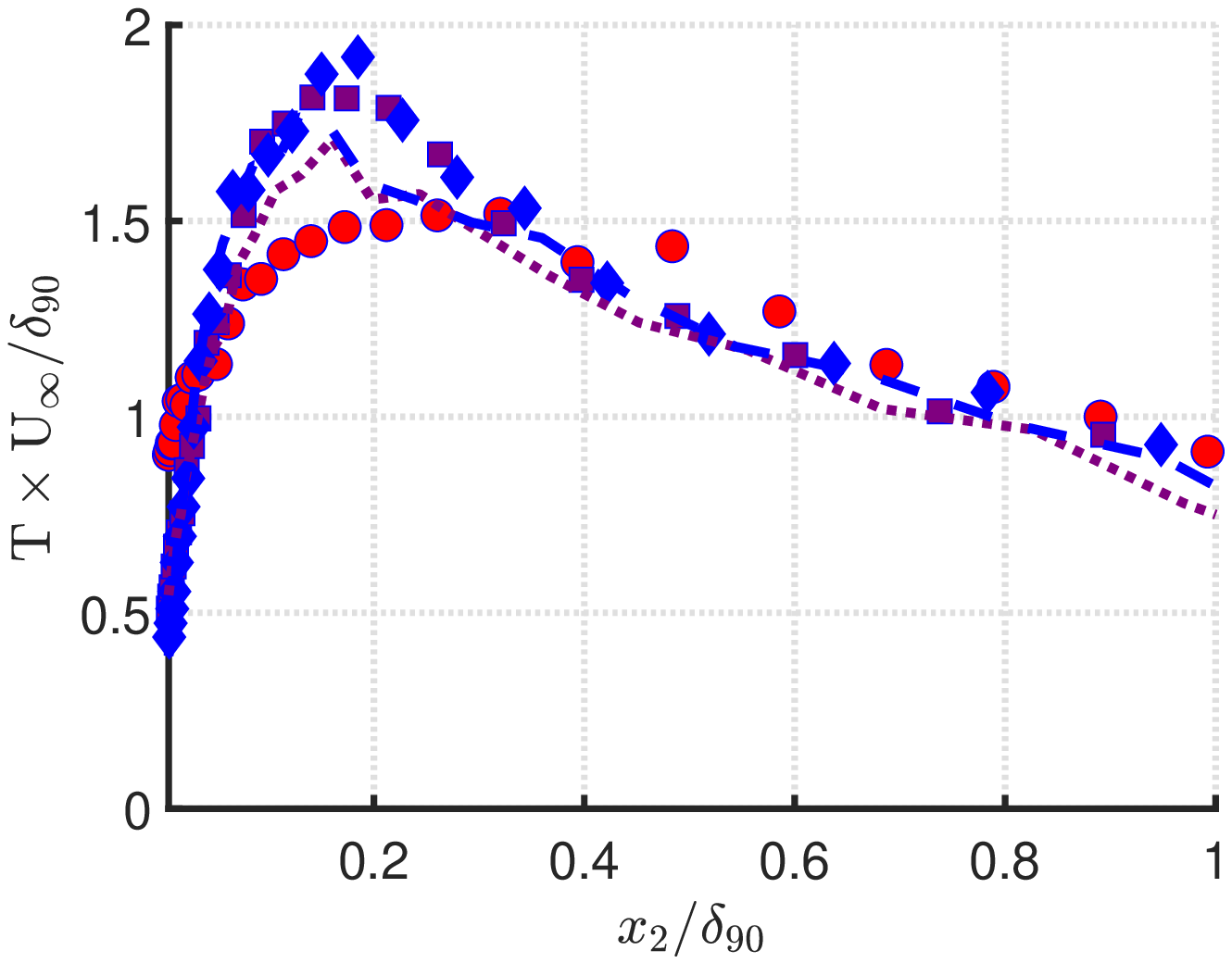} \\
    
    
    
  \end{tabular}
  \caption{Integral scales of turbulence. \revPJ{Integral scales estimated from HWA measurements performed at U$_{\infty} \approx 10$ m/s are shown in solid or empty symbols (circles, squares or diamonds).} 
  (a)~Integral time scale T for 3 \revPJ{mm} thick substrates (b)~Integral time scale T for 15 \revPJ{mm} thick substrates.
  Legends: Red circles 10 PPI, purple squares 45 PPI, and blue diamonds 90 PPI. Blue dashed and purple dotted lines correspond to $90$ PPI and $45$ PPI foams respectively, at similar $Re_{\tau}$ $\sim7000$ and $15$ \revPJ{mm} thick foam substrate.} 
  \label{fig:timescales}
\end{figure*}

The time scales thus calculated are plotted in figure \ref{fig:timescales}. The time scales have been normalised with outer-layer variables $U_{\infty}$ and $\delta_{90}$, as such the plots show time scale per unit boundary-layer turnover time. The figures \ref{fig:timescales} (\textit{a}-\textit{b}) show that the time-scales appear to collapse for the $3$ \revPJ{mm} substrate irrespective of the pore density (PPI). For the $15$ \revPJ{mm} substrate, the most porous foam ($10$ PPI) compares poorly in the near-wall region. More specifically, the $10$ PPI 15 \revPJ{mm} substrate has the shortest eddy turnover time. Since, the measurements were performed at different $Re_{\tau}$, therefore, two additional cases at similar $Re_{\tau}$ and thickness are plotted in figures \ref{fig:timescales} (\textit{b}). As can be seen from figure \ref{fig:timescales} (\textit{b}), the $10$ PPI 15 \revPJ{mm} substrate remains an outlier, as it has the shortest eddy turnover time. Nevertheless, at these Reynolds number \revPJ{$Re_{\tau} \sim 7000$ the large energetic structures are pushed away from the wall and roughness breaks the inner peak.} This also explains a slight reduction in eddy turnover time for $45$ and $90$ PPI foams at $Re_{\tau} \sim 7000$.  

The hot-wire data for all the cases can be further explored to examine the spectral content of the turbulent structures and the similarity (or lack thereof) between the different substrates can be further elucidated.

\subsection{Spectral analysis of deep and shallow flows over porous foam. \label{sec:level4c}}



\begin{figure*}
  \centering
  \begin{tabular}{@{}p{0.5\linewidth}@{\quad}p{0.5\linewidth}@{}}
   
     \hspace{3.5 cm}
  \subfigimg[width=65 mm,pos=ul,vsep=20pt,hsep=32pt]{}{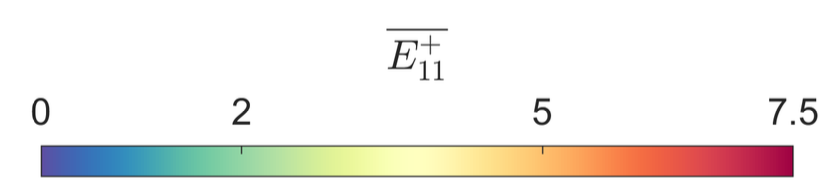} 
    \\
    
    \subfigimg[width=65 mm,pos=ul,vsep=2pt,hsep=38pt]{(a)}{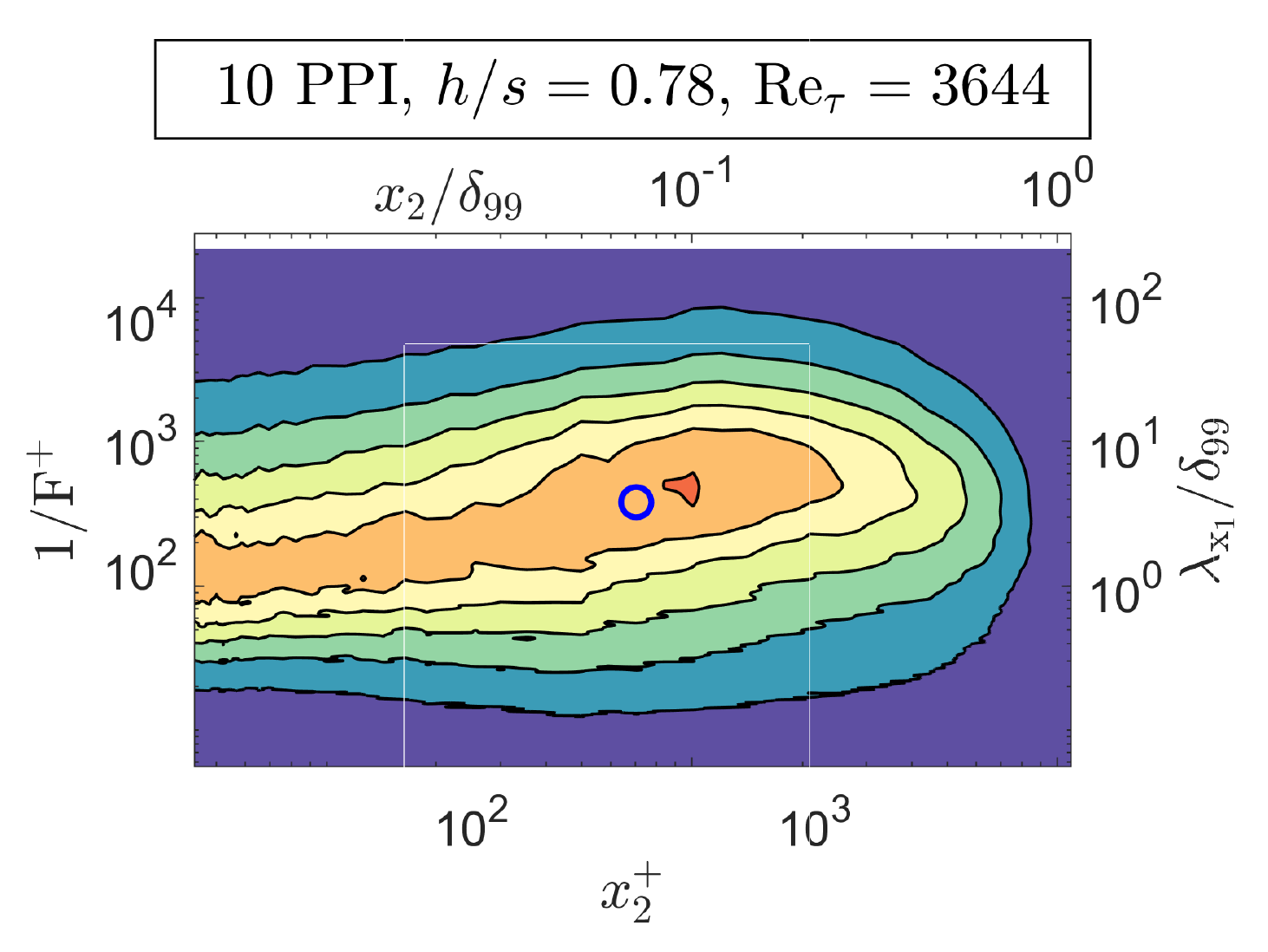} &
    \subfigimg[width=65 mm,pos=ul,vsep=2pt,hsep=38pt]{(b)}{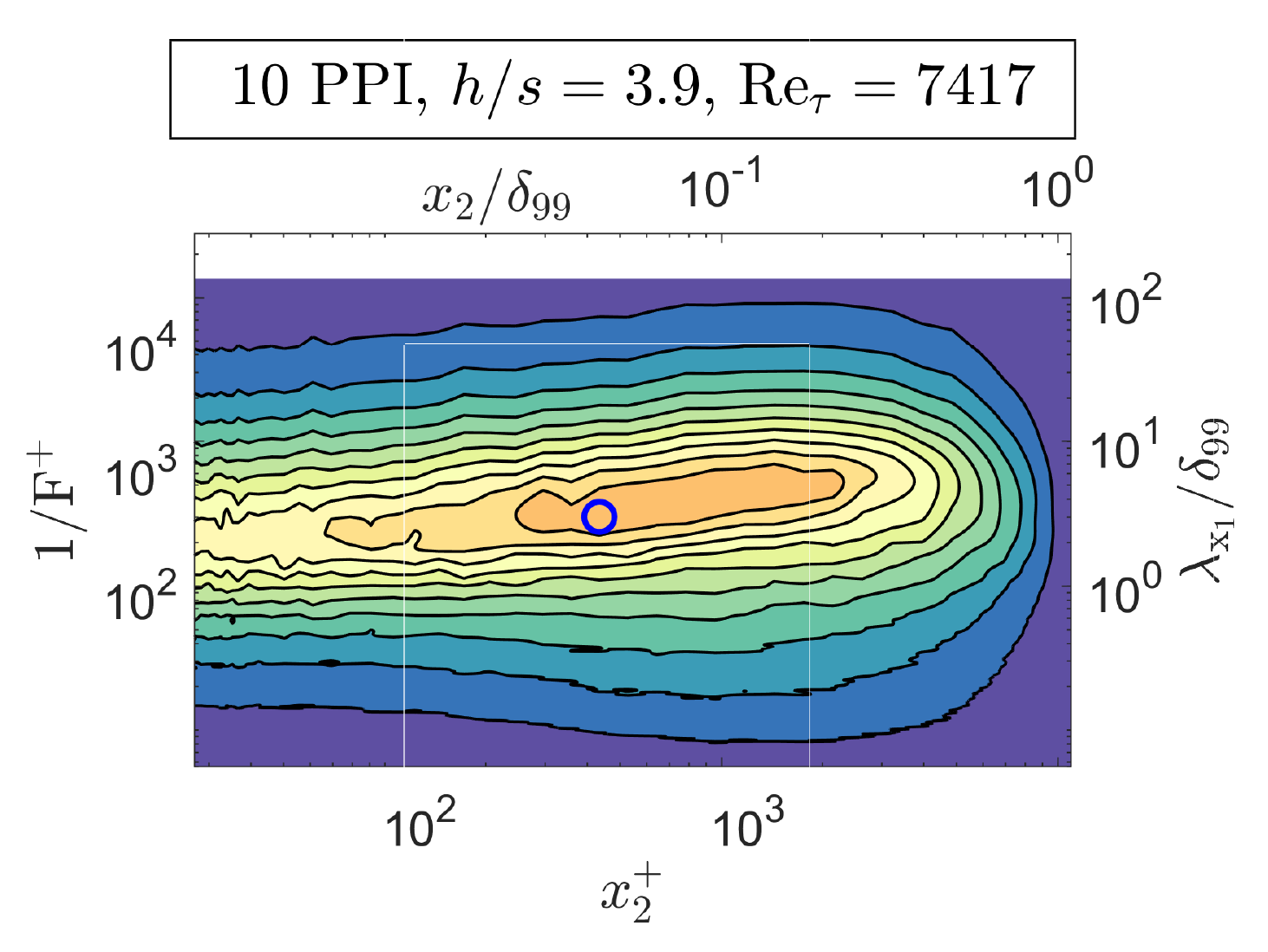} \\
     \subfigimg[width=65 mm,pos=ul,vsep=2pt,hsep=38pt]{(c)}{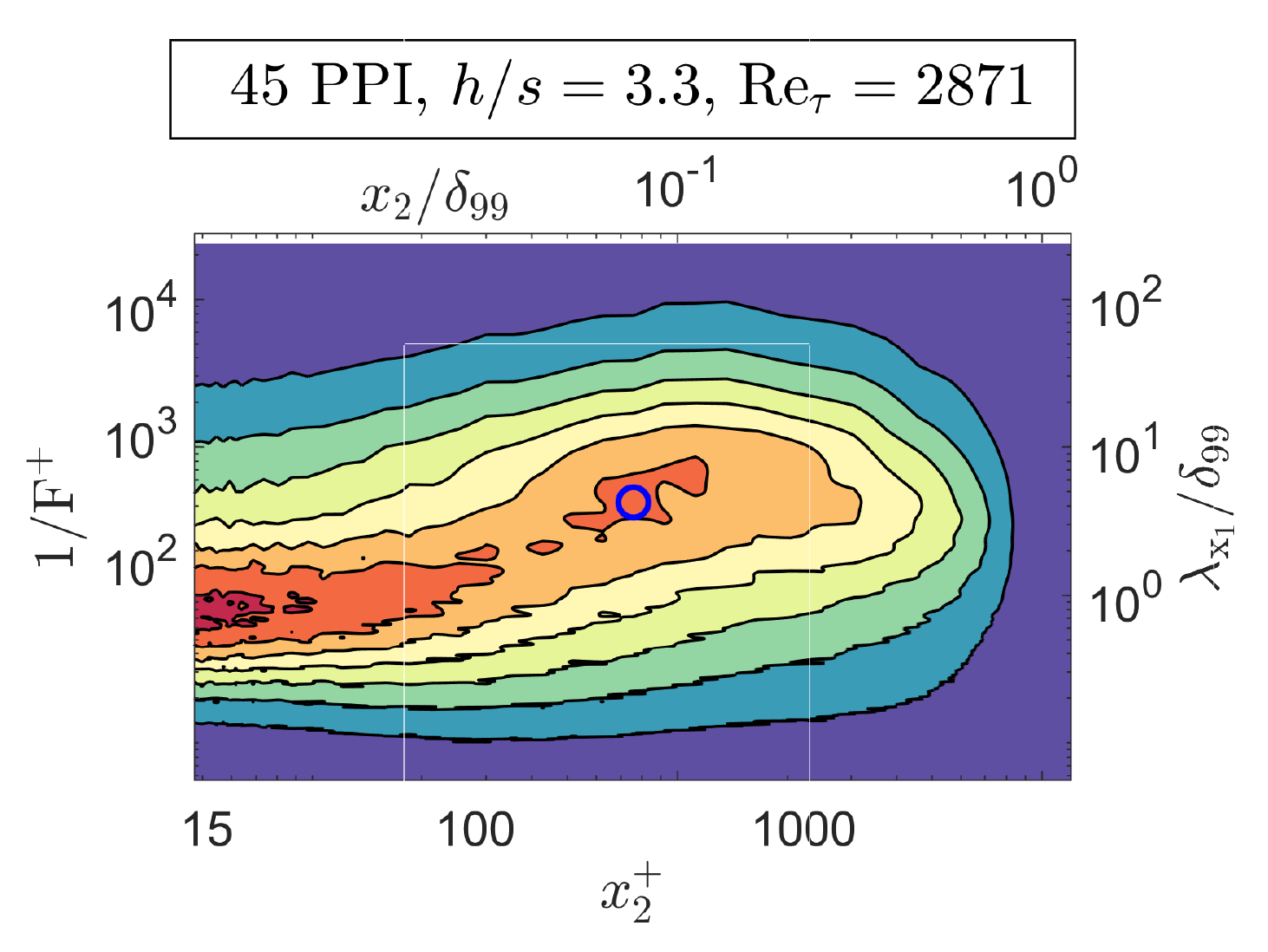} &
    \subfigimg[width=65 mm,pos=ul,vsep=2pt,hsep=38pt]{(d)}{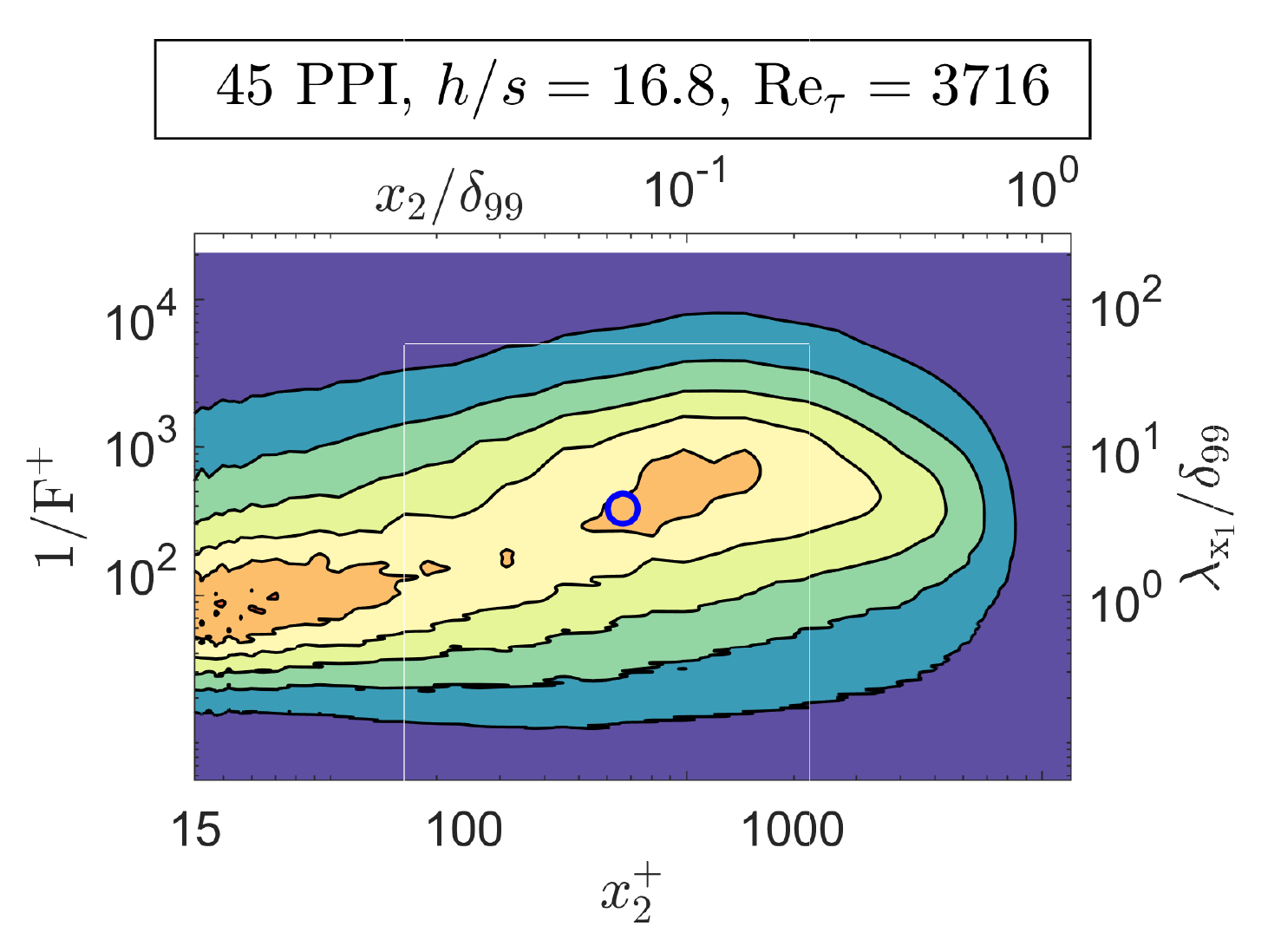} \\
     \subfigimg[width=65 mm,pos=ul,vsep=2pt,hsep=38pt]{(e)}{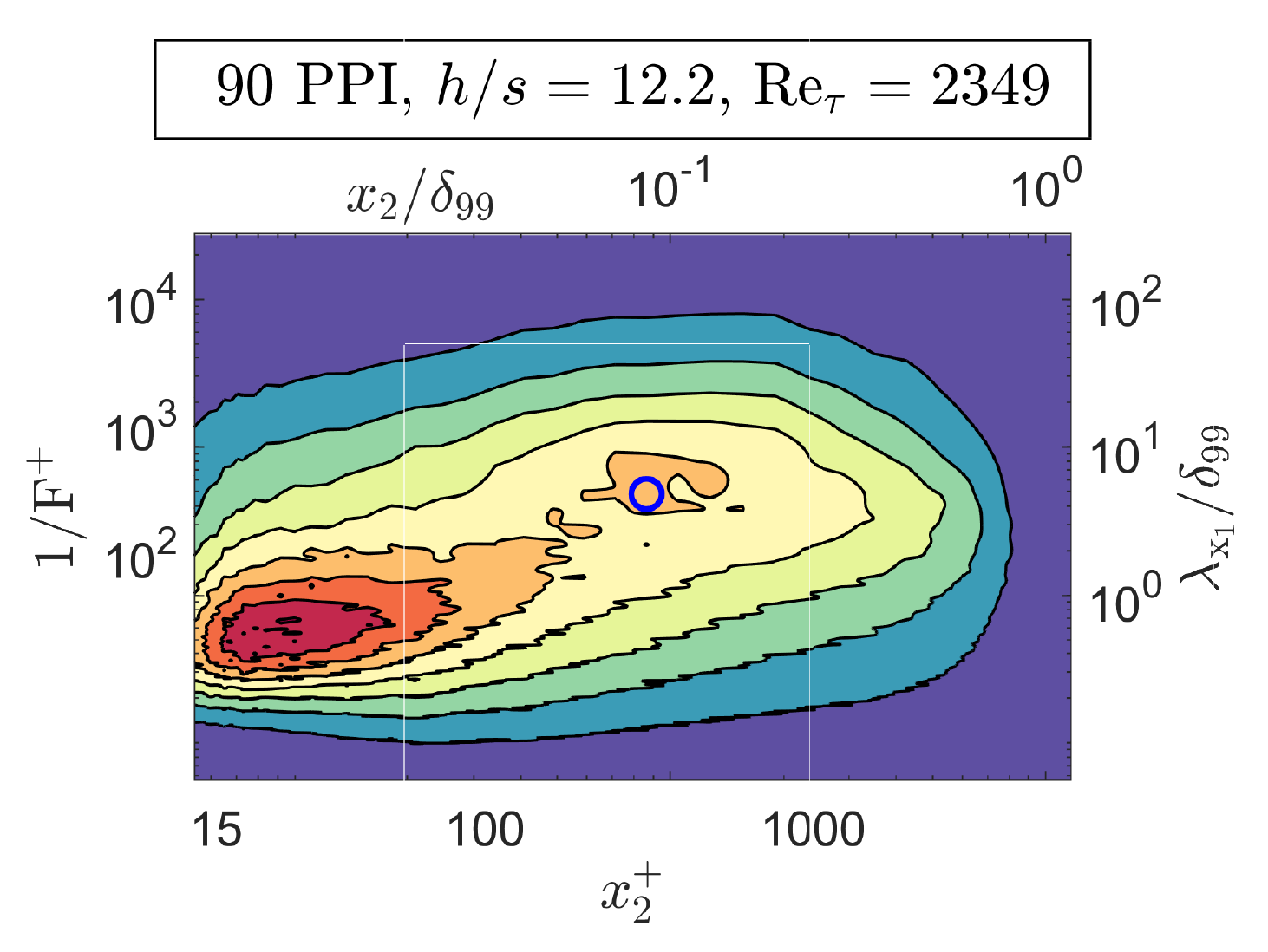} &
    \subfigimg[width=65 mm,pos=ul,vsep=2pt,hsep=38pt]{(f)}{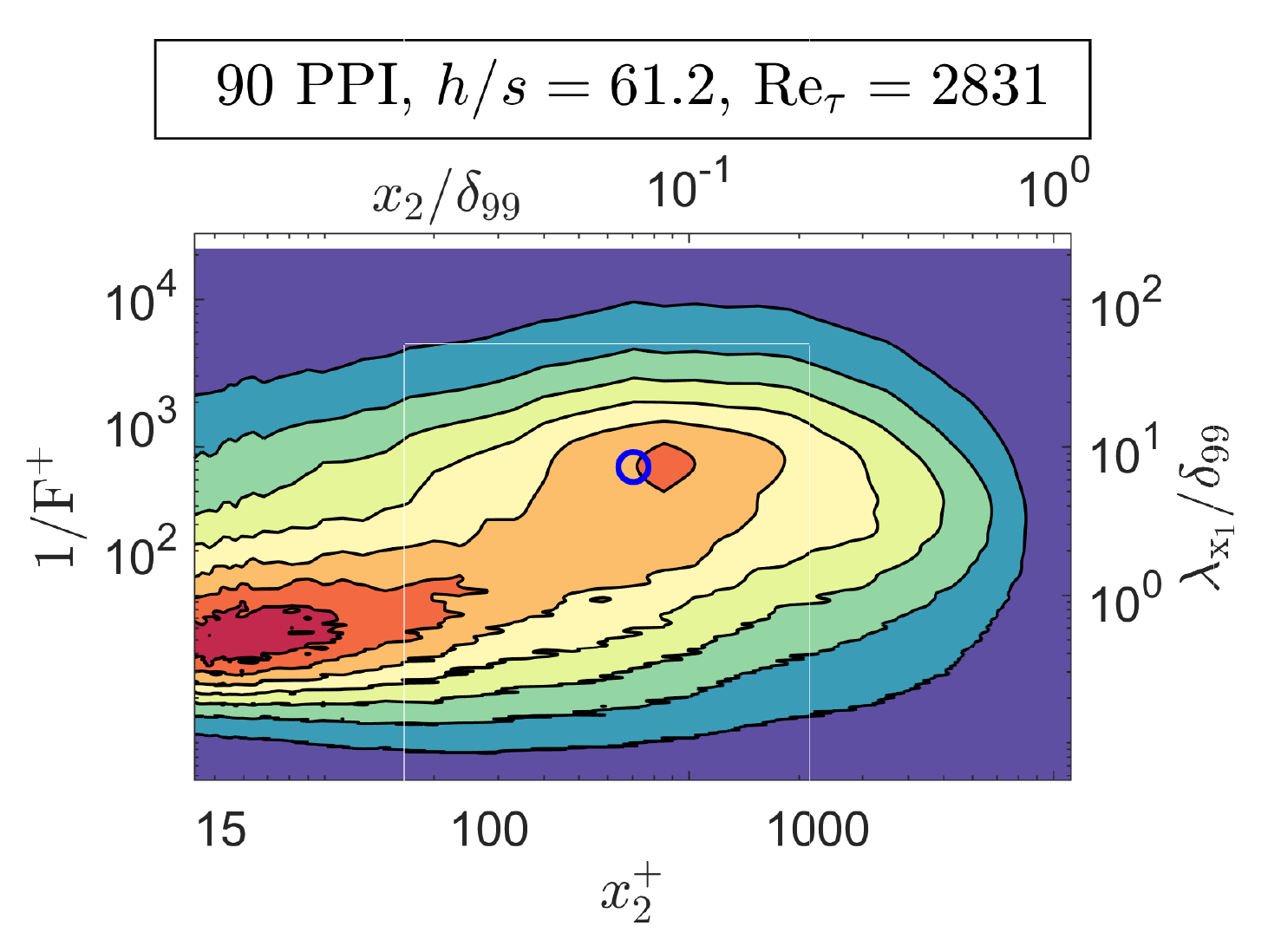} \\
  \end{tabular}
  \caption{Premultiplied 1-D \revPJ{streamwise} velocity energy spectra,\revPJ{ { $\overline{E_{11}^+}$}\bigg($\frac{2\pi\rm{F} \times E_{11}}{U_{\tau}^2}$\bigg),~measured at a distance of $x_1$ $=~ 3.3$ {m} downstream of inlet at $U_{\infty} \approx 10$ m/s.} (a) 3 \revPJ{mm} thick $10$ PPI foam, (b) 15 \revPJ{mm} thick $10$ PPI foam, (c) 3 \revPJ{mm} thick $45$ PPI foam, (d) 15 \revPJ{mm} thick $45$ PPI foam, (e) 3 \revPJ{mm} thick $90$ PPI foam, (f) 15 \revPJ{mm} thick $90$ PPI foam. 
  \revPJ{The blue circle represents ${x_2}^+=3.9\sqrt{\rm{Re}_{\tau}}^{} $~\citep{mathis2009large}.}} 
  \label{fig:E11_inner}
\end{figure*}

In order to obtain frequency related information of the turbulent kinetic energy associated with \revPJ{streamwise} velocity disturbances, pre-multiplied wall parallel turbulent energy spectra (\revPJ{$\overline{E_{11}^+}$}) were computed. 
\revPJ{Figure \ref{fig:E11_inner} shows contours of the pre-multiplied energy spectrogram.~In~figure \ref{fig:E11_inner}, the time scale ($1/\rm{F}^+$) and the wall-normal distance ($x_2^+$) in wall units are plotted on the left and bottom axis of the figure respectively. The top axis on the figure corresponds to the wall-normal distance normalised by $\delta_{99}$. Finally, the temporal axis ($1/\rm{F}^+$) is converted to the
spatial axis ($\lambda_{x_1}/\delta_{99}$), labeled on the right hand side of the figure,~assuming a constant convection velocity of $U_{\infty}$. Note that the x-axis limits are slightly different within the subplots.}

Various cases have been ordered based on the thickness to pore ratio ($h/s$) at same fetch-based Reynolds number ($Re_{x_1}$). The figures on the left column are for $3$~\revPJ{mm} thick substrate while figures on the right correspond to $15$~\revPJ{mm} thick substrate. The rows are arranged so that top row corresponds to porous substrates with highest permeability, while the lowest permeability substrates are at the bottom row. For the 3~\revPJ{mm} 90 PPI case, the \revPJ{near-wall} peak is observed \revPJ{around~${x_2}^+\approx 15$ (figure \ref{fig:E11_inner} (\textit{e})), which is consistent with the smooth wall literature \citep{mathis2009large}}. For the same substrate thickness, as permeability increases at first the \revPJ{near-wall} peak moves closer to the wall ($45$ PPI case), \revPJ{and appears to be more spread throughout the logarithmic region}. \revPJ{More importantly, the more permeable (45 PPI) substrate tends \revPJ{to}~break-up near-wall structures and the energy is \revPJ{reduced}.}~With a further increase in permeability ($10$ PPI case) the \revPJ{near-wall} peak in \revPJ{{$\overline{E_{11}^+}$}} (see figure \ref{fig:E11_inner} (\textit{a})) is smeared \revPJ{out}. \revPJ{Finally, for the thick foam substrates the near-wall energy peak is both smeared out and the near wall-peak ceases to exist for the thick 10 PPI foam (see figure \ref{fig:E11_inner} (\textit{b})).} \revPJ{For flows past rough wall \citep{Squire2016}, an increase in $k_s^+$, leads to a reduction in near-wall peak. While for an impermeable rough walls the inner-wall peak, associated with near wall cycle, is absent for $k_s+ \ge 70$, the absence of near-wall peak is only observed for $k_s+ > 350$ in the present study.}



\revPJ{Figures \ref{fig:E11_inner} (\textit{c})~and~\ref{fig:E11_inner}~(\textit{f}) show energy spectra for cases at~$Re_{\tau} \approx 2800$ where the flow is at the deep foam limit($h/s > 1$ )) and at a \revPJ{dense canopy flow regime}. The outer-layer peak energy does not show any significant dependence on $h/s$. This could be due to the fact that for dense (small $s^+$) porous surfaces, the substrate filaments shelter each other and the spectral shapes are closer to that of a smooth wall case.}


%

\revPJ{ Figures \ref{fig:E11_inner} (\textit{a})~and~\ref{fig:E11_inner}~(\textit{d}) represent perhaps a more interesting case at a matched $Re_{\tau} \approx 3700$. These two conditions not only have a similar values of $k_s^+$ but also the 10 PPI foam is at sparse ($s^+>100$) and shallow ($h/s<1$) substrate limits.~\cite{efstathiou2018mean} had argued that the outer-layer peak becomes weaker as the flow transitions from deep to shallow substrate. Based on their arguments \citep{efstathiou2018mean}, the intensity of the outer layer peak in the 10 PPI and the 3 mm thick substrate (figure \ref{fig:E11_inner} (\textit{a})) should be lower compared to the 45 PPI and 15 mm substrate (figure \ref{fig:E11_inner} (\textit{d})), which comparatively has a lower value of~$h/s$, lower values of sparsity and permeability based Reynolds number. However, the outer-layer peak for the 45 PPI and 15 mm thick foam is substantially lower than the 10 PPI and 3 mm thick substrate, which contradicts \cites{efstathiou2018mean} hypothesis.~Once again, we see the 45 PPI and 15 mm thick substrate remains an outlier, as also shown in previous sections.}




\revPJ{As already shown in section \ref{sec:level4b}, no evidence of KH-type instability was found in the present study. Therefore, the outer layer peak cannot be associated with a KH-type flow instability, as previously argued by~\citet{efstathiou2018mean} and \citet{manes2011turbulent}. In fact, the wavelength ($6-10\delta_{99}$) and time scale associated with outer-peak corresponds to very-large-scale motions~\citep{mathis2009large}.~While similar findings have been made in previous studies over impermeable rough wall \citep[see][for instance]{Squire2016}, yet the present study is the first to confirm the presence of these VLSMs over porous foams (see figure \ref{fig:E11_inner}). However, there is some evidence of weakening and shifting of VLSMs away from the substrate with increasing $Re_K$, compared to the smooth wall location marked in blue circle \citep{mathis2009large} in figure \ref{fig:E11_inner}.}

\begin{figure*}
  \centering
  \begin{tabular}{@{}p{0.5\linewidth}@{\quad}p{0.5\linewidth}@{}}
 
    \subfigimg[width=65 mm,pos=ul,vsep=2pt,hsep=38pt]{(a)}{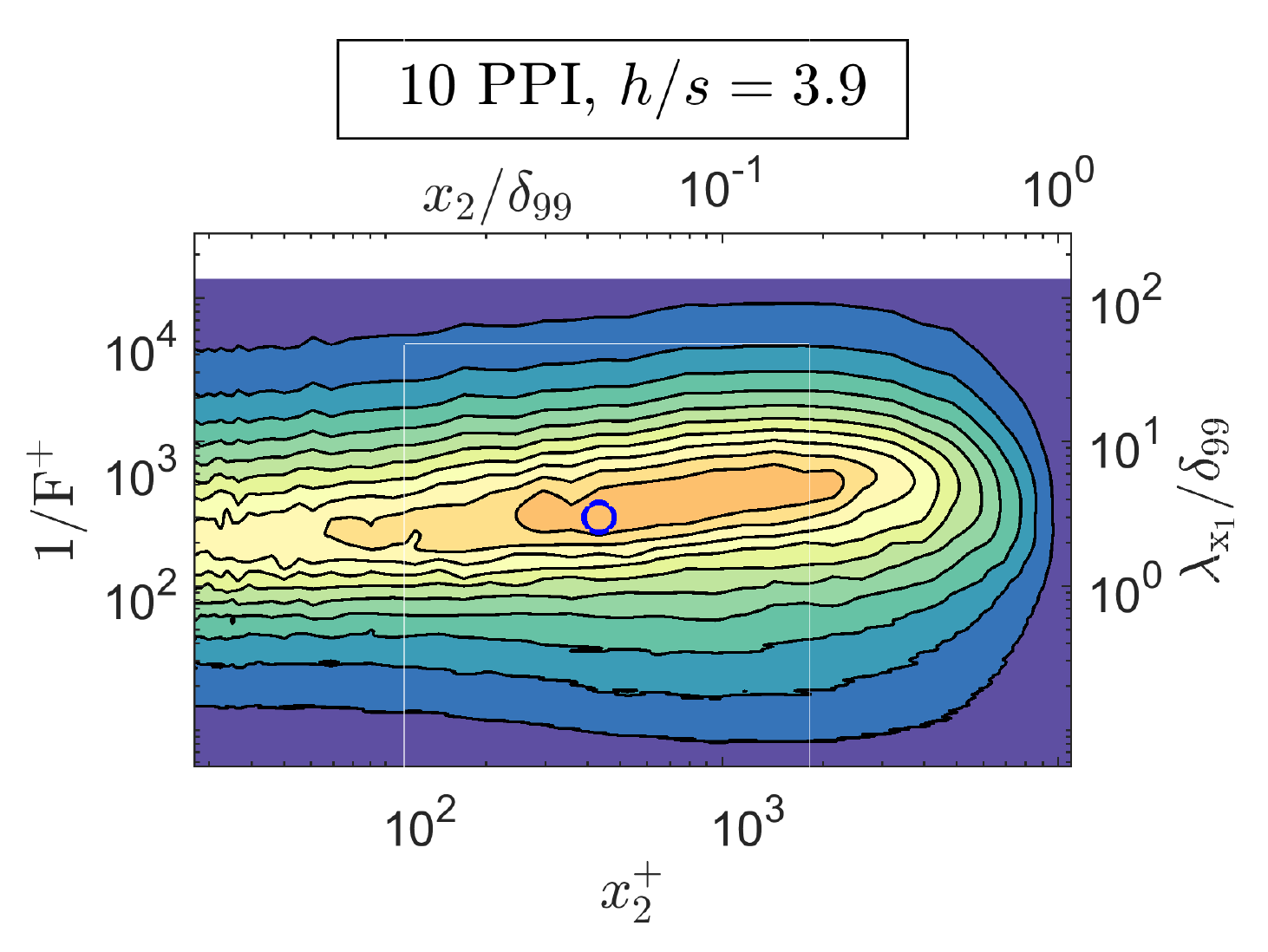} &
    \subfigimg[width=65 mm,pos=ul,vsep=2pt,hsep=38pt]{(b)}{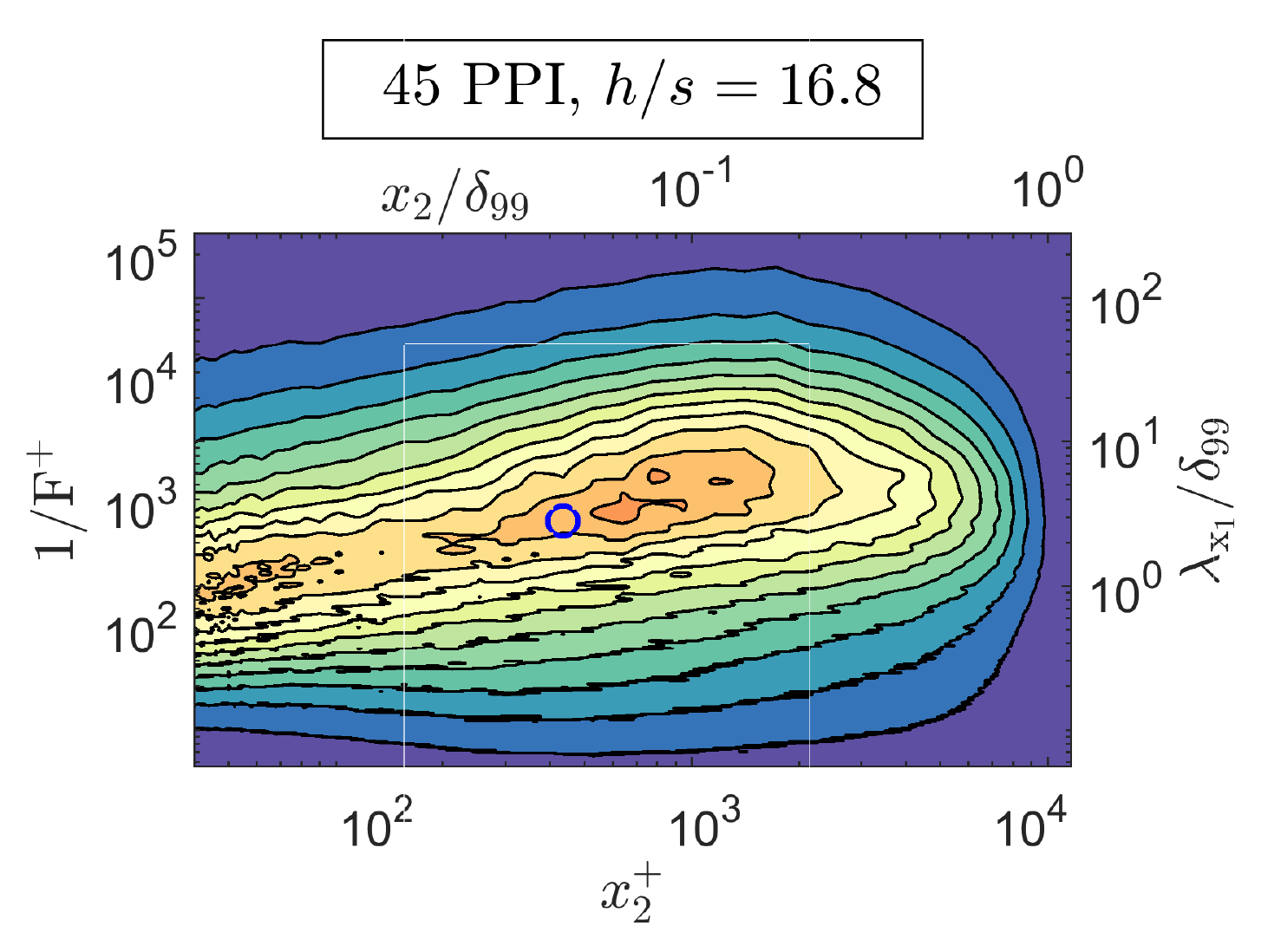} \\
    \hspace{3.6 cm}
    \subfigimg[width=65 mm,pos=ul,vsep=2pt,hsep=38pt]{(c)}{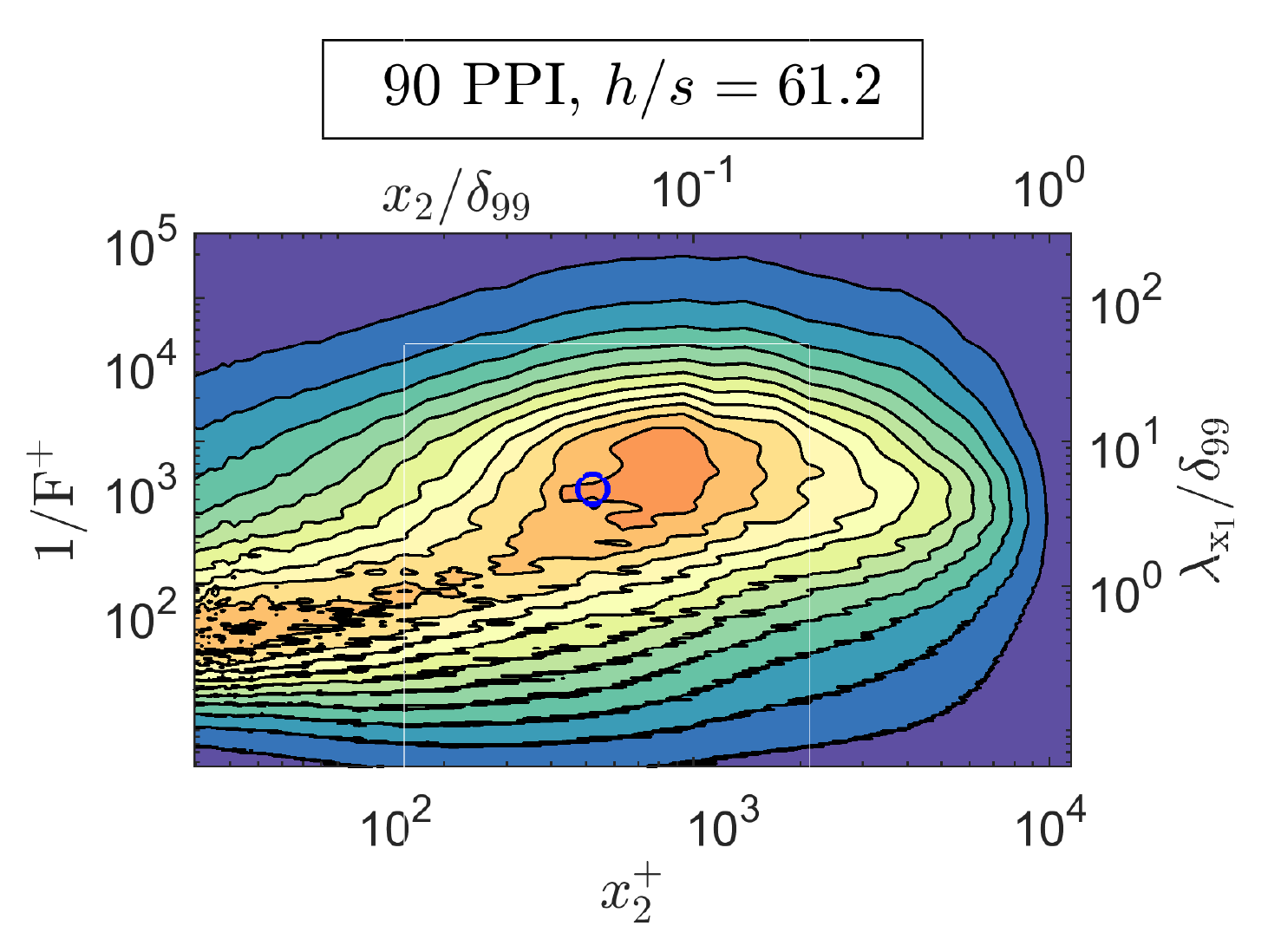}
    \\
  \end{tabular}
  \caption{\revPJ{ { $\overline{E_{11}^+}$} over 15 mm thick foam}~at~\revPJ{$Re_{\tau}\approx~7000$.}
  (a)~$10$ PPI foam. (b)~$45$ PPI foam. (c)~$90$ PPI foam. 
  \revPJ{The blue circle represents ${x_2}^+=3.9\sqrt{\rm{Re}_{\tau}}^{} $~\citep{mathis2009large}. See figure \ref{fig:E11_inner} for colorbar.}}
  \label{fig:E11_same_RET}
\end{figure*}



\revPJ{In order to decouple the effect of $Re_{\tau}$ and to quantify the impact of $Re_{K}$ on the outer-layer peak and VLSMs, figure \ref{fig:E11_same_RET} shows streamwise velocity spectra \revPJ{at $Re_{\tau} \approx 7000$}. While the $Re_{\tau}$ is kept constant, $k_s^+$ and $Re_{K}$ both increase at the same time obfuscating their relative importance. However, following \cite{Esteban} and \cite{wangsawijaya2023towards_JFM}, permeability and roughness can potentially be decoupled using the following relation:}

\revPJ{\begin{equation} 
\revPJ{k_s^+ = k_{sb}^+ Re_{K}}.
  \label{equation_roughness_permeability}     
\end{equation}}

\revPJ{Here, $k_{sb}^+$ is the roughness due to blockage, which would be the equivalent roughness of impermeable surface ($Re_{K}=1$). Therefore, the maximum achievable $k_{sb}^+$ is around $80$ for 10 PPI thick foam. For such surfaces ($k_s^{+} \ge 70$), \cite{Squire2016} has already shown that at $Re_{\tau} \ge 7000$, roughness has negligible impact on~$\overline{E_{11}^+}$ in the inertial layer and beyond. Therefore, the reduction in the peak intensity of $\overline{E_{11}^+}$ can be linked to an increase in $Re_{K}$. For 90 PPI foam, a reduction in peak values of $\overline{E_{11}^+}$ is achieved as $Re_{K}$ increases (see figures \ref{fig:E11_same_RET} (\textit{c}) and \ref{fig:E11_inner} (\textit{f})) for the same $h/s$.}


\revPJ{\cite{efstathiou2018mean} had hypothesised that the permeability based Reynolds could also be related to the reduction in the outer layer peak. However, the magnitude of $U_\tau$ used in \cites{efstathiou2018mean} study was obtained from the smooth wall region upstream of the porous substrate and does not include information about the effect of substrate permeability on flow. In the present study we are able to confirm their hypothesis only for substrates with the same thickness. This observation does not hold for the low speed case of the thick 45 PPI foam (Figure \ref{fig:E11_inner} \textit{d}), which shows an increase in the magnitude of the outer layer peak of $\overline{E_{11}^+}$ as $Re_{K}$ increases (Figure \ref{fig:E11_same_RET} \textit{ b}).}




\revPJ{Nevertheless, the spectral energy is contained within narrower bands of frequencies as the permeability based Reynolds number, $Re_{K}$, increases. This channeling of energy to narrow bands of frequencies leads to spectral shrinkage and flattening of the integral time scale close to the wall, especially for the 10 PPI and 15 mm thick foam at $Re_{\tau}~\approx~7000$.} A quantitative comparison of the energy distribution and \revPJ{spectral shrinkage} across different cases can be obtained by computing the Shannon entropy of the spectral content of the streamwise velocity. \cite{wesson2003quantifying} define Shannon entropy of the spectral content as:

{\begin{equation} 
\rm{SH} =  \sum_{i}^{N} \frac{-S_i \log{S_i}}{\log{N}} \\
  \label{Shannon}         
\end{equation}}



The Shannon entropy has also been used in the past by \cite{manes2011turbulent}; therefore, a direct comparison~\revPJ{with their results is possible}.
\begin{figure*}
  \centering
  \begin{tabular}{@{}p{0.5\linewidth}@{\quad}p{0.5\linewidth}@{}}
    \subfigimg[width=65 mm,pos=ur,vsep=15pt,hsep=12pt]{(a)}{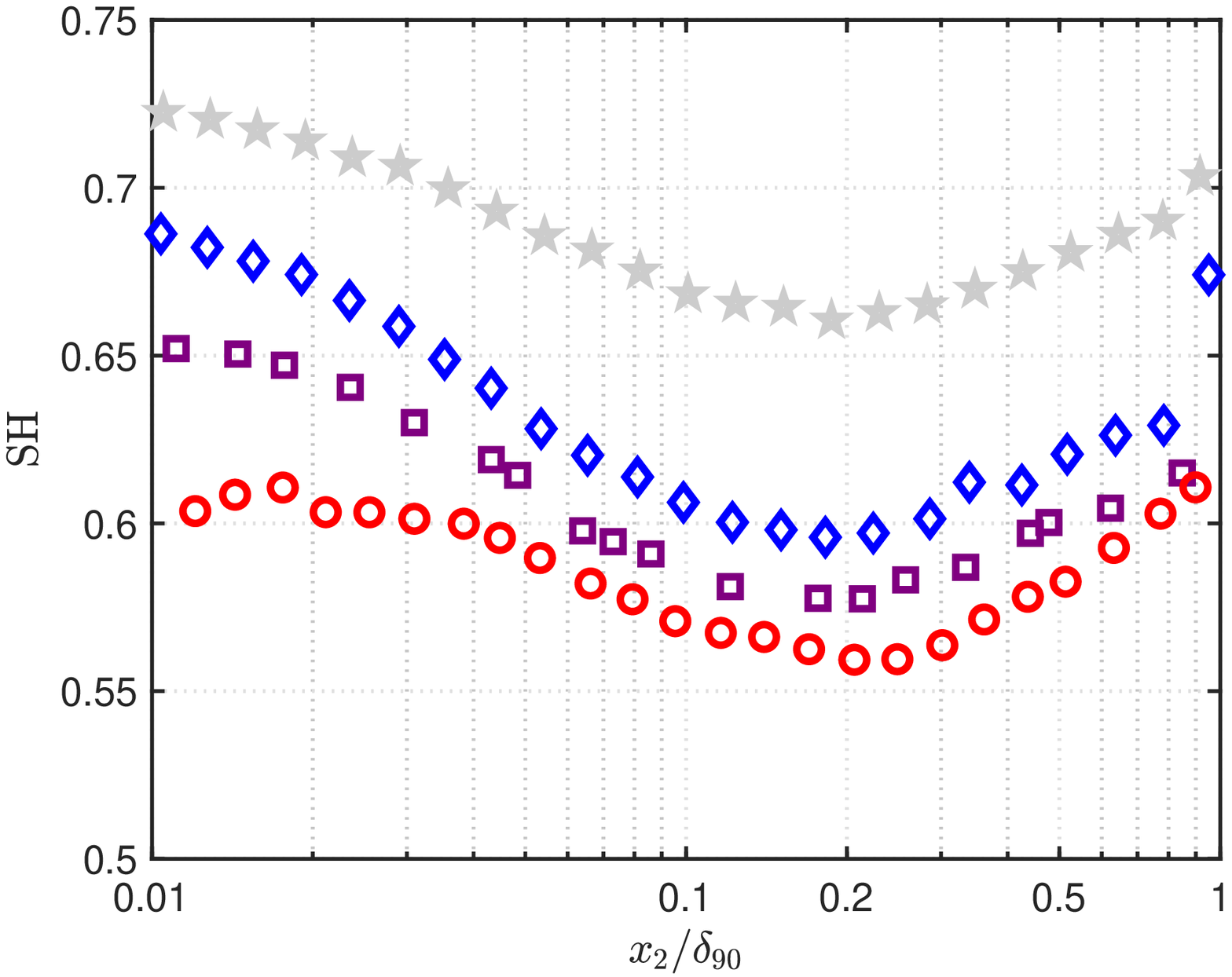} &
    \subfigimg[width=65 mm,pos=ur,vsep=15pt,hsep=12pt]{(b)}{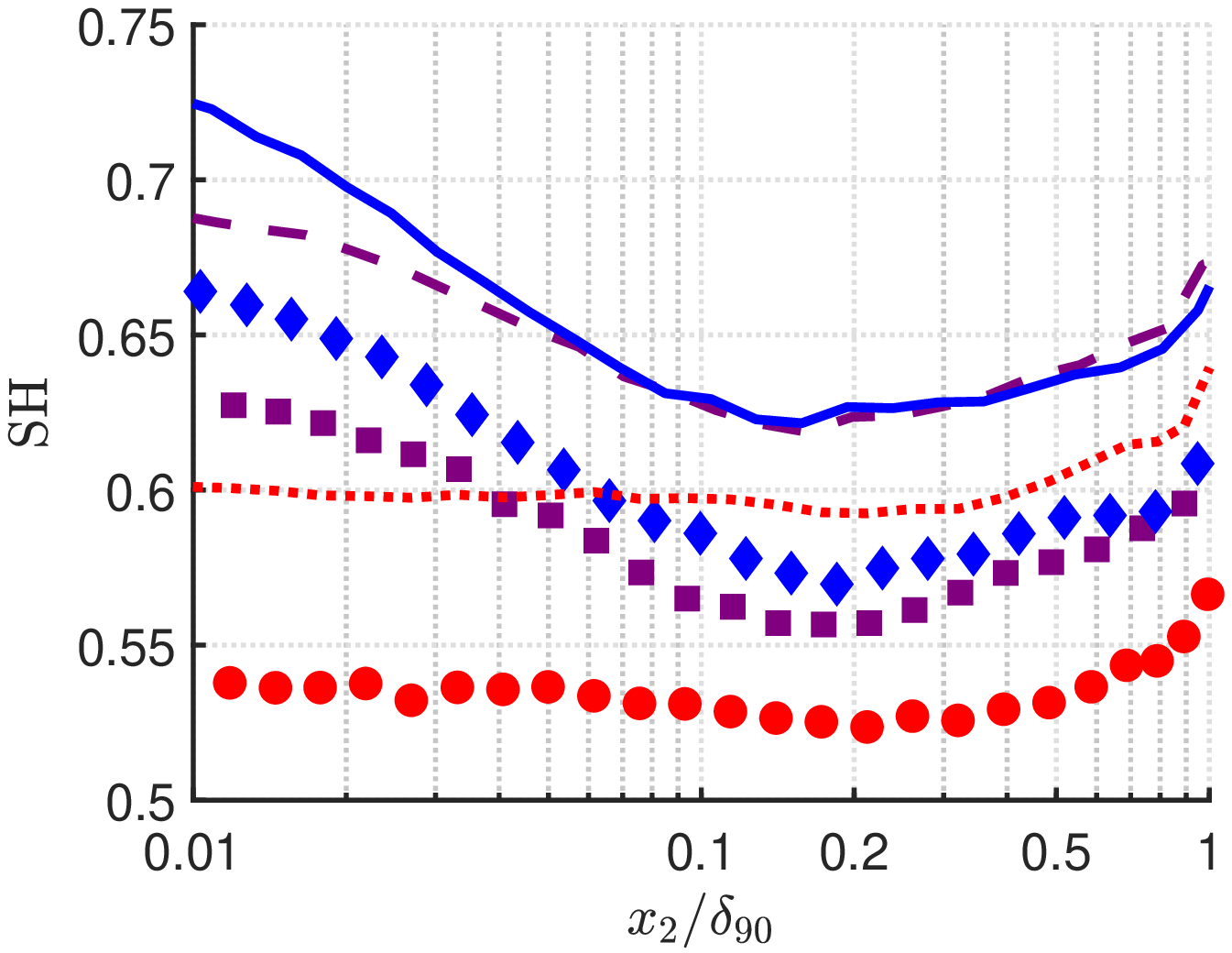}  \\
  \end{tabular}
  \caption{Shannon Entropy evaluated from streamwise velocity spectra. (a)~3~\revPJ{mm} thick substrate Legends: Red circles 10 PPI $Re_{\tau}=3644$,$Re_{K}=19.5$; Purple squares 45 PPI $Re_{\tau}=2871$,$Re_{K}=6.47$; Blue diamonds 90 PPI $Re_{\tau}=2349$,$Re_{K}=1.63$; and Gray pentagon 45 PPI $Re_{\tau}=8545$,$Re_{K}=17.2$.(b)~15 \revPJ{mm} thick substrate Legends: Red circles 10 PPI $Re_{\tau}=7417$,$Re_{K}=24.9$; Purple squares 45 PPI $Re_{\tau}=3716$,$Re_{K}=7.32$; Blue diamonds 90 PPI $Re_{\tau}=2831$,$Re_{K}=1.6$; Red dotted line 10 PPI $Re_{\tau}=13367$,$Re_{K}=45$; Purple dashed line 45 PPI $Re_{\tau}=7436$,$Re_{K}=13.55$ and Blue solid line 90 PPI $Re_{\tau}=6756$,$Re_{K}=3.25$.}
  \label{11111}
\end{figure*}

As mentioned by \cite{manes2011turbulent}, Shannon entropy is a measure of scale heterogeneity and spectral shrinkage. In presence of coherent structures, the energy is concentrated around fewer scales that results in shrinkage of spectra around the frequency (and hence wavenumber) of the corresponding coherent structure. As shown in figure \ref{11111}, normalised Shannon entropy increases with a decrease in permeability. Furthermore, Shannon entropy is not a function of $h/s$, which is inline with the observations made from figures \ref{fig:E11_inner} and \ref{fig:E11_same_RET}.~\revPJ{The spectral shrinkage is maximum for the 10 PPI and 15 mm thick foam substrate.} \cite{manes2011turbulent} had argued that the Shannon's entropy should scale with permeability ($Re_{K}$). \cite{manes2011turbulent} had associated this with the mixing-layer analogy. If $Re_K$ determines the permeability and the shear penetration depth is captured by the ratio $\delta_{99}/y_d$, then the $10$ PPI, $15$ \revPJ{mm} thick substrate at $Re_{K} \sim 45$ should have shown the lowest values of Shannon entropy~(Figure \ref{11111} dotted red line). Instead the same 10 PPI substrate at $Re_{K}=25$ has lower values of Shannon entropy compared to 10 PPI substrate at $Re_{K}=45$. Moreover, as shown in figure \ref{11111}, the Shannon entropy seems to be invariant to a single classical porous material parameters reported in the study. This is further reinforced by the fact that the $15$ \revPJ{mm} $45$ PPI case at $Re_{K}$ of $19.5$ and $s^+$ of $91.2$ shows much lower $\rm{SH}$ compared to $3$ \revPJ{mm} thick $10$ PPI porous substrate at same $Re_{K}$ but much higher $s^+$. 

For similar $Re_{K} \sim 19$, the Shannon entropy for the case of $45$ PPI foam is vastly greater than the $10$ PPI, for the 3 \revPJ{mm} thick substrate. Therefore, wall-permeability alone does \revJFM{not} determine existence of large coherent structures in the case of porous foams with a finite thickness. For instance, at the deep foam limit the~$Re_{K}$ determines scale heterogeneity only when $Re_{\tau}$ (at a given fetch distance) is similar. The independence of SH from wall-permeability ($Re_K$) can be due to increase in sparsity ($s^+$) with increasing Reynolds number ($Re_{\tau}$), which may \revPJ{alter} permeability effects of porous substrate.


\section{\label{sec:level5}Discussion}

In the present study, relative foam thickness, pore density and size were varied to assess their impact on turbulent boundary layer above a foam. For thick foam substrates, a deep foam limit is achieved for foams with higher pore density ($45$ PPI and $90$ PPI). Such deep foams remain at dense foam limit at low Reynolds number based on average pore size ($s^{+} < 50$), and differences in outer-layer similarity are observed, provided that the permeability based Reynolds number is high enough ($Re_K > 1$). In particular, velocity disturbances are substantially attenuated, and the extent of wall-normal velocity correlation, $R_{22}$, diminishes significantly. Therefore, 15 \revPJ{mm} thick $45$ PPI substrate has the lowest values of $\Lambda^{\vert 2+}_{22}(x_2)$. The 15 \revPJ{mm} thick $45$ PPI foam also has the smallest extent of streamwise velocity streaks at a given ejection or sweep event ($R_{12}$) compared to the rest of the cases. More importantly, these differences persists well into the outer-layer. The $45$ PPI foam has similar values of $Re_K$ and $s^+$ in deep and shallow substrate limits, and the only noticeable difference are measured in the values of $k_s^+$. In other words, at thin substrate limit, the effect of solid wall below the foam substrate is non-negligible, as it attenuates the the zero displacement plane and hence the equivalent sand grain roughness. Therefore, for porous foams  \say{thickness~induced~surface-roughness, $k_{sr}$} \citep[see][for details]{Esteban} can influence the outer-layer statistics. This is achieved by means of higher $Q_2^+$ and $Q_4^+$ events measured throughout the boundary layer, as such inner-layer is able to communicate with outer layer. Similar observations were made by \cite{krogstad1992comparison} for flows over and past impermeable rough wall. \revPJ{Nevertheless, the reduction in the intensity of the outer-layer peak observed for the 45 PPI and 15 mm thick foam compared to the 10 PPI and 3 mm thick foam implies that the frontal solidity (defined as h/s by \citep{efstathiou2018mean}) alone cannot explain the differences because the shelter solidity is comparable in both the studies. Therefore, the spatial arrangement of the pores and the shelter solidity \citep[see][for details]{placidi2018turbulent} may play an important role. The shelter solidity, $\lambda_s$, should be inversely proportional to the pore size because it represents the sheltered area, which together with the pitch of the filament represents the total planar area. Therefore, as the pore size increases, $\lambda_s$ decreases, which increases the total drag \citep{placidi2018turbulent}. Previously, \citet{placidi2018turbulent} had observed that at intermediate levels of shelter solidity, $\lambda_s$, the local morphology can influence turbulence statistics in the outer layer. Thus, at intermediate levels of sparsity ($25 \le s^+ \le 40$) and relative foam thickness ($h/s$), the impact of local foam morphology may influence outer-layer statistics. Additionally, a nonlinear coupling between roughness and permeability cannot be ruled out. However, to confirm these hypotheses, additional tests are required.}   

While it is tempting to draw an analogy between dense-deep foams and that of flow past dense canopy \citep{sharma2020_dense}, in present study no evidence of KH type flow instability is found for similar levels of sparsity and deep thickness limit. Nevertheless, it is hypothesized that the foam density limits required for the inception of KH instability could be lower in the case of porous substrate compared to the flow past dense canopy. This is backed by the findings of \cite{manes2011turbulent,efstathiou2018mean}, who found the peak in \revPJ{streamwise} velocity spectra associated with KH instability at lower Reynolds number ($s^+$) compared to the present study.
~As the wall-\revPJ{permeability} is further increased for the same substrate thickness, the pore size becomes same order of magnitude as the substrate thickness ($h/s < 10$) enabling access to shallow and sparse foam limits. \revPJ{At such conditions, spectral shrinkage is observed because of high $Re_{k}$ \citep[see also][for instance]{manes2011turbulent}. This leads to the flattening of the integral time-scale close to the wall $x_2<0.2\times\delta_{90}$. When the foam sparsity is increased further, i.e. $s^+>200$, it is hypothesized that as viscous scales shrink, spectral shrinkage is not observed.} This is because the pore size becomes substantially larger than the near-wall viscous and permeability scales ($Re_K$). This is also evidenced from the spectral heterogeneity, which no longer scales with wall-permeability ($Re_K$) at high sparsity limit ($s^+>200$). Therefore, flow at high $s^+$ becomes analogous to flow past sparse canopies  \citep{bailey2013turbulence,sharma2020_sparse}. The increase in velocity disturbances is also similar to ones observed in high sparsity limits for canopies \citep{sharma2020_sparse}. This limits the attenuation of wall-normal velocity disturbances that drive wall-pressure fluctuations \citep{carpio2019experimental}. \revPJ{{{More importantly, the spectral shrinkage plays an important role in demarcating the cases for which the VLSMs are present because with increase in the spectral shrinkage the spectral slope associated with large scale motions becomes shallow, which attenuates the wavelengths associated with the outer-layer peak.}}}

Similar to an impermeable rough wall, roughness sub-layer pushes the energetic flow away from the surface in the case of porous walls, as evidenced from streamwise kinetic energy spectra. Therefore, the wall-normal location of the velocity energy spectra peak depends on foam density ($s^+$) and \revPJ{$Re_K$}. An important distinction between flow past sparse porous substrates and roughness is that when the roughness is sparse, the wall becomes akin to smooth wall, while when the \revPJ{foam} is sparse, the flow becomes fully-rough. This is because permeability increases equivalent sand grain roughness \citep{Esteban}. Therefore, either when the substrate is very sparse ($s^+ \ge 60$) or when the substrate thickness becomes comparable to pore-size ($h \sim s$) the substrate acts as a rough wall, and \revPJ{permeability effects are limited to the spectral shrinkage, and reduction in the intensity of outer-layer peak and associated wavelengths.} 
In contrast, when the permeability-based or the pore-based Reynolds numbers are comparable to that of viscous scales, then changes in outer-layer velocity statistics are negligible. Therefore, the outer-layer similarity is achieved for two extreme cases when either the substrate is akin to rough walls or similar to a smooth wall ($Re_K$ $\sim 1$).

\section{\label{sec:level6}Conclusions}

The present study quantifies the effect of wall \revPJ{permeability}  and substrate thickness on flows past porous foams. 
For a broad range of Reynolds numbers, the turbulent statistics, the spatio-temporal scales and energy spectra were quantified above the porous substrate within the boundary layer. In particular, the present manuscript extends current state-of-the-art \citep{manes2011turbulent,efstathiou2018mean} to include the effects of foam density ($s^+$) and relative foam thickness ($h/s$) on turbulent boundary layer over porous walls over a range of $Re_K$. We cover both transitionally- and fully-rough regimes and quantify the turbulent flow structures through the use of two-point correlations. The foam thickness-to-pore size range from $h/s \approx 0.7-60$, while various Reynolds numbers range from $Re_{\tau} \approx 2000-13500$, $Re_K \approx 1 - 50$ and $s^+ \approx 75- 400$. 

Two research questions \revJFM{have} driven the present study: 1) Is the flow over such porous surfaces analogous to flows over rough surfaces away from the wall? If so, does the outer-layer similarity in velocity statistics holds for such porous foams? 2) For what values of pore thickness and size can we expect to reduce the correlation of wall-normal velocity fluctuations?

As it turns out these questions are interlinked for the case of flow past a permeable foam. In particular, the present study shows a substantial reduction in the correlations of the velocity fluctuations ($R_{12}$ and $R_{22}$), at deep-dense substrate limits with high permeability based on Reynolds numbers ($Re_{K} > 1$), which \revPJ{weakens} the Townsend's outer-layer similarity and provide an avenue for using porous walls in aerofoil self-noise reduction applications. In particular, this is achieved by an increased relative vertical momentum exchange by an increase in ejection $Q_2^+$ and sweep $Q_4^+$ events across the boundary layer. Therefore, \revPJ{the} wall-permeability boundary condition is felt across the boundary-layer, resulting in substantial reduction in velocity disturbance field above the porous wall. \revPJ{However, neither the existing framework for flows past rough walls nor for flows over porous walls can fully explain the differences in the outer-layer for permeable foams with intermediate values of pore density and relative foam thickness $(h/s)$.} As such, the present study shows that the \revPJ{relative} success of outer-layer similarity depends on \revPJ{the pore density ($s^+$)}, permeability ($Re_{K}$) and relative foam thickness $(h/s)$. Therefore, in the \revPJ{outer layer}, the flow over porous surfaces is analogous to flows over rough surfaces only at the shallow or sparse foam limits at high Reynolds number ($Re_{\tau}$). \revPJ{At a given $Re_{\tau}$, the effect of permeability is restricted to spectral shrinkage and a reduction~in~the wavelength of the outer~layer~peak. At dense and thick substrate limits the VLSMs are observed even at reasonably high permeability based Reynolds number ($Re_{k}>1$), and an increase in $Re_{k}$ pushes the outer-layer peak away from the wall.} 





The influence of permeability, surface roughness, and substrate structure, as well as their non-linear interactions, needs to be explored further. Future work should include systematic variations of surface roughness for a given permeability (and vice versa). This can potentially be achieved by adding a high permeability surface (that is rough) on top of a surface with a given permeability, which will enable a better understanding of the effects of roughness and permeability on turbulent flow over porous surfaces.

\backsection[Acknowledgements]
{We acknowledge the support from E. Rodr\'iguez-L\'opez and M.A Ferreira in the data acquistion phase as well as Luis Esteban-Blay and Tim Schoelle for their help in the initial data curation.}

\backsection[Funding]{We acknowledge the financial support from EPSRC (Grant Ref no: EP/S013296/1) and European Office for Airforce Research and Development (Grant No: FA9550-19-1-7022,Programme Manager: Dr. Doug Smith). PJ acknowledges the financial support from UK Research and Innovation (UKRI) under the UK government’s Horizon Europe funding guarantee [grant number EP/X032590/1)].}

\backsection[Declaration of interests]{The authors report no conflict of interest.}
 
\backsection[Data availability statement]{All data supporting this study have been made openly available from the University of Southampton repository at~\href{https://doi.org/10.5258/SOTON/D2925}{https://doi.org/10.5258/SOTON/D2925}}

\backsection[Author ORCID]{P. Jaiswal: \href{https://orcid.org/0000-0002-5240-9911}{0000-0002-5240-9911}, B. Ganapathisubramani: \href{https://orcid.org/0000-0001-9817-0486}{0000-0001-9817-0486}}



\scalebox{0}{%
\begin{tikzpicture}
    \begin{axis}[%
        hide axis
        ]
        \addplot [
            draw={rgb,255:red,55;green,126;blue,184},
            mark=triangle*,
            mark options={fill={rgb,255:red,55;green,126;blue,184}, draw=black, scale=2, line width=0.0pt, solid},
            mark size=1.5pt,
            line width=1pt,
            ]{0}; \label{leg:triblue}
        \addplot [
            draw={rgb,255:red,55;green,126;blue,184},
            mark=square*,
            mark options={fill={rgb,255:red,55;green,126;blue,184}, draw={rgb,255:red,55;green,126;blue,184}, scale=2, line width=0.0pt, solid},
            mark size=1.5pt,
            line width=1pt,
            solid,
            ]{0}; \label{leg:squareblue}
        \addplot [
            draw={rgb,255:red,55;green,126;blue,184},
            dashed,
            mark=square*,
            mark options={fill={rgb,255:red,55;green,126;blue,184}, draw={rgb,255:red,55;green,126;blue,184}, scale=2, line width=0.0pt, solid},
            mark size=1.5pt,
            line width=1pt,
            ]{0}; \label{leg:squarebluedashed}       
%
               
        \addplot [
            draw=red,
            mark=*,
            mark options={fill=red, draw=red, scale=2, line width=0.0pt, solid},
            mark size=1.5pt,
            line width=1pt,
            solid,
            ]{0}; \label{leg:circlered}
        \addplot [
            draw=red,
            mark=*,
            mark options={fill=red, draw=red, scale=2, line width=0.0pt, solid},
            mark size=1.5pt,
            line width=1pt,
            dashed,
            ]{0}; \label{leg:circlereddashed}                        
        \addplot [
            draw=red,
            mark=triangle*,
            mark options={fill=red, draw=black, scale=2, line width=0.0pt, solid},
            every mark/.append style={rotate=180},
            mark size=1.5pt,
            line width=1pt,
            ]{0}; \label{leg:trired}

        \addplot [
            draw=ForestGreen,
            mark=diamond*,
            mark options={fill=ForestGreen, draw=black, scale=2, line width=0.0pt, solid},
            mark size=1.5pt,
            line width=1pt,
            ]{0}; \label{leg:diagreen}

        \addplot [
            color=lightgray,
            solid,
            line width=1.pt,
            dashed,
            ]{0}; \label{leg:greydashed}
		
			 \addplot [
            color=gray,
            mark options={solid, scale=1.3},
            line width=1.pt,
            solid,
            ]{0}; \label{leg:greysolid}		
            \addplot [
            color=lightgray,
            mark options={solid, scale=1.3},
            line width=1.pt,
            solid,
            ]{0}; \label{leg:lightgraysolid}
            
			 \addplot [
            color=lightgray,
            mark=triangle,
            mark options={fill=lightgray, draw=lightgray, scale=1.5, line width=0.5pt, solid, rotate=0},
            line width=1.pt,
            solid,
            ]{0}; \label{leg:greyuptrianglesolid}            
            
             \addplot [
            color=lightgray,
            mark options={solid, scale=1.3},
            line width=1.5pt,
            solid,
            ]{0}; \label{leg:greysolidthick}

        \addplot [
            color=lightgray,
            dotted,
            line width=1.pt,
            ]{0}; \label{leg:greydotted}
		
        \addplot [
            color=red,
            solid,
            line width=1.pt,
            ]{0}; \label{leg:red}
        \addplot [
            color=black,
            solid,
            line width=1.pt,
            ]{0}; \label{leg:black}
        \addplot [
            color=black,
            dotted,
            line width=1.pt,
            ]{0}; \label{leg:blackdotted}

        \addplot [
            color=red,
            dashdotted,
            line width=1.pt,
            ]{0}; \label{leg:reddashdotted}            

        \addplot [
            color={rgb,255:red,55;green,126;blue,184},
            dashdotted,
            line width=1.pt,
            ]{0}; \label{leg:bluedashdotted}            

        \addplot [
            color={rgb,255:red,55;green,126;blue,184},
            solid,
            line width=1.pt,
            ]{0}; \label{leg:blue}

            \addplot [
            color={rgb,1:red,1;green,0.49;blue,0},
            solid,
            line width=1.pt,
            ]{0}; \label{leg:orange}
            
            \addplot [
            color={rgb,1:red,0;green,0.0;blue,0},
            dashed,
            line width=1.pt,
            ]{0}; \label{leg:blackdashline}

            \addplot [
            color={rgb,1:red,0;green,0.0;blue,0},
            dashdotted,
            line width=1.pt,
            ]{0}; \label{leg:black_dashdotted}

            \addplot [
            color={rgb,1:red,0;green,0.2;blue,1},
            dashed,
            line width=1.pt,
            ]{0}; \label{leg:bluedashline}
            
            \addplot [
            color={rgb,1:red,1;green,0.1;blue,1},
            dashed,
            line width=1.pt,
            ]{0}; \label{leg:purpledashedline}
            
            \addplot [
            color={rgb,1:red,0.1;green,0.5;blue,0},
            dashed,
            line width=1.pt,
            ]{0}; \label{leg:Greendashedline}
            
            \addplot [
            color={rgb,1:red,0.8;green,0.0;blue,0.0},
            dashed,
            line width=1.pt,
            ]{0}; \label{leg:reddashedline}

            \addplot [
            color={rgb,1:red,0.8;green,0.8;blue,0.8},
            dashed,
            line width=1.pt,
            ]{0}; \label{leg:greydashedline}
            
              \addplot [
            color={rgb,1:red,0.8;green,0.8;blue,0.8},
            dashed,
            line width=2.pt,
            ]{0}; \label{leg:greydashedline-thick}

        \addplot [
            color=black,
            mark options={solid, scale=1.3},
            line width=0.5pt,
            dashed,
            ]{0}; \label{leg:blackdashed}
        \addplot [
            color=red,
            mark options={solid, scale=1.3},
            line width=1.pt,
            dashed,
            ]{0}; \label{leg:reddashed}

        \addplot [
            color=black,
            mark=square,
            mark options={solid, scale=1.2},
            line width=1.pt,
            dashed,
            ]{0}; \label{leg:blacksquaredash}
        \addplot [
            color=black,
            mark=square*,
            mark options={fill=black, draw=black, scale=1.3, line width=0.pt, solid},
            line width=0.pt,
            only marks,
            ]{0}; \label{leg:blacksquare}

			    \addplot [
            color=black,
            mark=square*,
            mark options={fill=black, draw=black, scale=1.3, line width=0.pt, solid},
            line width=1.pt,
            solid,
            ]{0}; \label{leg:blacksquareline}
    		
			\addplot [
            color=black,
            mark=square*,
            mark options={fill=white, draw=black, scale=1.3, line width=0.pt, solid},
            line width=1.pt,
            solid,
            ]{0}; \label{leg:blacksquarelineempty}	
		
        \addplot [
            color=lightgray,
            mark=o,
            mark options={solid, scale=1.2},
            line width=1.pt,
            dashed,
            ]{0}; \label{leg:greycircledash}
            
             \addplot [
            color=lightgray,
            mark=o,
            mark options={solid, scale=1.2},
            line width=1.pt,
            solid,
            ]{0}; \label{leg:greycirclesolid}

        \addplot [
            color=black,
            mark=o,
            mark options={solid, scale=1.3},
            line width=1.pt,
            dashed,
            ]{0}; \label{leg:blackcircledash}

        \addplot [
            color=red,
            dashdotted,
            mark=square*,
            mark options={fill=red, draw=red, scale=1.3, line width=0.pt, solid},
            line width=1.pt,
            ]{0}; \label{leg:reddashdottedsquare}
        \addplot [
            color={rgb,255:red,55;green,126;blue,184},
            dashdotted,
            mark=square*,
            mark options={fill={rgb,255:red,55;green,126;blue,184}, draw={rgb,255:red,55;green,126;blue,184}, scale=1.3, line width=0.pt, solid},
            line width=1.pt,
            ]{0}; \label{leg:bluedashdottedsquare}

        \addplot [
            color={rgb,255:red,55;green,126;blue,184},
            solid,
            mark=square*,
            mark options={fill={rgb,255:red,55;green,126;blue,184}, draw={rgb,255:red,55;green,126;blue,184}, scale=1.3, line width=0.pt, solid},
            line width=1.pt,
            ]{0}; \label{leg:bluesquare}
            
            \addplot [
            color={rgb,255:red,0;green,0;blue,255},
            solid,
            mark=square*,
            mark options={fill=white, draw={rgb,255:red,0;green,0;blue,255}, scale=1.3, line width=0.pt, solid},
            line width=1.pt,
            ]{0}; \label{leg:bluesquare_empty}

        \addplot [
            color=red,
            solid,
            mark=square*,
            mark options={fill=red, draw=red, scale=1.3, line width=0.0pt, solid},
            line width=1.pt,
            ]{0}; \label{leg:redsquare}

        \addplot [
            color={rgb,255:red,55;green,126;blue,184},
            mark=*,
            mark options={fill={rgb,255:red,55;green,126;blue,184}, draw={rgb,255:red,55;green,126;blue,184}, scale=1.3, line width=0.0pt, solid},
            line width=1.pt,
            ]{0}; \label{leg:bluecircle}
        \addplot [
            color=red,
            mark=*,
            mark options={fill=red, draw=red, scale=1.3, line width=0.0pt, solid},
            line width=1.pt,
            ]{0}; \label{leg:redcircle}

        \addplot [
            color=red,
            dashdotted,
            mark=*,
            mark options={fill=red, draw=red, scale=1.3, line width=0.0pt, solid},
            line width=1.pt,
            ]{0}; \label{leg:reddashdottedcircle}

            \addplot [
            color=red,
            solid,
            mark=*,
            mark options={fill=white, draw=red, scale=1.3, line width=0.5pt, solid},
            line width=1.pt,
            ]{0}; \label{leg:redcontdottedcircle}
            
            \addplot [
            color=red,
            only marks, 
            mark=*,
            mark options={fill=red, draw=red, scale=1.3, line width=0.0pt, solid},
            line width=1.pt,
            ]{0}; \label{leg:redcirclenoline}
            
            \addplot [
            color=red,
            only marks, 
            mark=*,
            mark options={fill=white, draw=red, scale=1.3, line width=0.0pt, solid},
            line width=1.pt,
            ]{0}; \label{leg:redcirclenoline_empty}
            
        \addplot [
            color={rgb,255:red,55;green,126;blue,184},
            dashdotted,
            mark=*,
            mark options={fill={rgb,255:red,55;green,126;blue,184}, draw={rgb,255:red,55;green,126;blue,184}, scale=1.3, line width=0.0pt, solid},
            line width=1.pt,
            ]{0}; \label{leg:bluedashdottedcircle}

        \addplot [
            color=black,
            mark options={solid, scale=1.3},
            line width=1.pt,
            solid,
            ]{0}; \label{leg:blacksolid}

        \addplot [
            color=red,
            mark=square,
            mark options={fill=red, draw=red, scale=1.0, line width=0.5pt, solid, rotate=45},
            line width=1.pt,
            solid,
            ]{0}; \label{leg:reddiamondsolid}

        \addplot [
            color=red,
            mark=square,
            mark options={fill=red, draw=red, scale=1.0, line width=0.5pt, solid, rotate=45},
            line width=1.pt,
            dashed,
            ]{0}; \label{leg:reddiamonddashed}
            
            \addplot [
            color=red,
            mark=*,
            mark options={fill=white, draw=red, scale=1.0, line width=0.5pt, solid, rotate=0},
            line width=1.pt,
            dashed,
            ]{0}; \label{leg:redcircledashedline}

        \addplot [
            color={rgb,255:red,77;green,175;blue,74},
            mark=triangle,
            mark options={fill={rgb,255:red,77;green,175;blue,74}, draw={rgb,255:red,77;green,175;blue,74}, scale=1.5, line width=0.5pt, solid, rotate=0},
            line width=1.pt,
            solid,
            ]{0}; \label{leg:greenuptrianglesolid}

        \addplot [
            color={rgb,255:red,77;green,175;blue,74},
            mark=triangle,
            mark options={fill={rgb,255:red,77;green,175;blue,74}, draw={rgb,255:red,77;green,175;blue,74}, scale=1.5, line width=0.5pt, solid, rotate=0},
            line width=1.pt,
            dashed,
            ]{0}; \label{leg:greenuptriangledashed}

        \addplot [
            color={rgb,255:red,55;green,126;blue,184},
            mark=triangle,
            mark options={fill={rgb,255:red,55;green,126;blue,184}, draw={rgb,255:red,55;green,126;blue,184}, scale=1.5, line width=0.5pt, solid, rotate=180},
            line width=1.pt,
            solid,
            ]{0}; \label{leg:bluedowntrianglesolid}
			
			 \addplot [
            color={rgb,255:red,0;green,0;blue,255},
            mark=diamond*,
            mark options={fill=white, draw={rgb,255:red,0;green,0;blue,255}, scale=1.5, line width=0.5pt, solid, rotate=0},
            line width=1.pt,
            solid,
            ]{0}; \label{leg:bluedowndiamondsolid}

        \addplot [
            color={rgb,255:red,55;green,126;blue,184},
            mark=triangle,
            mark options={fill={rgb,255:red,55;green,126;blue,184}, draw={rgb,255:red,55;green,126;blue,184}, scale=1.5, line width=0.5pt, solid, rotate=180},
            line width=1.pt,
            dashed,
            ]{0}; \label{leg:bluedowntriangledashed}  
            
            \addplot [
            color={rgb,255:red,255;green,127;blue,0},
            mark=triangle,
            mark options={fill={rgb,255:red,255;green,127;blue,0}, draw={rgb,255:red,255;green,127;blue,0}, scale=1.3, line width=0.5pt, solid, rotate=0},
            line width=1.pt,
            only marks,
            ]{0}; \label{leg:orangetriangle} 
            
             \addplot [
            color={rgb,255:red,255;green,127;blue,0},
            mark=x,
            mark options={fill={rgb,255:red,255;green,127;blue,0}, draw={rgb,255:red,255;green,127;blue,0}, scale=1.3, line width=0.5pt, solid, rotate=0},
            line width=1.pt,
            only marks,
            ]{0}; \label{leg:orange_cross}

			 \addplot [
            color={rgb,255:red,0;green,0;blue,255},
            mark=x,
            mark options={fill={rgb,255:red,0;green,0;blue,255}, draw={rgb,255:red,0;green,0;blue,255}, scale=1.3, line width=0.5pt, solid, rotate=0},
            line width=1.pt,
            only marks,
            ]{0}; \label{leg:blue_cross}

             \addplot [
            color={rgb,255:red,0;green,0;blue,255},
            mark=x,
            mark options={fill={rgb,255:red,0;green,0;blue,255}, draw={rgb,255:red,0;green,0;blue,255}, scale=1.3, line width=0.5pt, solid, rotate=0},
            line width=1.pt,
            solid,
            ]{0}; \label{leg:bluecrossline}

            \addplot [
            color={rgb,255:red,0;green,0;blue,255},
            mark=*,
            mark options={fill=white, draw={rgb,255:red,0;green,0;blue,255}, scale=1.7, line width=0.5pt, solid, rotate=0},
            line width=1.pt,
            only marks,
            ]{0}; \label{leg:blue-circle}

 \addplot [
            color={rgb,255:red,255;green,0;blue,0},
            mark=square,
            mark options={fill={rgb,255:red,255;green,0;blue,0}, draw={rgb,255:red,255;green,0;blue,0}, scale=1.7, line width=0.5pt, solid, rotate=0},
            line width=1.pt,
            only marks,
            ]{0}; \label{leg:orange-square}

        \addplot [
            color={rgb,255:red,255;green,127;blue,0},
            mark=square,
            mark options={fill={rgb,255:red,255;green,127;blue,0}, draw={rgb,255:red,255;green,127;blue,0}, scale=1.3, line width=0.5pt, solid, rotate=0},
            line width=1.pt,
            solid,
            ]{0}; \label{leg:orangesquaresolid}
            \addplot [
            color={rgb,255:red,255;green,127;blue,0},
            mark=diamond*,
            mark options={fill={rgb,255:red,255;green,127;blue,0}, draw={rgb,255:red,255;green,127;blue,0}, scale=1.3, line width=0.5pt, solid, rotate=0},
            line width=1.pt,
            solid,
            ]{0}; 
	  
        \addplot [
            color={rgb,255:red,255;green,127;blue,0},
            mark=square,
            mark options={fill={rgb,255:red,255;green,127;blue,0}, draw={rgb,255:red,255;green,127;blue,0}, scale=1.3, line width=0.5pt, solid, rotate=0},
            line width=1.pt,
            dashed,
            ]{0}; \label{leg:orangesquaredashed}           
        \addplot [
            color={rgb,255:red,255;green,127;blue,0},
            mark=diamond*,
            mark options={fill=white, draw={rgb,255:red,255;green,127;blue,0}, scale=1.3, line width=0.pt, solid},
            line width=0.pt,
            only marks,
            ]{0}; \label{leg:orangediamond}  
            \addplot [
            color=blue,
            mark=triangle*,
            mark options={fill=white, draw=blue, scale=1.3, line width=0.pt, solid},
            line width=0.pt,
            only marks,
            ]{0}; \label{leg:triangle_blue}  
            
			\addplot [
            color=lightgray,
            mark=diamond*,
            mark options={fill=white, draw=lightgray, scale=1.7, line width=0.pt, solid},
            line width=1.pt,
            only marks,
            ]{0}; \label{leg:greydiamond}

            \addplot [
            color=black,
            mark=*,
            mark options={fill=white, draw=black, scale=1.3, line width=0.pt, solid},
            line width=0.pt,
            only marks,
            ]{0}; \label{leg:black_circle}  
            
           \addplot [
            color=red,
            mark=square*,
            mark options={fill=white, draw=red, scale=1.3, line width=0.pt, solid},
            line width=0.pt,
            only marks,
            ]{0}; \label{leg:square_red}  
            \addplot [
            color=black,
            mark=*,
            mark options={fill=black, draw=black, scale=1.3, line width=0.pt, solid},
            line width=0.pt,
            only marks,
            ]{0}; \label{leg:black_circle_fill}

              \addplot [
            color=blue,
            mark=square,
            mark options={fill=white, draw=blue, scale=1.7, line width=0.1pt, solid},
            line width=0.pt,
            only marks,
            ]{0}; \label{leg:blue_square_marks}  
                 
    \end{axis}
    \end{tikzpicture}}

\bibliographystyle{jfm}
\bibliography{jfm_my}

\providecommand{\noopsort}[1]{}\providecommand{\singleletter}[1]{#1}%
\begin{thebibliography}{47}
\expandafter\ifx\csname natexlab\endcsname\relax\def\natexlab#1{#1}\fi
\def\au#1{#1} \def\ed#1{#1} \def\yr#1{#1}\def\at#1{#1}\def\jt#1{\textit{#1}} \def\bt#1{#1}\def\bvol#1{\textbf{#1}} \def\vol#1{#1} \def\pg#1{#1} \def\publ#1{#1}\def\arxiv#1{#1}\def\org#1{#1}\def\st#1{\textit{#1}}

\bibitem[Bailey \& Stoll(2013)]{bailey2013turbulence}
{\sc \au{Bailey, Brian~N} \& \au{Stoll, Rob}} \yr{2013}  \at{Turbulence in sparse, organized vegetative canopies: a large-eddy simulation study}.  \jt{Boundary-layer meteorology}  \bvol{147}~(3),  \pg{369--400}.

\bibitem[Benedict \& Gould(1996)]{benedict1996towards}
{\sc \au{Benedict, L.~H.} \& \au{Gould, R.~D.}} \yr{1996}  \at{Towards better uncertainty estimates for turbulence statistics}.  \jt{Experiments in fluids}  \bvol{22}~(2),  \pg{129--136}.

\bibitem[Breugem {\em et~al.\/}(2006)Breugem, Boersma \& Uittenbogaard]{breugem2006influence}
{\sc \au{Breugem, WP}, \au{Boersma, BJ} \& \au{Uittenbogaard, RE}} \yr{2006}  \at{The influence of wall permeability on turbulent channel flow}.  \jt{Journal of Fluid Mechanics}  \bvol{562},  \pg{35}.

\bibitem[Carpio {\em et~al.\/}(2019)Carpio, Mart{\'\i}nez, Avallone, Ragni, Snellen \& van~der Zwaag]{carpio2019experimental}
{\sc \au{Carpio, Alejandro~Rubio}, \au{Mart{\'\i}nez, Roberto~Merino}, \au{Avallone, Francesco}, \au{Ragni, Daniele}, \au{Snellen, Mirjam} \& \au{van~der Zwaag, Sybrand}} \yr{2019}  \at{Experimental characterization of the turbulent boundary layer over a porous trailing edge for noise abatement}.  \jt{Journal of Sound and Vibration}  \bvol{443},  \pg{537--558}.

\bibitem[Chung {\em et~al.\/}(2014)Chung, Monty \& Ooi]{chung2014idealised}
{\sc \au{Chung, D.}, \au{Monty, J.P.} \& \au{Ooi, A.}} \yr{2014}  \at{An idealised assessment of townsend's outer-layer similarity hypothesis for wall turbulence}.  \jt{Journal of Fluid Mechanics}  \bvol{742}.

\bibitem[Clifton {\em et~al.\/}(2008)Clifton, Manes, R{\"u}edi, Guala \& Lehning]{clifton2008shear}
{\sc \au{Clifton, Andrew}, \au{Manes, Costantino}, \au{R{\"u}edi, Jean-Daniel}, \au{Guala, Michele} \& \au{Lehning, Michael}} \yr{2008}  \at{On shear-driven ventilation of snow}.  \jt{Boundary-layer meteorology}  \bvol{126}~(2),  \pg{249--261}.

\bibitem[Efstathiou \& Luhar(2018)]{efstathiou2018mean}
{\sc \au{Efstathiou, C.} \& \au{Luhar, M.}} \yr{2018}  \at{Mean turbulence statistics in boundary layers over high-porosity foams}.  \jt{Journal of Fluid Mechanics}  \bvol{841},  \pg{351--379}.

\bibitem[Esteban {\em et~al.\/}(2022)Esteban, Rodr{\'\i}guez-L{\'o}pez, Ferreira \& Ganapathisubramani]{Esteban}
{\sc \au{Esteban, LB}, \au{Rodr{\'\i}guez-L{\'o}pez, E}, \au{Ferreira, MA} \& \au{Ganapathisubramani, B}} \yr{2022}  \at{Mean flow of turbulent boundary layers over porous substrates}.  \jt{Physical Review Fluids}  \bvol{7}~(9),  \pg{094603}.

\bibitem[Ferreira {\em et~al.\/}(2018)Ferreira, Rodriguez-Lopez \& Ganapathisubramani]{ferreira2018alternative}
{\sc \au{Ferreira, M.~A.}, \au{Rodriguez-Lopez, E.} \& \au{Ganapathisubramani, B.}} \yr{2018}  \at{An alternative floating element design for skin-friction measurement of turbulent wall flows}.  \jt{Experiments in Fluids}  \bvol{59}~(10),  \pg{155}.

\bibitem[Finnigan(2000)]{Finnigan2000}
{\sc \au{Finnigan, J.}} \yr{2000}  \at{Turbulence in plant canopies}.  \jt{Annual Review of Fluid Mechanics}  \bvol{32}~(1),  \pg{519--571}.

\bibitem[Finnigan {\em et~al.\/}(2009)Finnigan, Shaw \& Patton]{finnigan2009turbulence}
{\sc \au{Finnigan, John~J}, \au{Shaw, Roger~H} \& \au{Patton, Edward~G}} \yr{2009}  \at{Turbulence structure above a vegetation canopy}.  \jt{Journal of Fluid Mechanics}  \bvol{637},  \pg{387--424}.

\bibitem[Foucaut {\em et~al.\/}(2004)Foucaut, Carlier \& Stanislas]{foucaut2004piv}
{\sc \au{Foucaut, J.-M.}, \au{Carlier, J.} \& \au{Stanislas, M.}} \yr{2004}  \at{{PIV} optimization for the study of turbulent flow using spectral analysis}.  \jt{Measurement Science and Technology}  \bvol{15}~(6),  \pg{1046}.

\bibitem[Ganapathisubramani {\em et~al.\/}(2005)Ganapathisubramani, Hutchins, Hambleton, Longmire \& Marusic]{ganapathisubramani2005investigation}
{\sc \au{Ganapathisubramani, B.}, \au{Hutchins, N.}, \au{Hambleton, W.T.}, \au{Longmire, E.K.} \& \au{Marusic, I.}} \yr{2005}  \at{Investigation of large-scale coherence in a turbulent boundary layer using two-point correlations}.  \jt{Journal of Fluid Mechanics}  \bvol{524},  \pg{57--80}.

\bibitem[Glegg \& Devenport(2017)]{Glegg_book}
{\sc \au{Glegg, S. A.~L} \& \au{Devenport, W.~J.}} \yr{2017} {\em Aeroacoustics of Low Mach Number Flow: Fundamentals, Analysis and Measurement\/}.  \publ{Academic Press Elsevier}.

\bibitem[Gul \& Ganapathisubramani(2021)]{gul2021revisiting}
{\sc \au{Gul, M.} \& \au{Ganapathisubramani, B.}} \yr{2021}  \at{Revisiting rough-wall turbulent boundary layers over sand-grain roughness}.  \jt{Journal of Fluid Mechanics}  \bvol{911}.

\bibitem[Hahn {\em et~al.\/}(2002)Hahn, Je \& Choi]{Hahn2002}
{\sc \au{Hahn, S.}, \au{Je, J.} \& \au{Choi, H.}} \yr{2002}  \at{Turbulent channel flow with permeable walls}.  \jt{J. Fluid Mech.}  \bvol{450},  \pg{259–285}.

\bibitem[Hutchins \& Marusic(2007)]{hutchins2007large}
{\sc \au{Hutchins, N.} \& \au{Marusic, I.}} \yr{2007}  \at{Large-scale influences in near-wall turbulence}.  \jt{Philosophical Transactions of the Royal Society A: Mathematical, Physical and Engineering Sciences}  \bvol{365}~(1852),  \pg{647--664}.

\bibitem[Jaiswal(2020)]{jaiswal2020etude}
{\sc \au{Jaiswal, Prateek}} \yr{2020}  \at{Etude exp{\'e}rimentale du bruit propre de profil a{\'e}rodynamique}. PhD thesis, Universit{\'e} de Sherbrooke.

\bibitem[Jaiswal {\em et~al.\/}(2020)Jaiswal, Moreau, Avallone, Ragni \& Pr{\"o}bsting]{jaiswal2020use}
{\sc \au{Jaiswal, Prateek}, \au{Moreau, St{\'e}phane}, \au{Avallone, Francesco}, \au{Ragni, Daniele} \& \au{Pr{\"o}bsting, Stefan}} \yr{2020}  \at{On the use of two-point velocity correlation in wall-pressure models for turbulent flow past a trailing edge under adverse pressure gradient}.  \jt{Physics of Fluids}  \bvol{32}~(10),  \pg{105105}.

\bibitem[Jim{\'e}nez(2004)]{jimenez2004turbulent}
{\sc \au{Jim{\'e}nez, Javier}} \yr{2004}  \at{Turbulent flows over rough walls}.  \jt{Annu. Rev. Fluid Mech.}  \bvol{36},  \pg{173--196}.

\bibitem[Kim \& Adrian(1999)]{kim1999very}
{\sc \au{Kim, Kyung~Chun} \& \au{Adrian, Ronald~J}} \yr{1999}  \at{Very large-scale motion in the outer layer}.  \jt{Physics of Fluids}  \bvol{11}~(2),  \pg{417--422}.

\bibitem[Krogstad \& Antonia(1994)]{krogstad1994structure}
{\sc \au{Krogstad, P.-{\AA}} \& \au{Antonia, R.A.}} \yr{1994}  \at{Structure of turbulent boundary layers on smooth and rough walls}.  \jt{Journal of Fluid Mechanics}  \bvol{277},  \pg{1--21}.

\bibitem[Krogstad {\em et~al.\/}(1992)Krogstad, Antonia \& Browne]{krogstad1992comparison}
{\sc \au{Krogstad, P-{\AA}}, \au{Antonia, RA} \& \au{Browne, LWB}} \yr{1992}  \at{Comparison between rough-and smooth-wall turbulent boundary layers}.  \jt{Journal of Fluid Mechanics}  \bvol{245},  \pg{599--617}.

\bibitem[Kuwata \& Suga(2017)]{kuwata2017direct}
{\sc \au{Kuwata, Y} \& \au{Suga, K}} \yr{2017}  \at{Direct numerical simulation of turbulence over anisotropic porous media}.  \jt{Journal of Fluid Mechanics}  \bvol{831},  \pg{41--71}.

\bibitem[Macdonald {\em et~al.\/}(1979)Macdonald, {El-Sayed}, Mow \& Dullien]{Macdonald1979}
{\sc \au{Macdonald, E.~F.}, \au{{El-Sayed}, M.~S.}, \au{Mow, K.} \& \au{Dullien, F. A.~L.}} \yr{1979}  \at{Flow through porous media-the ergun equation revisited}.  \jt{Ind. Eng. Chem. Fundam.}  \bvol{18},  \pg{199--–208}.

\bibitem[Manes {\em et~al.\/}(2011)Manes, Poggi \& Ridolfi]{manes2011turbulent}
{\sc \au{Manes, Costantino}, \au{Poggi, Davide} \& \au{Ridolfi, Luca}} \yr{2011}  \at{Turbulent boundary layers over permeable walls: scaling and near wall structure}.  \jt{Journal of Fluid Mechanics}  \bvol{687},  \pg{141--170}.

\bibitem[Mathis {\em et~al.\/}(2009)Mathis, Hutchins \& Marusic]{mathis2009large}
{\sc \au{Mathis, Romain}, \au{Hutchins, Nicholas} \& \au{Marusic, Ivan}} \yr{2009}  \at{Large-scale amplitude modulation of the small-scale structures in turbulent boundary layers}.  \jt{Journal of Fluid Mechanics}  \bvol{628},  \pg{311--337}.

\bibitem[Motlagh \& Taghizadeh(2016)]{motlagh2016pod}
{\sc \au{Motlagh, S.~Y.} \& \au{Taghizadeh, S.}} \yr{2016}  \at{Pod analysis of low reynolds turbulent porous channel flow}.  \jt{International Journal of Heat and Fluid Flow}  \bvol{61},  \pg{665--676}.

\bibitem[Nikuradse(1933)]{Nikuradse1933}
{\sc \au{Nikuradse, J.}} \yr{1933}  \at{Laws of flow in rough pipes}.  \jt{NACA TM}  \pg{p. 1292}.

\bibitem[Otsu(1979)]{otsu1979threshold}
{\sc \au{Otsu, Nobuyuki}} \yr{1979}  \at{A threshold selection method from gray-level histograms}.  \jt{IEEE transactions on systems, man, and cybernetics}  \bvol{9}~(1),  \pg{62--66}.

\bibitem[Placidi \& Ganapathisubramani(2018)]{placidi2018turbulent}
{\sc \au{Placidi, M.} \& \au{Ganapathisubramani, B.}} \yr{2018}  \at{Turbulent flow over large roughness elements: effect of frontal and plan solidity on turbulence statistics and structure}.  \jt{Boundary-layer meteorology}  \bvol{167}~(1),  \pg{99--121}.

\bibitem[Rosti {\em et~al.\/}(2015)Rosti, Cortelezzi \& Quadrio]{rosti2015direct1}
{\sc \au{Rosti, Marco~E}, \au{Cortelezzi, Luca} \& \au{Quadrio, Maurizio}} \yr{2015}  \at{Direct numerical simulation of turbulent channel flow over porous walls}.  \jt{Journal of Fluid Mechanics}  \bvol{784},  \pg{396--442}.

\bibitem[Sharma \& Garc{\'\i}a-Mayoral(2020{\natexlab{{\em a\/}}})]{sharma2020_sparse}
{\sc \au{Sharma, Akshath} \& \au{Garc{\'\i}a-Mayoral, Ricardo}} \yr{2020{\natexlab{{\em a\/}}}}  \at{Scaling and dynamics of turbulence over sparse canopies}.  \jt{Journal of Fluid Mechanics}  \bvol{888}.

\bibitem[Sharma \& Garc{\'\i}a-Mayoral(2020{\natexlab{{\em b\/}}})]{sharma2020_dense}
{\sc \au{Sharma, Akshath} \& \au{Garc{\'\i}a-Mayoral, Ricardo}} \yr{2020{\natexlab{{\em b\/}}}}  \at{Turbulent flows over dense filament canopies}.  \jt{Journal of Fluid Mechanics}  \bvol{888}.

\bibitem[Sillero {\em et~al.\/}(2014)Sillero, Jim{\'e}nez \& Moser]{sillero2014two}
{\sc \au{Sillero, Juan~A}, \au{Jim{\'e}nez, Javier} \& \au{Moser, Robert~D}} \yr{2014}  \at{Two-point statistics for turbulent boundary layers and channels at reynolds numbers up to $\delta+ ~\approx 2000$}.  \jt{Physics of Fluids}  \bvol{26}~(10),  \pg{105109}.

\bibitem[Spencer \& Hollis(2005)]{spencer2005correcting}
{\sc \au{Spencer, Adrian} \& \au{Hollis, David}} \yr{2005}  \at{Correcting for sub-grid filtering effects in particle image velocimetry data}.  \jt{Measurement Science and Technology}  \bvol{16}~(11),  \pg{2323}.

\bibitem[Squire {\em et~al.\/}(2016)Squire, {Morrill-Winter}, Hutchins, Schultz, Klewicki \& Marusic]{Squire2016}
{\sc \au{Squire, D.~T.}, \au{{Morrill-Winter}, C.}, \au{Hutchins, N.}, \au{Schultz, M.~P.}, \au{Klewicki, J.~C.} \& \au{Marusic, I.}} \yr{2016}  \at{Comparison of turbulent boundary layers over smooth and rough surfaces up to high reynolds numbers}.  \jt{Journal of Fluid Mechanics}  \bvol{795},  \pg{210--240}.

\bibitem[Suga {\em et~al.\/}(2010)Suga, Matsumura, Ashitaka, Tominaga \& Kaneda]{suga2010effects}
{\sc \au{Suga, K}, \au{Matsumura, Y}, \au{Ashitaka, Y}, \au{Tominaga, S} \& \au{Kaneda, M}} \yr{2010}  \at{Effects of wall permeability on turbulence}.  \jt{International Journal of Heat and Fluid Flow}  \bvol{31}~(6),  \pg{974--984}.

\bibitem[Townsend(1980)]{townsend1980structure}
{\sc \au{Townsend, AAR}} \yr{1980} {\em The structure of turbulent shear flow\/}.  \publ{Cambridge university press}.

\bibitem[Volino {\em et~al.\/}(2007)Volino, Schultz \& Flack]{volino2007turbulence}
{\sc \au{Volino, R.J.}, \au{Schultz, M.P.} \& \au{Flack, K.A.}} \yr{2007}  \at{Turbulence structure in rough-and smooth-wall boundary layers}.  \jt{Journal of Fluid Mechanics}  \bvol{592},  \pg{263--293}.

\bibitem[Wallace(2016)]{wallace2016quadrant}
{\sc \au{Wallace, J.~M.}} \yr{2016}  \at{Quadrant analysis in turbulence research: history and evolution}.  \jt{Annual Review of Fluid Mechanics}  \bvol{48},  \pg{131--158}.

\bibitem[Wangsawijaya {\em et~al.\/}(2023)Wangsawijaya, Jaiswal \& Ganapathisubramani]{wangsawijaya2023towards_JFM}
{\sc \au{Wangsawijaya, D.~D.}, \au{Jaiswal, P.} \& \au{Ganapathisubramani, B.}} \yr{2023}  \at{Towards decoupling the effects of permeability and roughness on turbulent boundary layers}.  \jt{Journal of Fluid Mechanics}  \bvol{967},  \pg{R2}.

\bibitem[Wesson {\em et~al.\/}(2003)Wesson, Katul \& Siqueira]{wesson2003quantifying}
{\sc \au{Wesson, K.~H.}, \au{Katul, G.~G.} \& \au{Siqueira, M.}} \yr{2003}  \at{Quantifying organization of atmospheric turbulent eddy motion using nonlinear time series analysis}.  \jt{Boundary-Layer Meteorology}  \bvol{106}~(3),  \pg{507--525}.

\bibitem[White \& Nepf(2007)]{white2007shear}
{\sc \au{White, Brian~L} \& \au{Nepf, Heidi~M}} \yr{2007}  \at{Shear instability and coherent structures in shallow flow adjacent to a porous layer}.  \jt{Journal of Fluid Mechanics}  \bvol{593},  \pg{1--32}.

\bibitem[Wu \& Christensen(2007)]{wu2007outer}
{\sc \au{Wu, Y} \& \au{Christensen, KT}} \yr{2007}  \at{Outer-layer similarity in the presence of a practical rough-wall topography}.  \jt{Physics of Fluids}  \bvol{19}~(8).

\bibitem[Wu \& Christensen(2010)]{wu2010spatial}
{\sc \au{Wu, Y} \& \au{Christensen, Kenneth~T}} \yr{2010}  \at{Spatial structure of a turbulent boundary layer with irregular surface roughness}.  \jt{Journal of Fluid Mechanics}  \bvol{655},  \pg{380--418}.

\bibitem[Yovogan \& Degan(2013)]{yovogan2013effect}
{\sc \au{Yovogan, J.} \& \au{Degan, G.}} \yr{2013}  \at{Effect of anisotropic permeability on convective heat transfer through a porous river bed underlying a fluid layer}.  \jt{Journal of Engineering Mathematics}  \bvol{81}~(1),  \pg{127--140}.

\end{thebibliography}



\end{document}